\documentclass[12pt]{article}
\usepackage{amssymb,amsmath,amsthm,mathrsfs,latexsym,amsxtra,graphicx,appendix,hyperref,setspace, subfig,
fix-cm,color,cite,hyperref,dsfont}
\numberwithin{equation}{section}
\usepackage{tikz}
\usetikzlibrary{positioning}
\usetikzlibrary{intersections} 
\newlength{\PicScale}
\newcommand{\red}[1]{\textcolor{red}{#1}}

\newcommand{\blue}[1]{\textcolor{blue}{#1}}

\newcommand{\msm}[1]{\mbox{\small$#1$}}

\newcommand{\mfn}[1]{\mbox{\footnotesize$#1$}}
\newcommand{\msc}[1]{\mbox{\scriptsize$#1$}}
\newcommand{\op}{\hspace{1pt}}

\usepackage{longtable, multirow, array}

\newcolumntype{M}[1]{>{\centering\arraybackslash}m{#1}}
\hypersetup{
    pdftitle={},
    pdfauthor={},
    pdfsubject={}
}

\begin{document} 

\newcommand{\comment}[1]{{\color{blue}[#1]}}

\newcommand{\bea}{\begin{eqnarray}}
\newcommand{\eea}{\end{eqnarray}}
\newcommand{\be}{\begin{equation}}
\newcommand{\ee}{\end{equation}}
\newcommand{\eq}[1]{(\ref{#1})}

\newcommand{\del}{\partial}
\newcommand{\delbar}{\overline{\partial}}
\newcommand{\zbar}{\overline{z}}
\newcommand{\wbar}{\overline{w}}
\newcommand{\vbar}{\overline{\varphi}}

\newcommand{\hf}{\frac{1}{2}}
\newcommand{\qrt}{\frac{1}{4}}
\newcommand{\bz}{{\mathbb Z}}
\newcommand{\R}{{\mathbb R}}
\newcommand{\C}{{\mathbb C}}
\newcommand{\A}{{\mathbb A}}
\newcommand{\N}{{\mathbb N}}
\newcommand{\bH}{{\mathbb H}}
\renewcommand{\P}{{\mathbb P}}
\newcommand{\Q}{{\mathbb Q}}
\newcommand{\tX}{\widetilde{X}}
\newcommand{\mO}{\Omega}
\newcommand{\mJ}{{\mathbb J}}
\def\taubar{\overline{\tau}}
\def\Tr{{\rm Tr}}
\def\qhat{\hat{q}_0}
\def\ap{\alpha^{\prime}}
\def\eg{{\emph{e.g.}~}}
\def\a{\alpha} 
\def\b{\beta} 
\def\g{\gamma} 
\def\G{\Gamma}
\def\e{\epsilon}
\def\h{\eta}
\def\th{\theta} 
\def\Th{\Theta}  
\def\k{\kappa}
\def\la{\lambda} 
\def\L{\Lambda} 
\def\m{\mu}
\def\n{\nu}
\def\r{\rho} 
\def\s{\sigma} 
\def\t{\tau}
\def\f{\phi} 
\def\F{\Phi} 
\def\w{\omega}
\def\W{\Omega} 
\def\v{\varphi} 
\def\z{\zeta}
\def\ts{\textstyle}

\def\lieg{{\mathfrak g}}

\newcommand{\cB}{{\mathcal B }}
\newcommand{\cW}{{\mathcal W }}            
\newcommand{\cM}{{\mathcal M }}            
\newcommand{\cF}{{\mathcal F }}            
\newcommand{\cC}{{\mathcal C }}
\newcommand{\cL}{{\mathcal L }}                
\newcommand{\cO}{{\mathcal O }}            
\newcommand{\cH}{{\mathcal H }}            
\newcommand{\cA}{{\mathcal A }}            
\newcommand{\cG}{{\mathcal G }}
\newcommand{\cN}{{\mathcal N }}            
\newcommand{\cY}{{\mathcal Y }}    
\newcommand{\cD}{{\mathcal D }} 
\newcommand{\cV}{{\mathcal V }}    
\newcommand{\cJ}{{\mathcal J }}    
\newcommand{\E}{{\mathcal E }}   
\newcommand{\B}{{\mathcal B}}
\newcommand{\cZ}{\mathcal Z}
\newcommand{\cZh}{\hat{\mathcal Z}}
\newcommand{\bt}{{\mathbb T}}
\newcommand{\tf}{\textstyle\frac}

\renewcommand{\d}{{\partial}}

\newcommand{\wl}{{\widetilde{\lambda}}}

\def\wt{\widetilde}

\newcommand{\Gr}{\ensuremath{\mbox{Gr}}}
\newcommand{\SG}{\ensuremath{\mbox{SG}}}
\newcommand{\TN}{\ensuremath{\mbox{TN}}}
\newcommand{\CY}{\ensuremath{\mbox{CY}}}

\newcommand{\vac}{|0\rangle}

\newcommand{\etc}{{\it etc.~}}

\newcommand{\one}{\mathbf{1}}
\newcommand{\two}{\mathbf{2}}
\newcommand{\three}{\mathbf{3}}
\newcommand{\eo}{\epsilon_1}
\newcommand{\et}{\epsilon_2}
\newcommand{\bp}{\mathbf{+}}
\newcommand{\bm}{\mathbf{-}}

\newcommand{\wb}{{\bar w}}
\newcommand{\zb}{{\bar z}}
\newcommand{\xb}{{\bar x}}
\newcommand{\hb}{{\bar h}}
\newcommand{\qb}{{\bar q}}

\begin{titlepage}
\begin{center}

\hfill \\
\hfill \\
\vskip 0.75in

{\Large 
	\bf Stringy Tachyonic Instabilities of Non-Supersymmetric Ricci Flat Backgrounds
}\\

\vskip 0.4in

{\large Bobby Samir Acharya${}^{a,b}$, Gerardo Aldazabal${}^{c,d,e}$, Eduardo Andr\'es${}^{c,e}$,\\Anamar\'ia Font${}^{f,g}$, Kumar Narain${}^{a}$, and Ida G.~Zadeh${}^{a}$
}\\
\vskip 4mm

${}^{a}$
{\it International Centre for Theoretical Physics, Strada Costiera 11, 34151 Trieste, Italy} \vskip 1mm
${}^{b}$
{\it Department of Physics, Kings College London, London, WC2R 2LS, UK} \vskip 1mm
${}^{c}$
{\it G. F\'isica CAB-CNEA, Centro At\'omico Bariloche, Av. Bustillo 9500, Bariloche, Argentina} \vskip 1mm
${}^{d}$
{\it Consejo Nacional de Investigaciones Cient\'ificas y T\'ecnicas {\rm(}CONICET{\rm)}} \vskip 1mm
${}^{e}$
{\it Instituto Balseiro, Universidad Nacional de Cuyo {\rm(}UNCUYO{\rm)}, Av. Bustillo 9500, R8402AGP,  Bariloche, Argentina} \vskip 1mm
${}^{f}$
{\it Fac. de Ciencias, Universidad Central de Venezuela, A.P.20513, Caracas 1020-A, Venezuela} \vskip 1mm
${}^{g}$
{\it Max-Planck-Institut f\"ur Gravitationsphysik, Albert-Einstein-Institut, 14476 Golm, Germany} \vskip 1mm

\end{center}

\vskip 0.35in

\begin{center} {\bf ABSTRACT } \end{center}
Superstring/M-theory compactified on compact Ricci flat manifolds have recently been conjectured to exhibit instabilities whenever the metrics do not have special holonomy. We use worldsheet conformal field theory to investigate instabilities of Type II superstring theories on compact, Ricci flat, spin 3-manifolds including a worldsheet description of their spin structures. The instabilities are signalled by the appearance of stringy tachyons at small radius and a negative (1-loop) vacuum energy density at large radius. We briefly discuss the extension to higher dimensions.
\vfill

\noindent \today

\end{titlepage}

\setcounter{page}{1}
\setcounter{tocdepth}{2}

\tableofcontents

\section{Introduction}
\label{section_intro}

Compactifying string and M-theory yields lower dimensional theories of quantum gravity whose physical features depend on
geometrical properties of the underlying compact manifolds. An important aspect of such a construction is the stability of the lower 
dimensional theory, an issue related to whether the theory is supersymmetric or not. 
Indeed, a key question about superstring/M-theory is: does it predict low energy 
supersymmetry, say, below the compactification or GUT scale ? In 
\cite{Acharya:2019mcu} this question was addressed for compactifications well 
approximated by Riemannian manifolds and it was pointed out that a necessary 
condition for an affirmative answer is that all physically stable, compact Ricci 
flat manifolds must have special holonomy. In 
real dimension three, this conjecture 
asserts that 
the 3-torus, $\bt^3$, with its supersymmetric spin structure is the only stable, 
compact Ricci flat manifold. There are six topological types of compact, 
orientable, Ricci flat 3-manifolds, labeled as $G1$-$G6$ in \cite{Pfaeffle}, and 
they admit a total of 28 spin structures. The 3-torus with totally periodic spin 
structure is the only supersymmetric case, see \cite[Theorem 2]{McInnes:1998wq}. 
 It was shown in \cite{Acharya:2019mcu} that 26 out of the 27  
non-supersymmetric classes of compactifications suffer from generalised Witten 
{\it Bubble of Nothing} instabilities \cite{Witten:1981gj}. The last case was 
also shown to admit such an instability in \cite{GarciaEtxebarria:2020xsr} which 
extended the analysis of \cite{Acharya:2019mcu} considerably. This proves the 
conjecture in dimension three.

One of the main motivations for the present paper is to further investigate these dimension three examples from the string theory 
worldsheet point of view. After all, all of these examples are locally flat and, hence, in principle should be amenable to exact worldsheet 
methods. Whilst the analysis of \cite{Acharya:2019mcu, GarciaEtxebarria:2020xsr} was valid at compactification lengths larger than the string length, we will be able to investigate here the small radius region, complementing the previous studies. We will find certain universal properties of the non-supersymmetric backgrounds including the appearance of tachyonic stringy winding states at small radius, which also seems very much consistent with the conjecture. At large radius the tachyons disappear and the 1-loop cosmological constant is negative, reflecting the existence of more massless bosons than fermions in the spectrum.

Spin structures played a crucial role in the analyses of \cite{Acharya:2019mcu, GarciaEtxebarria:2020xsr} and we will have to investigate how different spacetime spin structures are manifested at the level of the worldsheet conformal field theory and we will see the interplay between these and modular invariance. Spin structures modify the GSO projection leading to a dependence on string winding modes in addition to fermion number operators. Modular invariance then entails half-integer 
shifts in momentum modes of spacetime fermions, corresponding to anti-periodic boundary conditions.
More generally, one would like to construct worldsheet theories with target spaces being compact manifolds which admit multiple
spin structures. We consider the above three dimensional target spaces in the present paper and leave a detailed analysis of some higher 
dimensional cases for upcoming work \cite{upcoming}. 

Compactifications with anti-periodic spin structure on the circle have been already studied in the 
literature \cite{Rohm:1983aq, Kounnas:1989dk,Atick:1988si}. 
In dimension two, the only compact, orientable, closed Ricci flat manifold is the torus, $\bt^2$.  In dimension three one has Ricci flat orientable manifolds which are not diffeomorphic to tori. These manifolds are all of the form $(\bt^2\times S^1)/G$, where $G$ is a finite group which acts by rotating $\bt^2$, together with translations along the $S^1$. As such, $G$ acts freely on $\bt^2\times S^1$. Considering the anti-periodic spin structure only along the $S^1$ and periodic spin structures along the directions of the torus yields a higher dimensional generalisation of the models mentioned above. We refer to these as class I models. New models are constructed when we consider putting anti-periodic spin structures along all three directions. We denote these as class II models. As we will see, orbifold invariance imposes constraints on spin structures. We will determine the worldsheet realisation of class I and class II models and analyse the stability behaviour associated with their spin structures.
The resulting theories exhibit typical tachyonic instabilities as in many other non-supersymmetric string compactifications studied previously, see e.g.
\cite{Rohm:1983aq, Seiberg:1986by, Dixon:1986iz, AlvarezGaume:1986jb,
Nair:1986zn, Ginsparg:1986wr, Itoyama:1986ei, Kounnas:1989dk, Blum:1997gw, Kachru:1998hd,
Harvey:1998rc, Kachru:1998yy, Kachru:1998pg,Font:2002pq, Blaszczyk:2014qoa, Abel:2015oxa, Aaronson:2016kjm, Kaidi:2019tyf,Itoyama:2020ifw} 
for constructions in the NSR formalism.

The plan of this paper is as follows. In section \ref{section_T3} we first construct the modular invariant
partition function of the $(\bt^2\times S^1)/G$ orbifold theories and then compute their spectrum.
We also look into the operator interpretation and the consistency of the operator product expansion (OPE) in class II models.
In section \ref{section_lambda} we consider the 1-loop cosmological constant $\Lambda$ of the worldsheet theories. To begin 
we make an analytical estimation of $\Lambda$ in the limit where the radius of the circle is large and tachyons are absent in the spectrum. 
We then compute $\Lambda$ numerically for finite radius, which enables us to assess the onset of instability due to emergence of tachyons for each model. In section \ref{section_Td} we present some salient features of higher dimensional counterparts of these worldsheet theories, leaving the details of the analysis and results for upcoming work. Section \ref{conc} is devoted to a summary of our results and a discussion of potential interesting directions to pursue. Appendix \ref{app_thetas_lattices} contains definitions and properties of the Jacobi theta functions and lattice sums which we use throughout the paper.

\section{\texorpdfstring{$\bt^3/\bz_N$}{T3/ZN} backgrounds}
\label{section_T3}

In this section we consider compactification of type II strings on $\bt^3/\bz_N$, including all possible spin structures.
To begin, in subsection \ref{sub_T3} we discuss the purely toroidal case, corresponding to  
the models denoted as $G1$ in tables 2 and 3 of \cite{Acharya:2019mcu}. We will then construct toroidal $\bz_N$ orbifolds, which are 
the $G2$-$G5$ models in those tables, in subsection \ref{sub_T3_Z}. We classify these toroidal orbifolds into two classes: class I 
which can have anti-periodic spin structure only along one direction (concretely along the circle in $\bt^2\times S^1$), 
and class II which can have anti-periodic spin structures along all three directions. 
The latter are apparently new models in the literature and as such, we study their properties in more detail in subsection \ref{sub_opi}. 
We first analyse the operator interpretation in class II models and then show that the algebra of the worldsheet conformal field theory is consistent, 
namely that the operator product expansion (OPE) is closed.
Finally, the matter spectrum of our theories and in particular, the tachyonic states, will be  studied in section \ref{sub_T3_spec}.

\subsection{\texorpdfstring{$\bt^3$}{T3} with generic spin structure}\label{sub_T3}

The target space $\bt^3$ is taken to be the product of three orthogonal circles of radii $R_1$, $R_2$ and $R_3$, i.e.
the torus lattice generators are $a_1=(R_1,0,0)$, $a_2=(0,R_2,0)$, and $a_3=(0,0,R_3)$. The $B$-field background is
set to zero. By $SO(3,3)$ transformations we can reach a generic point in the moduli space of the toroidal compactification. 
The spin structures along the lattice generators are denoted
$(s_1,s_2,s_3)$. They can take the values $s_i=0$, or $s_i=1$, corresponding respectively
to space-time fermions being periodic or anti-periodic along the $a_i$ cycles. There are then in total 8 possible spin structures.
The generalization to the d-torus $\bt^d$ is straightforward and we will consider it in the following.

We want to write down the partition function for type II string theory compactified on $\bt^d$ with generic spin structures. 
As usual, the contribution of worldsheet NSR fermions will be expressed in terms of Jacobi Theta functions $\vartheta_{(a,b)}(z,\tau)$,
whose definition and properties are summarised in Appendix \ref{app_thetas}. Relevant combinations are
\be
{\rm V}= \frac{1}{2}\,\frac{(\vartheta_3^4-\vartheta_4^4)}{\eta^{12}}\ , \quad{\rm Sc} = \frac{1}{2}\,\frac{(\vartheta_3^4+\vartheta_4^4)}{\eta^{12}}\ ,\qquad
{\rm Sp} = \frac{1}{2}\,\frac{(\vartheta_2^4+\vartheta_1^4)}{\eta^{12}}\ ,\qquad{\rm Sp}' = \frac{1}{2}\,\frac{(\vartheta_2^4-\vartheta_1^4)}{\eta^{12}}\ ,
\label{V}
\ee
where the $\vartheta_i(\tau)$ are defined in \eqref{theta1234}, and $\eta(\tau)$ is the Dedekind eta function,
cf. \eqref{etadef}. We have dropped the $\tau$ dependence for conciseness.  The labels refer to $SO(8)$ conjugacy classes: scalar (Sc), vector (V), spinor (Sp), and conjugate spinor (${\rm Sp}'$). 

On the other hand, the contribution of worldsheet bosons will include a lattice sum over quantised winding and momenta.
It is useful to introduce the general sum over a $(d,d)$ lattice $\Gamma$ given by
\be\label{ZGamma}
 Z_{\Gamma}(\tau,x,y;u,v):=
 \sum_{k,w\in\bz^d} q^{\frac{1}{2}\sum\limits_{i=1}^d\big(\frac{k_i+x_i}{2 R_i}+(w_i+y_i)R_i\big)^2}
 \bar{q}^{\frac{1}{2}\sum\limits_{i=1}^d\big(\frac{k_i+x_i}{2 R_i}-(w_i+y_i)R_i\big)^2}
 e^{2\pi i \sum\limits_{i=1}^d\big(u_i(k_i+x_i)+v_i(w_i+y_i)\big)}\ ,
\ee
where $q=e^{2\pi i\tau}$, $x,y,u,v$ are $d$-dimensional real vectors, and we have set $\alpha'=\tf12$. 
Here we have chosen the $d$-dimensional target space torus to be a product orthogonal circles with radii $R_i$, and have
turned off the $B$-field background, but of course by $SO(d,d)$ transformations we can extend this definition to the entire moduli space of 
$d$-dimensional toroidal compactification. Properties of $Z_{\Gamma}(\tau,x,y;u,v)$ are collected in Appendix \ref{app_lattice}. 

The spacetime spin structures are realised by modifying the GSO projection as
\be\label{gso}
(-1)^{F_{L,R}} \longrightarrow (-1)^{F_{L,R}}(-1)^{\sum\limits_{i=1}^d s_i w_i}\ ,
\ee
where $w_i$ are the winding numbers along the $a_i$ cycles and $F_{L}$ and $F_{R}$ are the left and right moving fermion number operators, respectively. Modular invariance then implies that for space-time fermions the quantised momentum $k_i$ will be shifted by $\frac{s_i}2$.

This can be understood from a more general sigma model point of view{\footnote{We thank E. Witten for explaining this --- see also \cite{Atick:1988si}.}} as follows: A sigma model is based on maps $\Phi: \Sigma \rightarrow M$ from the worldsheet $\Sigma$ to spacetime $M$. Let $\Phi$ be a fixed map to $M$ with some chosen spin structure $S$. The pullback to $\Sigma$ of $S$ using $\Phi$ is a spin structure $s$ on $\Sigma$. Then the worldsheet theory will have a GSO projection $P_{GSO}$ associated to $s$.
Now, if $H^1(M, \bz_2)$ is non-trivial then there exists more than one spin structure on $M$. If $S^{\prime}$ is such a spin structure, then $S^\prime = S\otimes \varepsilon$, where $\varepsilon$ is a real line bundle and, pulled back to the worldsheet $s^\prime=s\otimes\epsilon$, where $\epsilon$ is some real line bundle on $\Sigma$. $\epsilon$ provides additional signs as fermions go around the 1-cycles of $\Sigma$ and this will be reflected in a modified GSO projection:
\be
P_{GSO} \longrightarrow P_{GSO} \otimes \epsilon
\ee
This is how \eqref{gso} arises.

Returning now to our specific models,
schematically, the type IIB partition function will contain pieces
\be
\begin{split}
{\rm(NS,NS)} &: ~{\rm(V,V)}\, Z^{(e)}_{\Gamma} + {\rm(Sc,Sc)}\, Z^{(o)}_{\Gamma},\\
{\rm(R,R)} &: ~~{\rm(Sp,Sp)}\, Z^{(e)}_{\Gamma}  + {\rm(Sp',Sp')}\, Z^{(o)}_{\Gamma}, \\
{\rm(NS,R)} &: ~~{\rm(V,Sp)}\, \hat{Z}^{(e)}_{\Gamma} + {\rm(Sc,Sp')}\, \hat{Z}^{(o)}_{\Gamma},\\
{\rm(R,NS)} &: ~~ {\rm(Sp,V)}\, \hat{Z}^{(e)}_{\Gamma} + {\rm(Sp',Sc)}\, \hat{Z}^{(o)}_{\Gamma},
\end{split}
\label{G1p}
\ee
where $Z^{(e)}_{\Gamma}$ and  $Z^{(o)}_{\Gamma}$ stand for the lattice sum  $Z_{\Gamma}(\tau,0,0;0,0)$
with $\sum_i s_i w_i$ being even and odd, respectively, whereas $\hat{Z}^{(e)} _{\Gamma}$ and $\hat{Z}^{(o)} _{\Gamma}$ are  
$Z^{(e)}_{\Gamma}$ and $Z^{(o)}_{\Gamma}$ with shifted momenta $ k_i \rightarrow k_i+ \tf{s_i}{2}$. More precisely,
\be
\begin{split}
\hspace*{-2mm}
Z^{(e)}_{\Gamma}&= \tfrac12\left[Z_\Gamma (\tau,0,0;0,0) +  Z_\Gamma (\tau,0,0;0,\tf{s}2)\right], \
Z^{(o)}_{\Gamma}= \tfrac12\left[Z_\Gamma (\tau,0,0;0,0) -  Z_\Gamma (\tau,0,0;0,\tf{s}2)\right], \\
\hspace*{-2mm}
\hat{Z}^{(e)}_{\Gamma}&= \tfrac12\left[Z_\Gamma (\tau,\tf{s}2,0;0,0) +  Z_\Gamma (\tau,\tf{s}2,0;0,\tf{s}2) \right], \ 
\hat{Z}^{(o)}_{\Gamma}= \tfrac12\left[Z_\Gamma (\tau,\tf{s}2,0;0,0) -  Z_\Gamma (\tau,\tf{s}2,0;0,\tf{s}2) \right],
\end{split}
\label{zgeo}
\ee
where $s$ is a vector with components $s_i \in \{0,1\}$, and $Z_{\Gamma}(\tau,x,y;u,v)$ is the lattice sum defined in \eqref{ZGamma}.

The spin-structures on the $d$-cycles of $\bt^d$ are encoded in the vector $s$.
As expected, spacetime supersymmetry is broken unless $s=0$, in which case the odd lattice sums vanish while the even ones
reduce to $Z_\Gamma (\tau,0,0;0,0)$. On the contrary, the spectrum is non-supersymmetric whenever one, or several, 
of the spin structures takes the value $s_i=1$. In particular, there are tachyons for generic radii.

In order to write a compact expression for the full partition function we start by
substituting \eqref{V} and \eqref{zgeo} in eq. (\ref{G1p}) to obtain
\be
\begin{split}
{\rm(NS,NS)} &: ~\tfrac{1}{4}\, \big[(3,3)+(4,4)\big] \,Z_\Gamma (\tau,0,0;0,0)
-\tfrac{1}{4}\, \big[(3,4)+(4,3)\big] \, Z_\Gamma (\tau,0,0;0,\tf{s}2), \\
{\rm(R,R)} &: ~~\tfrac{1}{4}\, \big[(2,2)+(1,1)\big]  \,Z_\Gamma (\tau,0,0;0,0)
+ \tfrac{1}{4}\, [(2,1)+(1,2)] \, Z_\Gamma (\tau,0,0;0,\tf{s}2), \\
{\rm(NS,R)}&: ~~\tfrac{1}{4}\, \big[(3,2)-(4,1)\big] \, Z_\Gamma (\tau,\tf{s}2,0;0,0)
+ \tfrac{1}{4} \, \big[(3,1)-(4,2)\big] \, Z_\Gamma (\tau,\tf{s}2,0;0,\tf{s}2), \\
{\rm(R,NS)} &: ~~\tfrac{1}{4}\, \big[(2,3)-(1,4)\big] \, Z_\Gamma (\tau,\tf{s}2,0;0,0)
+ \tfrac{1}{4}\, \big[(1,3)-(2,4)\big] \, Z_\Gamma (\tau,\tf{s}2,0;0,\tf{s}2).
\end{split}
\label{G1}
\ee
where $(i,j):=\vartheta_i^4(\tau)\,\vartheta_j^4(\bar\tau)/|\eta(\tau)|^{24}$.

From \eqref{G1} it then follows that the partition function for the $d$-dimensional toroidal models is given by
\bea\label{ZTd}
&&\cZ_{\bt^d}=\frac{1}{4}\int_{\cal F}  \frac{d^2 \tau}{\tau_2^{6-\frac{d}{2}}}\;\frac{1}{|\eta^{12}(\tau)|^2}
\sum_{\alpha_L,\beta_L,\alpha_R,\beta_R=\{0,\frac12\}} \!\!\!\! {\mathcal C}(\alpha_L, \beta_L, \alpha_R, \beta_R)
\times\qquad\;\;\\
&&\qquad\qquad\qquad\qquad\qquad\;\times\;\vartheta^4_{(\alpha_L,\beta_L)}(\tau)\;\bar\vartheta^4_{(\alpha_R,\beta_R)}(\bar\tau)\,\;
Z_{\Gamma}\big(\tau,(\alpha_L+\alpha_R)s,0;0,(\beta_L+\beta_R)s\big)\ ,\nonumber
\eea
where $\cal F$ is the $SL(2,\bz)$ fundamental domain, and the 
$\vartheta_{(\alpha,\beta)}(\tau):=\vartheta_{(\alpha,\beta)}(0,\tau)$ 
are defined in Appendix \ref{app_thetas}.
To avoid cluttering we introduced the constant ${\mathcal C}(\alpha_L, \beta_L, 
\alpha_R, \beta_R)$, which reads{\footnote{The factor 
$(-1)^{4(\alpha_L\beta_L+\alpha_R\beta_R)}$ is associated with the choice of 
$\vartheta^4_2+\vartheta^4_1$ and $\bar\vartheta^4_2+\bar\vartheta^4_1$ in the 
GSO projection in the Ramond sector for type IIB models. The other choice, 
namely a relative minus sign between $\vartheta^4_1$ and $\vartheta^4_2$ in both 
left \emph{and} right moving Ramond sectors would just replace this factor with 
(+1). For type IIA models, the GSO projection has opposite relative signs in the 
left and right moving Ramond sectors which would then result in factors 
$(-1)^{4(\alpha_L\beta_L)}$ or $(-1)^{4(\alpha_R\beta_R)}$ and corresponds to 
exchanging ${\rm Sp} \leftrightarrow {\rm Sp}'$ in the right moving sector of 
eq. (\ref{G1p}). Note that this factor is modular 
invariant.}\label{GSOfootnote}} 
\be\label{ccdef}
{\mathcal C}(\alpha_L, \beta_L, \alpha_R, \beta_R):=(-1)^{2(\alpha_L+\alpha_R)}(-1)^{2(\beta_L+\beta_R)}(-1)^{4(\alpha_L\beta_L+\alpha_R\beta_R)}\, .
\ee
Modular invariance of the partition function is then straightforward to establish using transformation properties of the $\vartheta$ functions, as well as of 
the lattice sum $Z_{\Gamma}$, outlined in Appendix \ref{app_thetas_lattices}.

Although our main focus is on $\cZ_{\bt^3}$, we would like to briefly
comment on the case of circle compactification, i.e. $d=1$. For non-standard spin structure $s_1=1$,
the partition function in \eqref{ZTd} agrees with well known results obtained from Scherk-Schwarz reductions 
\cite{Rohm:1983aq, Kounnas:1989dk}, or by modifying the GSO projection \cite{Atick:1988si}. It has also been considered as
an interpolating model \cite{Blum:1997gw}, which breaks supersymmetry at finite $R_1$ but, as one
can easily check, tends to type IIB at $R_1 \to \infty$, and to type 0B at $R_1 \to 0$.

\subsection{\texorpdfstring{$(\bt^2\times S^1)/\bz_N$}{T2xS1/ZN} with general spin structure}
\label{sub_T3_Z}

We now consider quotients of $\cZ_{\bt^3}$ by a discrete symmetry group $\bz_N$ generated by $g$ which involves a rotation by $2\pi/N$
in the plane spanned by basis vectors $a_1$ and $a_2$, together with a translation by $a_3/N$ in the orthogonal direction, 
see \cite[table 2]{Acharya:2019mcu}. The data in our notation is conveniently summarised in Table \ref{table3bobby}.
Notice that the orbifolds are freely acting due to the translation.
In momentum space the translation part of $g$ will be of the form $e^{2 \pi i \frac{k_3}N}$ on the bosons. 
On fermions, the rotational part of $g$, say ${\mathcal R}_g$, is lifted to  $Spin(3)=SU(2)$ according to 
${\mathcal R}_g \to (-1)^{s_g}\, e^{\frac{i\pi}N \sigma_3}$, where $\sigma_3$ is the third Pauli matrix and $s_g\in \{0,1\}$. 
The translational part of $g$ acting on fermions is $e^{\frac{2\pi i}{N}(k_3+\frac{s_3}{2})}$. Altogether, the action on 
fermions is generated by
\begin{equation}\label{gdef}
g\Big|_{\text{fermions}} =e^{\frac{2\pi i}{N}(k_3+\frac{s_3}{2})} e^{i\pi s_g} e^{i\frac{\pi}N \sigma_3}\ .
\end{equation}
Thus, in all these $\bz_N$ orbifolds, the requirement that $g^N=1$ leads to the condition
\begin{equation}
(-1)^{N s_g+s_3}=-1\ .
\label{condition}
\end{equation}
We conclude that necessarily $s_3=1$ for $N$ even, while $s_g=1+s_3$ mod 2 for $N=3$ .

The rotational part of $g$  must be an automorphism of the projected lattices in 
$Z^{(e)}_{\Gamma}$, $ Z^{(o)}_{\Gamma}$, $\hat{Z}^{(e)}_{\Gamma}$, and  $\hat{Z}^{(o)}_{\Gamma}$. 
This means it must be  an automorphism of the original lattice $\Gamma$ together with a grading defined by $s_i$. 
This is because in the spectrum \eqref{G1}, $\sum_i s_i w_i$ even and odd appears with different $SO(8)$ conjugacy classes.
For $\bz_2$, as $(a_1,a_2)\rightarrow (-a_1,-a_2)$, the grading defined by $s_i$ is automatically satisfied for any $s_i$. 
However, $g^2=1$ on fermions implies that $s_3=1$,  but $s_1$, $s_2$, and $s_g$ are arbitrary. Thus there are 8 possibilities. 
For $\bz_4$, as $(a_1,a_2)\rightarrow (a_2,-a_1)$ invariance of the  grading defined by $s_i$ requires $s_1 =s_2$, whereas 
$g^4=1$ again enforces $s_3=1$, so there are 4 possible spin structures. For $\bz_3$ the basis vectors are transformed 
as $(a_1,a_2) \rightarrow (a_2,-a_1-a_2)$, which implies $s_2=s_1 $ and $s_2=s_1+s_2$ mod 2. Therefore, in the $\bz_3$ case 
$s_1=s_2=0$, and $s_g=1+s_3$ mod 2 from \eqref{condition}. 
For $\bz_6$, actually equal to $\bz_2 \times \bz_3$, we again have $s_1=s_2=0$, and necessarily
$s_3=1$. So again there are 2 possibilities given by $s_g$ equal to 0 or 1. The results for all possible 
spin structures are collected in Table \ref{table3bobby}.
They completely agree with Theorem 3.3 in \cite{Pfaeffle}, see also
\cite[table 3]{Acharya:2019mcu}{\footnote{The case of the 3-manifold with holonomy group $\bz_2 \times \bz_2$ is more complicated (as in the RNS formulation, we will not be able to bosonise the worldsheet fermions and  will have to deal with  Majorana fermions) and will not be discussed here.}}.

\begin{table}[ht]
\renewcommand{\baselinestretch}{1.5}
\setlength\tabcolsep{3pt}
\centering
\footnotesize{
\begin{tabular}{|c|c|c|c|c|c|}
\hline
\scriptsize{Model} & \scriptsize{Group} & \scriptsize{Lattice} & \scriptsize{Spin structures} & \scriptsize{\#}& \scriptsize{Class} \\
\hline
$G1$ & $\mathds 1$ & $a_1=(R_1,0,0)$, $a_2=(0,R_2,0)$, $a_3=(0,0,R_3)$ &$s_1, s_2, s_3\in \{0,1\}$  & 8 &  ---\\
\hline
$G2$ & $\bz_2$ &$a_1=(R_1,0,0)$, $a_2=(0,R_2,0)$, $a_3=(0,0,R_3)$ & $s_1,s_2 \in \{0,1\},\ s_3=1, \ s_g \in \{0,1\}$  & 8 & I, II\\
\hline
$G3$ & $\bz_3$ & $a_1=(L,0,0)$, $a_2=(-\tf{1}2L,\tf{\sqrt3}2L,0)$, $a_3=(0,0,R_3)$ & $s_1=s_2=0,\ s_3 \in\{0,1\},\ s_g=s_3+1$  & 2 & I\\
\hline
$G4$ & $\bz_4$ & $a_1=(L,0,0)$, $a_2=(0,L,0)$, $a_3=(0,0,R_3)$ & $s_1=s_2\in \{0,1\},\ s_3=1,\ s_g\in \{0,1\}$ & 4 &  I, II\\
\hline
$G5$ & $\bz_6$ &$a_1=(L,0,0)$, $a_2=(\tf{1}2L,\tf{\sqrt3}2L,0)$, $a_3=(0,0,R_3)$ &  $s_1=s_2=0,\ s_3=1,\ s_g \in\{0,1\}$ & 2  & I\\
\hline
\end{tabular}
}
\caption{Spin structures on $(\bt^2\times S^1)/\bz_N$. $s_i=0$ and $s_i=1$ correspond respectively to periodic and anti-periodic spin structures along the $a_i$ cycles, whereas $s_g$ appears in the lift of the rotation by $2\pi/N$ to $Spin(3)$. 
For class I models $(s_1,s_2)=(0,0)$, while $(s_1,s_2)\ne(0,0)$ for class II .}
\label{table3bobby}
\end{table}

The orbifold partition function must include all sectors $(g^r,g^p)$, $r,p\in\{0,\cdots ,N-1\}$,
where the first and second entries refer to the boundary conditions along the worldsheet $\sigma$ and $t$ directions, respectively 
\cite{Dixon:1985jw, Dixon:1986jc}. More precisely, the full partition function $\cZ$ takes the form
\be\label{T3ZNclass12}
\cZ=\frac{1}{N}\sum_{r=0}^{N-1} \sum_{p=0}^{N-1} \cZ_{(g^r,g^p)}\ .
\ee
The sum in $r$ is over twisted sectors while the sum in $p$ implements the projection over the $\bz_N$ action.
In operator language, the term $\cZ_{(g^r,g^p)}$
computes ${\rm Tr}_{{\cal H}_r} g^p q^{L_0} \bar{q}^{\bar{L}_0}$, where ${\cal H}_r$ is the $g^r$  twisted Hilbert space. 
Since we already know how $ g^p$ acts on the untwisted Hilbert space ${\cal H}_0$, we can unambiguously calculate the partition function in the $(1,g^p)$ sectors. By the $S$-transformation, $\tau \rightarrow-\tf{1}{\tau}$, we can  obtain the results for $(g^p,1)$. Repeatedly
applying the $T$-transformation, $\tau \rightarrow \tau+1$, gives the sectors  $(g^p, g^{m p})$. 
Now, if $m$ is the least positive integer such that $m\,r= 0~{\rm{mod}}~ N$, then the partition function in the $(g^r,1)$ sector should be invariant under $\tau \rightarrow \tau+m$. This requirement gives the  level matching condition $m(L_0-\bar{L}_0) \in\bz$. It follows that if 
$N$ is prime, then in all the twisted Hilbert spaces (i.e. $r \neq 0$),~$N(L_0-\bar{L}_0) \in\bz$ ensures modular invariance \cite{Vafa:1986wx}.

If $N$ is not prime, \eg $N=6$, the $g^2, g^3, g^4$ twisted sectors can have some subtleties. For instance,
in $\bz_6$ the $(g^3,1)$ sector is unambiguously obtained by $S$ transformation on $(1,g^3)$ sector, but $\tau\rightarrow \tau+n$ can only determine $(g^3,g^3)$. 
The sectors $(g^3,g^p)$, with $p=1,2$, can be found acting with $(TST)^{(3-p)}$ transformations on $(g,g^p)$. 
Then $T$-transformations will give  $(g^3,g^4)$ and $(g^3,g^5)$. 
In all the three cases, $(g^3,g)$, $(g^3,g^2)$ and $(g^3,g^3)$, the level matching condition $2(L_0-\bar{L}_0) \in\bz$ 
ensures modular invariance. However, one still has to check that there is a proper operator interpretation,
namely, that there exists a definition of the $g$-action on ${\cal H}_3$ such that the partition function in the
$(g^3, g^p)$ sectors obtained by modular transformations indeed computes 
${\rm Tr}_{{\cal H}_3 } g^p  q^{L_0} \bar{q}^{\bar{L}_0}$ for $p=1,2,3$. 
For the well studied  toroidal Abelian orbifolds with supersymmetric spin structure, such an operator interpretation exists \cite{AlvarezGaume:1986es}. 
Since $\bz_N$ acts as a rotation only on $\bt^2$, these issues will not arise  when the spin structures $s_1$ and $s_2$ are
trivial. We will refer to models with $(s_1,s_2)=(0,0)$ as class I, and with $(s_1,s_2)\ne(0,0)$ as class II.
Class I models can occur for all $N$, but class II only for $N=2,4$, according to previous findings condensed in 
Table \ref{table3bobby}.
In the next sections we will discuss in more detail the differences between the two classes. For class II we will also
address the question of the $g$-action in twisted sectors as well as the closure of their operator algebra in subsection \ref{sub_opi}.

\subsubsection{$(1,1)$ sector}\label{T3_ZN_11}
When $r=0$ and $p=0$ we have the purely toroidal partition function with periodic or antiperiodic fermions
along the cycles $a_i$ according to whether $s_i=0$ or $s_i=1$, respectively. Thus, $\cZ_{(1,1)}$ is obtained
by setting $d=3$ in eq. \eqref{ZTd}. Actually, the result in  \eqref{ZTd} is valid when $\bt^3$ is the orthogonal product of
3 circles. However, for the orbifolds with $N=3,6$, this is not the case since $\bz_N$ has to be an automorphism of the
$\bt^2$ lattice whose basis vectors $\{a_1, a_2\}$ are given in Table \ref{table3bobby}. In these two orbifolds we just have to replace
$ Z_{\Gamma}(\tau,x,y;u,v)$ defined in \eqref{ZGamma} by a generalization that can be expressed as
\be
Z_{\Gamma}(\tau,x,y;u,v)= Z_{\Gamma_{(2,2)}}(\tau, {\bf x}, {\bf y}; {\bf u}, {\bf v}) \, Z_{\Gamma_{(1,1)}}(\tau,x_3,y_3;u_3,v_3) .
\ee
The 1-dimensional sum $Z_{\Gamma_{(1,1)}}$ conforms to \eqref{ZGamma}, while ${\bf x}$, ${\bf y}$, ${\bf u}$,
${\bf v}$ are 2-dimensional vectors and 
\be
Z_{\Gamma_{(2,2)}}(\tau,{\bf x}, {\bf y}; {\bf u}, {\bf v})= 
\sum_{k_i, w_i \in \bz^2} q^{\frac12 P_L^2} \bar q^{\frac12 P_R^2}
 e^{2\pi i \sum\limits_{i=1}^2\big(u_i(k_i+x_i)+v_i(w_i+y_i)\big)} . 
 \label{zt2}
\ee
Turning off the $B$-field background to simplify,  $P_{L,R} = \left(\frac{k_i + x_i}{2} \pm a_i \cdot a_j (w_j+ y_j) \right) a^{*i}$, 
where the $a^{*i}$ are the dual lattice vectors.

It is important to note that we are considering backgrounds of the form $\bt^3=\bt^2\times S^1$, 
where the rotational part ${\mathcal R}_g$ acts only on $\bt^2$. We have therefore assumed {\it factorised} lattices for 
$\bt^3$ such that in all $(g^r,g^p)$ sectors the invariant sublattices of $\Gamma$ are even and self-dual.
A systematic analysis of higher dimensional cyclic orbifold models of the form $(\bt^d\times S^1)/\bz_N$ will be 
presented in upcoming work \cite{upcoming}; see section \ref{section_Td} for more discussion on this.

\subsubsection{$(1,g^p)$ sector}\label{T3_ZN_1gr}
Using the action of $g$ on bosons and fermions, we can now construct the partition functions in these sectors. 
Upon complexification, $g^p$ acts on the $\bt^2$ bosons by a phase $e^{2\pi i \frac{p}{N}}$. Thus, instead of $(\eta\bar\eta)^{-2}$,
these bosons will give:
\begin{equation}
|q^{-\frac1{12}}|^2\prod_{n=1}^\infty\Big|(1-e^{\frac{2\pi i p}{N}}q^n)^{-1}(1-e^{-\frac{2\pi i p}{N}}q^n)^{-1}\Big|^2=
\Big|2\sin(\tf{\pi p}N)\,\frac{\eta(\tau)}{\vartheta_{(\frac{1}{2},\frac{1}{2})}(\frac{p}{N},\tau)}\Big|^2\ ,
\end{equation}
where $p\in\bz$ and $0<p<N$.
Next, we can group the 8 lightcone fermions into 4 complex fermions and bosonise them. Then the rotational part of $g^p$ acts as:
\begin{eqnarray}
\vartheta_{(0,\beta)}(\tau)^4& \longrightarrow &  \vartheta_{(0,\beta)}(\tau)^3\;\vartheta_{(0,\beta)}(\tf{p}{N},\tau)\ ,\nonumber\\
\vartheta_{(\frac{1}{2},\beta)}(\tau)^4& \longrightarrow & e^{i\pi ps_g} \vartheta_{(\frac{1}{2},\beta)}(\tau)^3\;\vartheta_{(\frac{1}{2},\beta)}(\tf{p}{N},\tau)\ ,
\end{eqnarray} 
where in the second line $e^{i\pi p s_g}$ appears because $\vartheta_{(\frac12,\beta)}(\tau)^4$ arises from the spacetime spinor
$\text{spinor}'$ and  $g$ includes this factor; see the discussion above eq. (\ref{gdef}). 
The translational part of $g^p$ acts as $e^{2\pi i \frac{p}{N} k_3}$ on bosons and $e^{2\pi i \frac{p}{N}(k_3 +\frac{s_3}2)}$
on fermions. Finally, the GSO projection is modified according to \eqref{gso}.

Combining all ingredients, the partition function in the $(1,g^p)$ sector is found to be:
\begin{align}\label{Z0r}
&\cZ_{(1,g^p)}=\frac{1}{4}\int_{\cal F}  \frac{d^2 \tau}{\tau_2^{\frac{9}{2}} }\,\bigg|\frac{2\sin(\frac{\pi p}N)}{\eta^9(\tau)\,
\vartheta_{(\frac{1}{2},\frac{1}{2})}(\frac{p}{N},\tau)}\bigg|^2
\!\!\! \sum_{\alpha_L,\beta_L,\alpha_R,\beta_R=\{0,\frac12\}} 
\hspace*{-10mm} {\mathcal C}(\alpha_L, \beta_L, \alpha_R, \beta_R) \, 
[\epsilon \, (-1)^{ps_g}]^{2(\alpha_L+\alpha_R)}\times 
\notag \\[-5mm]
&{} \\
&\times\;\vartheta^3_{(\alpha_L,\beta_L)}(\tau)\vartheta_{(\alpha_L,\beta_L)}(\tf pN,\tau)
\;\bar\vartheta^3_{(\alpha_R,\beta_R)}(\bar\tau)\bar\vartheta_{(\alpha_R,\beta_R)}(\tf pN,\bar\tau)\;
Z_{\Gamma}\big(\tau,(\alpha_L+\alpha_R)s_3,0;\tf{p}{N},(\beta_L+\beta_R)s_3\big), \notag
\end{align}
where  $\Gamma$ is now the $(1,1)$ lattice corresponding to the $S^1$ and $\mathcal C$ is defined in eq. (\ref{ccdef}). 
The $\bt^2$ lattice drops out because the $g^p$ boundary condition forbids both quantised momenta and windings.
This further implies that in class II models with $(s_1,s_2) \neq (0,0)$, fermions, i.e. (NS,R) and (R,NS), will not contribute. 
The reason is that for fermions the quantised momenta $(k_1, k_2)$ cannot be zero because they have to be shifted 
by $\frac{1}{2}(s_1,s_2)$. This is taken into account by introducing
the parameter $\epsilon$ to keep track of spacetime bosons and fermions. By  
expanding $\cZ_{(1,g^p)}$ in powers of $\epsilon$ we see that even powers (i.e. 1 and $\epsilon^2$) come with bosons, and the odd powers come with spacetime fermions.
In class I models, $(s_1,s_2)=(0,0)$, both spacetime bosons and fermions contribute to $\cZ_{(1,g^p)}$ so we obtain the result 
by just setting $\epsilon=1$. On the other hand, in class II,  the absence of spacetime fermions  in $\cZ_{(1,g^p)}$
follows by expanding in powers of $\epsilon$, discarding the odd power terms, and then setting $\epsilon=1$.

One notable feature of class II models is that the dependence on $s_g$ drops out completely, as
can be easily seen from \eqref{Z0r}, where $s_g$ only appears with the linear terms in $\epsilon$.
The physical reason is simply that in the $(1,g^p)$ sectors there are no spacetime fermions.
Notice also that in class II, the factor $\epsilon^{2(\alpha_L+\alpha_R)}$ can be replaced by a
projector $\frac12(1+(-1)^{2(\alpha_L+\alpha_R)})$. In other words, the partition function of class II 
models can be obtained by inserting this projector into the partition function of class I models in which $\epsilon=1$.

One can check that $\cZ_{(1,g^p)}$ is invariant under $p \rightarrow p+N$ provided the condition in eq. \eqref{condition} is satisfied, 
i.e.  $(-1)^{N s_g+s_3}=-1$. With the same condition it is also invariant under $p \rightarrow N-p$, 
apart from $\vartheta_1$ terms (which pick a minus sign) but the latter anyway vanish at the level of the partition function.

\subsubsection{$(g^r,1)$ sectors}\label{T3_ZN_gr1}

The partition function in the $(g^r,1)$ sectors is obtained from the $S$ modular transformation of $\cZ_{(1,g^r)}$. 
Applying properties of $Z_\Gamma$ in eq. \eqref{SonZ} and of the $\vartheta$ functions in eq. \eqref{thetamod} leads to:
\begin{align}\label{Zr0}
&\!\!\! \!\! \cZ_{(g^r,1)}=\frac{1}{4}\int_{\cal F}  \frac{d^2\tau}{\tau_2^{\frac{9}{2}} }\,\bigg|\frac{2\sin(\frac{\pi r}N)}{\eta^9(\tau)\,
\vartheta_{(\frac{1}{2},\frac{1}{2})}(\frac{r\tau}{N},\tau)}\bigg|^2
\!\!\!\!  \sum_{\alpha_L,\beta_L,\alpha_R,\beta_R=\{0,\frac12\}}
\hspace*{-10mm}   {\mathcal C}(\alpha_L, \beta_L, \alpha_R, \beta_R) 
[\epsilon\, (-1)^{rs_g}\, e^{-\pi i \frac rNs_3}]^{2(\beta_L+\beta_R)} \times \notag \\[-5mm]
&{} \\
&\times\;\vartheta^3_{(\alpha_L,\beta_L)}(\tau)\vartheta_{(\alpha_L,\beta_L)}(\tf{r\tau}N,\tau)
\;\bar\vartheta^3_{(\alpha_R,\beta_R)}(\bar\tau)\bar\vartheta_{(\alpha_R,\beta_R)}(\tf{r\bar\tau}N,\bar\tau)\;
Z_{\Gamma}\big(\tau,(\alpha_L+\alpha_R)s_3,\tf{r}{N};0,(\beta_L+\beta_R)s_3\big)\ , \notag
\end{align}
where we used \eqref{period}.

It is instructive to write $\cZ_{(g^r,1)}$ in a way such that the spectrum of states is manifest. To this end
we introduce $Z^{(e)}_{\Gamma}(x,\frac{r}{N};0,0)$ and $Z^{(o)}_{\Gamma}(x,\frac{r}{N};0,0)$, defined to be the 1-dimensional 
lattice sum $Z_{\Gamma}(\tau,x,\frac{r}{N};0,0)$ restricted to windings $w_3$ with $(-1)^{w_3 s_3} =\pm 1$, respectively 
(if $s_3=0$ then $Z^{(o)}_{\Gamma}$ is null and $Z^{(e)}_{\Gamma}= Z_\Gamma$). We can then recast \eqref{Zr0} as:

\be
\begin{split}
&\cZ_{(g^r,1)}=  \frac{1}{4}\int_{\cal F}  \frac{d^2 \tau}{\tau_2^{\frac{9}{2}} }\,
\bigg|\frac{2\sin(\frac{\pi r}N)}{\eta^9(\tau)\,\vartheta_{(\frac{1}{2},\frac{1}{2})}(\frac{r\tau}{N},\tau)}\bigg|^2\times \\
&\times\,\bigg[{\Big\{}(|\cA|^2+|\cB|^2)\,Z^{(e)}_{\Gamma}(\tau,0,\tf{r}{N};0,0)+(|\cA'|^2+|\cB'|^2)\,Z^{(o)}_{\Gamma}(\tau,0,\tf{r}{N};0,0){\Big\}}+ \\
&\;\;-\,\Big\{ (\cA \bar{\cB}+\cB \bar{\cA})\,Z^{(e)}_{\Gamma}(\tau,\tf{s_3}{2},\tf{r}{N};0,0)+(\cA' \bar{\cB}'+\cB' \bar{\cA}'\,Z^{(o)}_{\Gamma}(\tau,\tf{s_3}{2},\tf{r}{N};0,0)\Big{\}}\bigg]\ , 
\end{split}
\label{Zr0explicit}
\ee
where 
\be\label{ABdef}
\begin{split}
\cA&:=\vartheta_{(0,0)}^3(\tau) \vartheta_{(0,0)}(\tf{r\tau}{N},\tau)-
\epsilon (-1)^{r s_g} \vartheta_{(0,\frac{1}{2})}^3(\tau)\vartheta_{(0,\frac{1}{2})}(\tf{r\tau}{N},\tau)\ ,\\
\cA'&:=\vartheta_{(0,0)}^3(\tau)\vartheta_{(0,0)}(\tf{r\tau}{N},\tau)+
\epsilon (-1)^{r s_g} \vartheta_{(0,\frac{1}{2})}^3(\tau)\vartheta_{(0,\frac{1}{2})}(\tf{r\tau}{N},\tau)\ ,\\
\cB&:=\vartheta_{(\frac{1}{2},0)}^3(\tau)\vartheta_{(\frac{1}{2},0)}(\tf{r\tau}{N},\tau)+
\epsilon (-1)^{r s_g} \vartheta_{(\frac{1}{2},\frac{1}{2})}^3(\tau)\vartheta_{(\frac{1}{2},\frac{1}{2})}(\tf{r\tau}{N},\tau)\ ,\\
\cB'&:=\vartheta_{(\frac{1}{2},0)}^3(\tau)\vartheta_{(\frac{1}{2},0)}(\tf{r\tau}{N},\tau)-
\epsilon (-1)^{r s_g} \vartheta_{(\frac{1}{2},\frac{1}{2})}^3(\tau)\vartheta_{(\frac{1}{2},\frac{1}{2})}(\tf{r\tau}{N},\tau)\ .
\end{split}
\ee
The first curly bracket in \eqref{Zr0explicit} is the partition function of the spacetime bosons and the second curly bracket that of the
spacetime fermions. Notice that the latter comes with an overall minus sign.

Several comments are in order:
\begin{trivlist}
\item[a)]
In the operator formulation, the partition function $\cZ_{(g^r,1)}$ should equal ${\rm Tr}_{{\cal H}_r} q^{L_0} \bar{q}^{\bar{L}_0}$,
with the trace computed over the Hilbert space ${\cal H}_r$ in the sector twisted by $g^r$. 
In particular, this means that there must be no phases and every spacetime boson must come with a positive 
integer while spacetime fermions come with a negative integer.

\item[b)]
Using properties in Appendix \ref{app_thetas_lattices} one can check that $\cZ_{(g^r,1)} = \cZ_{(g^{N-r},1)}$, 
apart from a minus sign for $\vartheta_1$ terms that vanishes at the level of the partition function. This flip in the sign for $\vartheta_1$ 
terms means that the $SO(8)$ spinor and ${\rm spinor}'$ get exchanged in the ${\cal H}_{r}$ and ${\cal H}_{N-r}$ Hilbert spaces 
both in the left and right moving sectors, but this will not change the partition functions.

\item[c)]
The factor $|2 \sin(\frac{\pi r}N)|^2$ gives the number of ``fixed points'' in the $g^r$ or $g^{N-r}$ twisted sectors. 
There are of course, strictly speaking, no fixed points because we have a freely acting orbifold. 
The appearance of the number of fixed points of the rotational part of the orbifold group can be understood as follows.
Writing the worldsheet coordinate fields for $\bt^2$ and $S^1$ as $Y$ and $Y_3$, the boundary conditions in the sector twisted by $g$
are $(Y(\sigma+2\pi), Y_3(\sigma+2\pi))=( {\mathcal R}_g Y(\sigma),Y_3(\sigma)+ {2\pi}\tf{R_3}N)$. Then, the minimum length 
that a string can acquire occurs when the $Y$ field takes values at a fixed point of the rotation ${\mathcal R}_g $ acting on $\bt^2$. We will loosely refer to such states as being localised at fixed points in what follows.
Such strings, localised at the fixed points on $\bt^2$ will give rise to a Hilbert space coming from fluctuations (i.e. oscillator modes). 
Thus, even though the orbifold action is fixed point free geometrically, there are nonetheless twisted Hilbert spaces associated with the fixed points of the discrete rotation of the torus $\bt^2$.

\item[d)]
In class I models, the partition function is $\cZ^{\rm I}_{(g^r,1)}= \cZ_{(g^r,1)}\big|_{\epsilon=1}$. 
In class II, the prescription is to expand in powers of $\epsilon$, discard linear terms 
and then set $\epsilon=1$. Thus, in class II, the parameter $s_g$ drops out, as expected.
As mentioned earlier, the partition function $\cZ^{\rm II} _{(g^r,1)}$ of class II 
can be obtained by inserting the projector $\frac12(1+(-1)^{2(\beta_L+\beta_R)})$ in $\cZ^{\rm I}_{(g^r,1)}$.

\item[e)]
It can be verified that $\cZ_{(g^r,1)}$ is invariant under $\tau \to \tau + m$, where $m$ is the smallest positive integer
such that $mr$ is a multiple of $N$.  Thus, the level matching condition $m(L_0-\bar{L}_0) \in\bz$ is satisfied. 

\item[f)]
Although $SO(8)$ is broken by the orbifold action, it is still convenient to speak of V, Sc, Sp and ${\rm Sp}'$ classes
related to the combinations $\cA$, $\cA'$, $\cB$ and $\cB'$.
For instance, in class I,  in (NS,NS) $|\cA|^2$ corresponds to (V,V) or (Sc,Sc) depending on whether $rs_g$ is even or odd. 

\end{trivlist}

\subsubsection{$(g^r,g^p)$ sectors}\label{T3_ZN_grgp}
We next compute the partition function in $(g^r,g^p)$ sectors, $r,p\in\bz$ and $0<r,p<N$. 
To this end we first consider the $(g^r,g)$ sector, which is particularly important because from $\cZ_{(g^r,g)}$
we can learn how $g$ acts on the $g^r$-twisted Hilbert space ${\mathcal H}_r$ and check consistency with its interpretation as
${\rm Tr}_{{\mathcal H}_r } g  q^{L_0} \bar{q}^{\bar{L}_0}$.
The partition function $\cZ_{(g^r,g)}$ is obtained from $\cZ_{(1,g)}$, cf. \eqref{Z0r}, by applying the modular transformation
$(TST)^r$. On the basis of the action of $g$ in ${\mathcal H}_r$ we can then determine $\cZ_{(g^r,g^p)}$.
We shall dispense with the details of the computation and outline the results for the two classes of models I and II.

\vskip 15pt
\noindent{\bf{Class I models: $(s_1,s_2)=(0,0)$}}

\vskip 10pt
\noindent 
Comparing the partition function in the sectors $(g^r,1)$ and $(g^r,g)$ we verify that the action of $g$ on 
${\mathcal H}_r$ amounts to the expected phases that arise as shifts by $1/N$ in the arguments of $\vartheta$ functions
and $Z_\Gamma$. We can also check that the action of $g$ on the $\bt^2$  fixed points of $g^r$ is properly taken into account.
In particular, the fixed points that survive in the $(g^r,g)$ sector are the ones fixed with respect to $g$.
We then conclude that the class I partition function is 
\begin{align}\label{ZrpI}
&\cZ^{{\rm I}}_{(g^r,g^p)}=\frac{1}{4}\int_{\cal F}\frac{d^2\tau}{\tau_2^{\frac{9}{2}} }\,
\bigg|\frac{2\sin(\frac{{\rm gcd}(r,p)\pi}N)}{\eta^9(\tau)\,\vartheta_{(\frac{1}{2},\frac{1}{2})}(\frac{r\tau}{N}+\frac{p}{N},\tau)}
\bigg|^2
\!\! \sum_{\alpha_L,\beta_L,\alpha_R,\beta_R=\{0,\frac12\}} 
\hspace*{-10mm} {\mathcal C}(\alpha_L, \beta_L, \alpha_R, \beta_R)\; \times \notag \\
& \times \, 
(-1)^{2s_g\big(p(\alpha_L+\alpha_R)+r(\beta_L+\beta_R)\big)} \; 
\vartheta^3_{(\alpha_L,\beta_L)}(\tau)\;\vartheta_{(\alpha_L,\beta_L)}(\tf{r\tau+p}N,\tau)
\;\bar\vartheta^3_{(\alpha_R,\beta_R)}(\bar\tau)\;\bar\vartheta_{(\alpha_R,\beta_R)}(\tf{r\bar\tau+p}N,\bar\tau)\;\times \notag \\
& \hspace*{2cm} \times\; e^{-2\pi i(\beta_L+\beta_R)\frac rNs_3}\; 
Z_{\Gamma}\big(\tau,(\alpha_L+\alpha_R)s_3,\tf{r}{N};\tf{p}{N},(\beta_L+\beta_R)s_3\big)\ ,
\end{align}
where ${\rm gcd}(r,p)$ is the greatest common divisor\footnote{We are using the convention ${\rm gcd}(r,0) = {\rm gcd}(0,r)=r$.} 
of $r$ and $p$. Note that ${\rm gcd}(r,p)$  is a modular invariant quantity, i.e. if under a modular transformation $(r,p)\rightarrow (r',p')$, then 
${\rm gcd}(r',p')={\rm gcd}(r,p)$. 
The level matching condition can be proven since we have already shown that $Z_{(g^r,1)}$ satisfies it.
$Z_{(g^r,g^p)}$ is obtained by applying $g^p$ on the Hilbert space of $g^r$ twisted states which gives only phases to different states and does not alter the values of $L_0$ and $\bar{L}_0$.

From \eqref{ZrpI} it follows that the $g^r$ and $g^{N-r}$ twisted sectors are equivalent (for the same $s_g$).
More precisely, one can prove the property
\begin{equation}\label{equivr}
\cZ^{{\rm I}}_{(g^r, g^p)}=\cZ^{{\rm I}}_{(g^{N-r}, g^{-p})}=\cZ^{{\rm I}}_{(g^{N-r}, g^{N-p})}\, .
\end{equation}
There are also some useful relations between $\cZ^{{\rm I}}_{(g^r, g^p)}$ with $s_g=0$ or
$s_g=1$, in the same orbifold. In particular, in the $\bz_2$ orbifold we find 
$\cZ^{{\rm I}}_{(g^r, g^p)}(s_g=0) = \cZ^{{\rm I}}_{(g^r, g^p)}(s_g=1)$, and it can be shown
that the spectrum with $s_g=0$ and $s_g=1$ is the same.
For the $\bz_4$ and $\bz_6$ orbifolds, the bosonic (NS,NS) and (R,R) terms in the $g^r$ twisted sectors, with even $r$, coincide for 
$s_g=0,1$. However this is not true for fermions where $s_g$ enters in the projection over orbifold invariant states.

To study the spectrum it is helpful to recast  $\cZ^{{\rm I}}_{(g^r, g^p)}$ as in \eqref{Zr0explicit}, with the auxiliary
definitions in \eqref{ABdef}. We refrain from writing the explicit formulas. Comparing with \eqref{Zr0explicit}, we just make replacements
$\vartheta_{(\alpha,\beta)}(\tf{r\tau}N, \tau) \to \vartheta_{(\alpha,\beta)}(\tf{r\tau+p}N,\tau)$, 
$Z^{(e)}_{\Gamma}(\tau,0,\tf{r}{N};0,0) \to Z^{(e)}_{\Gamma}(\tau,0,\tf{r}{N};\tf{p}{N},0)$, and so on.
There is also an additional factor $(-1)^{p s_g}$ multiplying the fermionic (NS,R)/(R,NS) terms.
The advantage of rewriting $\cZ^{{\rm I}}_{(g^r, g^p)}$ explicitly is twofold. First, we immediately identify the type of
lattice sum. Second, the combinations $\cA$, $\cA'$, $\cB$ and $\cB'$ readily indicate the corresponding $SO(8)$ classes, e.g. 
in (R,R) $|\cB|^2$ signifies (Sp,Sp) or $({\rm Sp}',{\rm Sp}')$ depending on whether $rs_g$ is even or odd. 
As already mentioned, although $SO(8)$ is broken by the orbifold action, it is still useful to label states in terms of  
V, Sc, Sp and ${\rm Sp}'$ classes. Our conventions for these classes is analogous to \eqref{V}, namely
\be\label{brokenVSp}
\begin{split}
{\rm V} &= \frac{\vartheta_{(0,0)}^3(\tau) \vartheta_{(0,0)}(\tf{r\tau+p}{N},\tau)-
\vartheta_{(0,\frac{1}{2})}^3(\tau)\vartheta_{(0,\frac{1}{2})}(\tf{r\tau+p}{N},\tau) }
{2\op \eta^9(\tau)\,\vartheta_{(\frac{1}{2},\frac{1}{2})}(\frac{r\tau + p}{N},\tau) } \, , 
\\
{\rm Sp} &= \frac{\vartheta_{(\frac12,0)}^3(\tau) \vartheta_{(\frac12,0)}(\tf{r\tau+p}{N},\tau) +
\vartheta_{(\frac12,\frac{1}{2})}^3(\tau)\vartheta_{(\frac12,\frac{1}{2})}(\tf{r\tau+p}{N},\tau) }
{2\op \eta^9(\tau)\,\vartheta_{(\frac{1}{2},\frac{1}{2})}(\frac{r\tau + p}{N},\tau) } \, .
\end{split}
\ee 
Changing the relative sign between the two terms in the numerator gives the expressions for Sc  and ${\rm Sp}'$. 
We did not include the arguments $(r,p)$ for the classes to streamline notation.

\vskip15pt
\noindent{\bf{Class II models: $(s_1,s_2)\ne(0,0)$}}

\vskip 10pt
\noindent 
We recall that in $(1,g^p)$ sectors, for $p \neq 0$, the spacetime fermions are absent, while for spacetime bosons the partition function 
is the same as that in class I models with $s_g=0$, see below eq. \eqref{Z0r}. Therefore, we can write the class II  partition function in these sectors as the projector 
$\frac{1}{2}(1+ (-1)^{2(\alpha_L+\alpha_R)})$ applied on the corresponding class I partition function. 
Now, the partition function in the $(g^r,g^p)$ sector can be obtained from that in the $(1,g^{{\rm gcd}(r,p)})$ sector
by the modular transformation
\be
{\begin{pmatrix}
 \frac{p}{{\rm gcd}(r,p)}& b\\
  \frac{r}{{\rm gcd}(r,p)}&d
 \end{pmatrix}}\ ,\qquad 
p d- r b ={\rm gcd}(r,p) \ .
\ee
Under this transformation, 
$(-1)^{2(\alpha_L+\alpha_R)} \longrightarrow (-1)^{2{\frac{p(\alpha_L+\alpha_R)+r(\beta_L+\beta_R)}{{\rm gcd}(p,r)}}}$,
as can be seen from \eqref{thetamod}. Thus, in the $(g^r, g^p)$ partition function there will be a projector
\be
{\mathcal P}(\alpha_L,\alpha_R,\beta_L,\beta_R; r,p)=
\frac{1}{2}\left(1+(-1)^{2{\frac{p(\alpha_L+\alpha_R)+r(\beta_L+\beta_R)}{{\rm gcd}(p,r)}}}\right). 
\label{projdef}
\ee 
The effect of ${\mathcal P}(\alpha_L,\alpha_R,\beta_L,\beta_R; r,p)$ is exactly equivalent to the $\epsilon $ prescription 
used in eqs. \eqref{Z0r} and \eqref{Zr0}.

Inserting the projector  ${\mathcal P}(\alpha_L,\alpha_R,\beta_L,\beta_R; r,p)$ in \eqref{ZrpI} leads to
\begin{align}\label{ZrpII}
&\cZ^{{\rm II}}_{(g^r,g^p)}=\frac{1}{4}\int_{\cal F}\frac{d^2\tau}{\tau_2^{\frac{9}{2}} }\,
\bigg|\frac{2\sin(\frac{{\rm gcd}(r,p)\pi}N)}{\eta^9(\tau)\,
\vartheta_{(\frac{1}{2},\frac{1}{2})}(\frac{r\tau}{N}+\frac{p}{N},\tau)}\bigg|^2
\sum_{\alpha_L,\beta_L,\alpha_R,\beta_R=\{0,\frac12\}} \hspace*{-10mm} {\mathcal C}(\alpha_L, \beta_L, \alpha_R, \beta_R) \;
\times \notag \\
& 
\times \, \tfrac{1}{2}\Big(1+(-1)^{2{\frac{p(\alpha_L+\alpha_R)+r(\beta_L+\beta_R)}{{\rm gcd}(p,r)}}}\Big) \;
\vartheta^3_{(\alpha_L,\beta_L)}(\tau)\;\vartheta_{(\alpha_L,\beta_L)}(\tf{r\tau+p}N,\tau)
\;\bar\vartheta^3_{(\alpha_R,\beta_R)}(\bar\tau)\;\bar\vartheta_{(\alpha_R,\beta_R)}(\tf{r\bar\tau+p}N,\bar\tau)\; \times \notag \\
& \hspace*{2cm} \times \; e^{-2\pi i(\beta_L+\beta_R)\frac rNs_3}\;
Z_{\Gamma}\big(\tau,(\alpha_L+\alpha_R)s_3,\tf{r}{N};\tf{p}{N},(\beta_L+\beta_R)s_3\big)\ .
\end{align}
Notice that there is no $s_g$ dependence, since the projector
enforces $(-1)^{2 s_g( p(\alpha_L+\alpha_R)+r (\beta_L+\beta_R))}=1$ for $s_g=0$ or $1$. 
This can be understood from the fact that, in the $(1,g^{{\rm gcd}(r,p)})$ sector, there are no spacetime fermions in class II models 
as fermions carry some momenta shifted by half along the directions rotated by $g^{{\rm gcd}(r,p)}$. As a result, $s_g$ never enters in the
$(1,g^{{\rm gcd}(r,p)})$ sector and consequently in none of the other sectors that are related to it by a modular transformation.
The level matching condition is clearly satisfied because these states are a subset of corresponding class I states where we have already shown that the level matching condition is satisfied.

It can be shown that $\cZ^{{\rm II}}_{(g^r, g^p)}=\cZ^{{\rm II}}_{(g^{N-r}, g^{N-p})}$, in analogy with eq.~\eqref{equivr} 
for class I. Moreover, comparing the partition functions of class I, in \eqref{ZrpI}, and class II, in \eqref{ZrpII}, we deduce the relation
\be
\cZ^{{\rm II}}_{(g^r, g^p)}=\frac12\left[\cZ^{{\rm I}, s_g=0}_{(g^r, g^p)}+ \cZ^{{\rm I}, s_g=1}_{(g^r, g^p)}\right],
\qquad {\rm for}\ {\rm gcd}(r,p)=1\, .
\label{average01}
\ee
Class II models arise only for $\bz_2$ and $\bz_4$ orbifolds.
In the $\bz_2$ case, taking into account an identity mentioned above, we readily conclude that
$\cZ^{{\rm II}}_{(g^r, g^p)}=\cZ^{{\rm I}, s_g=0}_{(g^r, g^p)}= \cZ^{{\rm I}, s_g=1}_{(g^r, g^p)}$,
for $(r,p)\not=(0,0)$ (since $s_1$ and $s_2$ are not both zero this is obviously not true
for $\cZ^{{\rm II}}_{(1,1)}$). The relation \eqref{average01} proves helpful in computations of the spectrum and the 
cosmological constant in class II models, given the results in class I --- see subsection \ref{sub_T3_spec}. 

\subsection{Operator interpretation in class II models}
\label{sub_opi}

We now address the important question of how the twisted sector contributions to  the class II partition function, i.e. $\cZ^{{\rm II}}_{(g^r,g^p)}$,
$r\not=0$, can be understood at the operator level, namely as arising from ${\rm Tr}_{{\cal H}_r} g^p q^{L_0} \bar{q}^{\bar{L}_0}$, 
where ${\mathcal H}_r$ is the $g^r$-twisted Hilbert space. Actually, we only need to consider $r=0,1,2$ since class II models exist only for the 
$\bz_2$ and $\bz_4$ cases.

For $r=1$, an explicit form of $\cZ^{{\rm II}}_{(g^r,g^p)}$ can be neatly derived from
the average relation \eqref{average01} so let us first go back to class I.
For $p=0$, $\cZ^{{\rm I}}_{(g,g^p)}$, is given in \eqref{Zr0explicit}, setting $\epsilon=1$, and for $p\not=0$ the
generalization is straightforward.  
For $r=1$, the $SO(8)$ classes corresponding to $\cA$, $\cA'$, $\cB$ and $\cB'$ are respectively  V, Sc, Sp and ${\rm Sp}'$
for $s_g=0$, but Sc, V, ${\rm Sp}'$ and Sp, for $s_g=1$.
Schematically, in the $g$-twisted sector in class I we then have
\begin{align}
\label{r1I}
\msm{
\cZ^{{\rm I}, s_g=0}_{(g,g^p)}
} &
\msm{
\sim \ \chi(g,g^p) \Big\{
\big[\big({\rm (V,V)} + {\rm (Sp,Sp)}\big) Z^{(e)}_{\Gamma_{(1,1)}}(0,\tfrac1{N};\tfrac{p}{N},0) + 
\big({\rm (Sc,Sc)} + ({\rm Sp}',{\rm Sp}')\big)  Z^{(o)}_{\Gamma_{(1,1)}}(0,\tfrac1{N};\tfrac{p}{N},0) \big] 
} 
\notag \\
&{} \hspace*{-5mm}
\msm{
-\big[\big({\rm (V,Sp)} + {\rm (Sp,V)}\big) Z^{(e)}_{\Gamma_{(1,1)}}(\tfrac{s_3}2,\tfrac1{N};\tfrac{p}{N},0)  + 
\big(({\rm Sc},{\rm Sp}')+ ({\rm Sp}',{\rm Sc})\big) Z^{(o)}_{\Gamma_{(1,1)}}(\tfrac{s_3}2,\tfrac1{N};\tfrac{p}{N},0) \big] \Big\}, 
}
\\[3mm]
\msm{\cZ^{{\rm I}, s_g=1}_{(g,g^p)} 
}
&
\msm{
\sim \ \chi(g,g^p) \Big\{
\big[\big({\rm (Sc,Sc)} + ({\rm Sp}',{\rm Sp}')\big) Z^{(e)}_{\Gamma_{(1,1)}}(0,\tfrac1{N};\tfrac{p}{N},0) + 
\big({\rm (V,V)} + ({\rm Sp},{\rm Sp})\big)  Z^{(o)}_{\Gamma_{(1,1)}}(0,\tfrac1{N};\tfrac{p}{N},0) \big] 
} 
\notag \\
&{} \hspace*{-5mm}
\msm{
-(-1)^p \big[\big(({\rm Sc},{\rm Sp}') + ({\rm Sp}',{\rm Sc})\big) Z^{(e)}_{\Gamma_{(1,1)}}(\tfrac{s_3}2,\tfrac1{N};\tfrac{p}{N},0)  + 
\big(({\rm V},{\rm Sp})+ ({\rm Sp},{\rm V})\big) Z^{(o)}_{\Gamma_{(1,1)}}(\tfrac{s_3}2,\tfrac1{N};\tfrac{p}{N},0) \big] \Big\}. 
}
\notag
\end{align}
Here we dropped the argument $\tau$ in the lattice sums and wrote the number of simultaneous fixed points of $g^r$ and $g^p$ as 
$\chi(g^r,g^p) := \big|2\sin(\frac{{\rm gcd}(r,p)\pi}N)\big|^2$. The label in $\Gamma_{(1,1)}$ stresses that it is the $S^1$
lattice, in distinction to the $(3,3)$ lattice $\Gamma$ appearing in $\cZ_{\bt^3}$, cf. \eqref{ZTd}.  
The $SO(8)$ classes are defined in \eqref{brokenVSp}.

Substituting the above results in the average relation \eqref{average01} yields
\begin{align}
\label{r1II}
\msm{
\cZ^{{\rm II}}_{(g,g^p)}
}  
& \msm{
\sim \ \tfrac12\op \chi(g,g^p) \Big\{
\big[{\rm (V,V)} + {\rm (Sc,Sc)} + {\rm (Sp,Sp)} + ({\rm Sp}',{\rm Sp}')\big)]
\big(Z^{(e)}_{\Gamma_{(1,1)}}(0,\tfrac1{N};\tfrac{p}{N},0) + Z^{(o)}_{\Gamma_{(1,1)}}(0,\tfrac1{N};\tfrac{p}{N},0) \big) 
}
\notag \\
&{} \hspace*{-10mm}
\msm{
-\big[{\rm (V,Sp)} + {\rm (Sp,V)} + (-1)^p\big( ({\rm Sc},{\rm Sp}')+ ({\rm Sp}',{\rm Sc}) \big)\big]
\big[Z^{(e)}_{\Gamma_{(1,1)}}(\tfrac{s_3}2,\tfrac1{N};\tfrac{p}{N},0) + (-1)^p
Z^{(o)}_{\Gamma_{(1,1)}}(\tfrac{s_3}2,\tfrac1{N};\tfrac{p}{N},0) \big] \Big\}, 
}
\end{align}
For the $\bz_4$ orbifold we also need results for the $g^2$-sector. It turns out that for $p=0,2$,
$\cZ^{{\rm II}}_{(g^2,g^{p})}$ has the same structure as $\cZ^{{\rm II}}_{(g,g^{\frac{p}2})}$, except for obvious changes. Instead, for $p=1,3$, 
\be
\label{r2II}
\msm{
\cZ^{{\rm II}}_{(g^2,g^p)}
\sim \ \op \chi(g^2,g^p) \Big\{
\big[{\rm (V,V)} + {\rm (Sp,Sp)}]
Z^{(e)}_{\Gamma_{(1,1)}}(0,\tfrac2{N};\tfrac{p}{N},0) 
+\big[{\rm (Sc,Sc)} + ({\rm Sp}',{\rm Sp}')\big]
Z^{(o)}_{\Gamma_{(1,1)}}(0,\tfrac2{N};\tfrac{p}{N},0) \Big\}
}
\ee
Notice that the fermionic pieces (R-NS)/(NS-R) are absent, as implied by the projector in eq.~\eqref{projdef}.
This also occurs in $\cZ^{{\rm II}}_{(1,g^p)}$, $p\not=0$, in both $\bz_2$ and $\bz_4$, as we already observed.

We now want  to verify that the action of $g^p$ on ${\mathcal H}_r$ is consistent with $\cZ^{{\rm II}}_{(g^r,g^p)}$.
The action of the rotational part of $g$ is certainly captured correctly in the arguments of $\vartheta$ functions and the
lattice sums, but this alone is not enough to reproduce for instance the fact that in the $\bz_4$ orbifold
there are no fermions in $\cZ^{{\rm II}}_{(g^2,g)}$. The missing ingredient is the action of $g$ on the fixed points on $\bt^2$. 
Under $g$, fixed points go into fixed points modulo vectors in the lattice generated by $(e_1,e_2)$. For spacetime
bosons the shift by lattice vectors does not matter because bosons are periodic along $e_1$ and $e_2$. On the
contrary,  in presence of non-trivial spin structures $(s_1,s_2)$ spacetime fermions are anti-periodic along one or two of the cycles 
and the shifts do matter. 

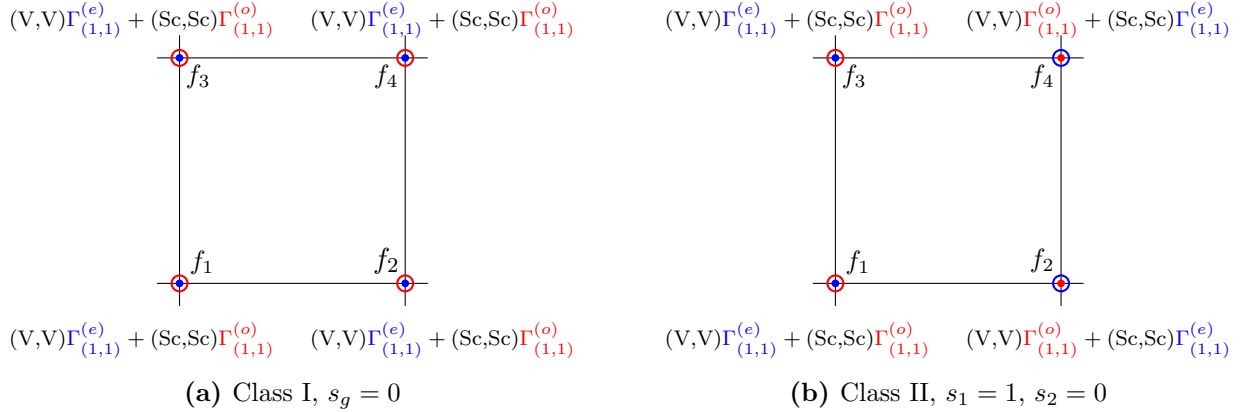
\begin{figure}[htb]
\centering
\subfloat[Class I, $s_g=0$]{
\begin{tikzpicture}
[scale=3,
	back/.style={thin},
	edge/.style={color=blue!95!black, thick},
         point/.style={inner sep=.75pt,circle,draw=blue,fill=blue,thick,anchor=base}]
\draw[back](-.1,0)--(1.1,0);
\draw[back](0,-.1)--(0,1.1);
\draw[back](-.1,1)--(1.1,1);
\draw[back](1,-.1)--(1,1.1);
\draw[thick, red] (0,0) circle (1pt);
\node[point] at (0, 0) {} node [shift={(0.3,0.275)}] {\mfn{f_1}}
node [shift={(-0.5,-0.75)}] {\msc{\text{(V,V)} \blue{\Gamma_{(1,1)}^{(e)}} + \text{(Sc,Sc)} \red{\Gamma_{(1,1)}^{(o)}}}};
\draw[thick, red] (0,1) circle (1pt);
\node[point] at (0, 1) {} 
node [shift={(0.25,2.75)}] {\mfn{f_3}} 
node [shift={(-0.5,3.5)}] {\msc{\text{(V,V)} \blue{\Gamma_{(1,1)}^{(e)}} + \text{(Sc,Sc)}\red{\Gamma_{(1,1)}^{(o)}}}};
\draw[thick, red] (1,0) circle (1pt);
\node[point] at (1, 0)   {} node [shift={(2.75,0.3)}] {\msm{f_2}}
node [shift={(3.5,-0.75)}] {\msc{\text{(V,V)} \blue{\Gamma_{(1,1)}^{(e)}} + \text{(Sc,Sc)} \red{\Gamma_{(1,1)}^{(o)}}}};
 \draw[thick, red] (1,1) circle (1pt);
\node[point] at (1, 1)   {} node [shift={(2.75,2.75)}] {\mfn{f_4}}
node [shift={(3.5,3.5)}] {\msc{\text{(V,V)} \blue{\Gamma_{(1,1)}^{(e)}} + \text{(Sc,Sc)} \red{\Gamma_{(1,1)}^{(o)}}}};
\end{tikzpicture} 
}
\hspace{5mm}
\subfloat[Class II, $s_1=1$, $s_2=0$]{
\begin{tikzpicture}
[scale=3,
	back/.style={thin},
	edge/.style={color=blue!95!black, thick},
         point/.style={inner sep=.75pt,circle,draw=blue,fill=blue,thick,anchor=base}]
\draw[back](-.1,0)--(1.1,0);
\draw[back](0,-.1)--(0,1.1);
\draw[back](-.1,1)--(1.1,1);
\draw[back](1,-.1)--(1,1.1);
\draw[thick, red] (0,0) circle (1pt);
\node[point] at (0, 0) {} node [shift={(0.3,0.275)}] {\mfn{f_1}}
node [shift={(-0.5,-0.75)}] {\msc{\text{(V,V)} \blue{\Gamma_{(1,1)}^{(e)}} + \text{(Sc,Sc)} \red{\Gamma_{(1,1)}^{(o)}}}};
\draw[thick, red] (0,1) circle (1pt);
\node[point] at (0, 1) {} 
node [shift={(0.25,2.75)}] {\mfn{f_3}}
node [shift={(-.5,3.5)}] {\msc{\text{(V,V)} \blue{\Gamma_{(1,1)}^{(e)}} + \text{(Sc,Sc)} \red{\Gamma_{(1,1)}^{(o)}}}};
\draw[thick, blue] (1,0) circle (1pt);
\node[inner sep=.75pt,circle,draw=red,fill=red,thick,anchor=base] at (1, 0)   {} 
node [shift={(2.75,0.3)}] {\mfn{f_2}}
node [shift={(3.5,-0.75)}] {\msc{\text{(V,V)} \red{\Gamma_{(1,1)}^{(o)}} + \text{(Sc,Sc)} \blue{\Gamma_{(1,1)}^{(e)}}}};
 \draw[thick, blue] (1,1) circle (1pt);
\node[inner sep=.75pt,circle,draw=red,fill=red,thick,anchor=base] at (1, 1)   {} 
node [shift={(2.75,2.75)}] {\mfn{f_4}}
node [shift={(3.5,3.5)}] {\msc{\text{(V,V)} \red{\Gamma_{(1,1)}^{(o)}} + \text{(Sc,Sc)} \blue{\Gamma_{(1,1)}^{(e)}}}};
\end{tikzpicture} 
}
\caption{Lattices associated to (NS,NS) states sitting at fixed points in class I and class II ${\mathbb Z}_2$ orbifolds. 
V and Sc refer respectively to vector and scalar $SO(8)$ classes.}
\label{fig_fix}
\end{figure}

The combinations $(Z^{(e)}_{\Gamma_{(1,1)}} + Z^{(o)}_{\Gamma_{(1,1)}})$ in \eqref{r1II} are entire lattice sums 
$Z_{\Gamma_{(1,1)}}$, but it is important to understand the action on the fixed points. 
Roughly speaking, what the partition function for class II says is that only half of the fixed points
contribute for each of the $SO(8)$ classes with even and odd windings.  
To be more precise, let us examine the fixed points for the $\bz_2$ orbifold. Choosing $e_1=(1,0)$ and $e_2=(0,1)$ as a
basis (note that the notation is slightly simplified here as compared to table \ref{table3bobby}), we have 4 fixed points with coordinates 
\be\label{z2fxpt}
f_1 = (0, 0), \quad f_2 = (\tfrac12, 0), \quad  f_3 = (0, \tfrac12), \quad f_4 = (\tfrac12, \tfrac12). 
\ee
These fixed points obey $(1-g)f_a = 0, e_1, e_2, e_1 + e_2$, respectively. 
Therefore, they carry a $\bz_2$ grading associated to the spin structures $(s_1, s_2)$ given
by $(1, (-1)^{s_1}, (-1)^{s_2}, (-1)^{s_1+s_2})$, respectively. For any choice of $(s_1, s_2) \not= (0, 0)$, there are 
two fixed points with grading $(+1)$ and two with grading $(-1)$. This
grading is linked to the grading of the $S^1$ lattice, which is the lattice appearing in \eqref{r1II},
because the projection as defined in the original toroidal theory 
(i.e. in the untwisted sector), as given in \eqref{gso}, comes with $(-1)^{\sum_{i=1}^3 s_i w_i}$.
This can be seen more physically by looking at the OPE of two twisted states, the right hand
side of which will necessarily be given by untwisted states, or equivalently computing a 3-point amplitude
involving two twisted and one untwisted state. If the two twisted states are localized at  fixed
points $f_a$ and $f_b$ respectively, then there is a selection rule forcing the untwisted
state to have $\bt^2$ winding $(1-g)(f_a-f_b)~{\rm{mod}}~2v$, with $v$ a lattice vector --- see e.g. \cite{Dixon:1986qv}. 
Since the untwisted states carry the grading given in \eqref{G1p}, this selection associates gradings to the
fixed points. The upshot is that each $SO(8)$ conjugacy class will appear with $\Gamma_{(1,1)}^{(e)}$ for two
fixed points with identical grading, and $\Gamma_{(1,1)}^{(o)}$ for the remaining two fixed points. This is illustrated in
Figure \ref{fig_fix}.

For the $\bz_4$ orbifold the story is similar. In this case the generator acts on the $\bt^2$ basis as 
$g:(e_1; e_2) \to (e_2,-e_1)$. There are two fixed points in the $g$-twisted sector, namely $f_1$ and $f_4$, cf. \eqref{z2fxpt},
which satisfy $(1-g)f_1 =0$ and  $(1-g)f_4 = e_1$. Now, recall that in the $\bz_4$ class II, the spin structures along the $\bt^2$ cycles
are $s_1=s_2=1$. Hence, the gradings  for $f_1$ and $f_4$ are  $(+1)$ and $(-1)$ respectively. 
One can check that the same fixed points and gradings appear in the $g^3$-sector.
The $g^2$-twisted sector is totally analogous to the twisted sector in the $\bz_2$ orbifold. Thus, in all
cases, half of the fixed points come with grading (+1) and the other half with  $(-1)$.

The proposal for the twisted Hilbert space is then the following. We start with the original untwisted Hilbert space given in 
\eqref{G1p}. In any given \mbox{$g^r$-twisted} sector there will be some fixed points on $\bt^2$, denoted by $f_a$, with 
$a$ ranging over the number of fixed points. Each $f_a$ satisfies the condition 
$(1-g^r)f_a = n^{(a)}_1 e_1 + n^{(a)}_2 e_2$, with $n^{(a)}_1, n^{(a)}_2 \in \bz$. 
We assign a grading $d_a$ to the fixed point $f_a$ defined by $d_a=(-1)^{n^{(a)}_1 s_1 + n^{(a)}_2 s_2}$.
Now, in \eqref{G1p}, $\Gamma^{(e)}$ and $\Gamma^{(o)}$ are respectively 
$\frac12(1 \pm (-1)^{w_1 s_1 + w_2 s_2} (-1)^{w_3 s_3})$ inserted in the full signature (3,3) lattice $\Gamma$.
In the twisted sectors, momenta and windings in $\bt^2$ are zero, but
there is a remnant action of $d_a$ on the fixed points. Concretely, the above projector operator on $\Gamma$
becomes $\frac12(1 \pm d_a (-1)^{w_3 s_3})$ inserted in the $S^1$ lattice $\Gamma_{(1,1)}$. 
Hence, for fixed points $f_a$ with even grading $d_a = 1$, the pairing of $SO(8)$ classes with even/odd lattice sums
is as in \eqref{G1p}, but for fixed points $f_b$ with odd grading $d_b =-1$, even and odd sums over $\Gamma_{(1,1)}$ are 
exchanged. We will shortly check that this proposal is consistent with the OPE. 

We now explain how the action on fixed points manifests itself in $\cZ^{{\rm II}}_{(g^r,g^p)}$.
We will first consider the simpler $\bz_2$ orbifold in which the fixed points $f_a$ are shown in \eqref{z2fxpt}. With the
concrete choice $(s_1,s_2)=(1,0)$, the gradings for the $f_a$ are $d_1=d_3=1$ and $d_2=d_4=-1$.
Then, in $\cZ^{{\rm II}}_{(g,1)}$ the assignment of lattices and states sitting at the fixed points is
\be\label{statesz2}
\begin{tabular}{ccc}
states & lattice for $f_1, f_3$ & lattice for $f_2, f_4$ \\
\ \, \msm{{\rm (V,V)} + {\rm (Sp,Sp)}} & \msm{\Gamma_{(1,1)}^{(e)}}  & \msm{\Gamma_{(1,1)}^{(o)}} \\
\ \, \msm{{\rm (Sc,Sc)} +  ({\rm Sp}',{\rm Sp}')}  & \msm{\Gamma_{(1,1)}^{(o)}}  & \msm{\Gamma_{(1,1)}^{(e)}} \\
\msm{ {\rm (V,Sp)} + {\rm (Sp,V)} }  & \msm{\hat\Gamma_{(1,1)}^{(e)}}  & \msm{\hat\Gamma_{(1,1)}^{(o)}} \\
\msm{ ({\rm Sc},{\rm Sp}')+ ({\rm Sp}',{\rm Sc}) }  & \msm{\hat\Gamma_{(1,1)}^{(o)}}  & \msm{\hat\Gamma_{(1,1)}^{(e)}}
\end{tabular}
\ee
Since $g: (f_1, f_2, f_3, f_4) \to (f_1, f_2 - e_1, f_3 - e_2, f_4- e_1 - e_2)$, $g$ acts as the identity on bosons
and as ${\rm diag}(1,-1,1,-1)$ on fermions.  Thus, $\cZ^{{\rm II}}_{(g,g)}$ contains all the bosons with two copies 
each of $\Gamma_{(1,1)}^{(e)}$ and $\Gamma_{(1,1)}^{(o)}$, while for fermions we have
$\msm{-2[{\rm (V,Sp)} + {\rm (Sp,V)} - ({\rm Sc},{\rm Sp}') - ({\rm Sp}',{\rm Sc}) \big]
\big[\hat Z^{(e)}_{\Gamma_{(1,1)}} - \hat Z^{(o)}_{\Gamma_{(1,1)}}\big]}$, which agrees with \eqref{r1II}.  
To simplify the expressions we are using the shorthand notation 
$\Gamma_{(1,1)}^{(e)}=\Gamma_{(1,1)}^{(e)}(0,\tfrac{r}{N};\tfrac{p}{N},0)$, 
$\hat\Gamma_{(1,1)}^{(e)}=\Gamma_{(1,1)}^{(e)}(\tfrac{s_3}2,\tfrac{r}{N};\tfrac{p}{N},0)$, and similarly for the odd
lattices.

Let us now look at the $\bz_4$ orbifold in which $s_1=s_2=1$. The fixed points in the $g$-twisted sector are $f_1$ and $f_4$, 
with gradings $d_1=1$ and $d_4=-1$. The spectrum in $\cZ^{{\rm II}}_{(g,1)}$ is also given in \eqref{statesz2}
(obviating $f_2$ and $f_3$). Now $g: (f_1,f_4) \to (f_1, f_4- e_1)$ and $g^2: (f_1,f_4) \to (f_1, f_4-e_1-e_2)$.
Therefore, $g$ acts trivially on bosons and as ${\rm diag}(1,-1)$ on fermions, while $g^2$ acts as the identity
on both. It is easy to check that this action of $g^p$, $p=1,2$, is properly encoded in $\cZ^{{\rm II}}_{(g,g^p)}$ in \eqref{r1II}.
We finally come to the more interesting $g^2$-twisted sector, which has fixed points $f_1$, $f_2$, $f_3$ and $f_4$ with
gradings $d_1=d_4=1$ and $d_2=d_3=-1$. The spectrum can be read from \eqref{statesz2}, exchanging $f_3$ and $f_4$.
In particular, the spectrum is the same for $f_2$ and $f_3$, consistent with $g: (f_1, f_2, f_3, f_4) \to (f_1, f_3, f_2 - e_1, f_4- e_1)$.
Since $f_2 \leftrightarrow  f_3$, for bosons and fermions these two sectors vanish upon inserting $g$ and taking the trace.
For bosons we are left with the fixed points $f_1$ and $f_4$, which altogether give two copies of $\Gamma_{(1,1)}^{(e)}$ 
for \msm{{\rm (V,V)} + {\rm (Sp,Sp)}} and two copies of $\Gamma_{(1,1)}^{(o)}$ for
\msm{ ({\rm Sc},{\rm Sc}) + ({\rm Sp}',{\rm Sp}')}. For fermions there is an additional 
minus sign for $f_4$ but not for $f_1$. Therefore these two sectors cancel among themselves and fermions are absent from the $(g^2, g)$ sector.
The results for bosons and fermions agree exactly with $\cZ^{{\rm II}}_{(g^2,g)}$ in \eqref{r2II}.
One may verify that the action of $g$ is also consistent with $\cZ^{{\rm II}}_{(g^2,g^2)}$. We skip the details
because this is completely analogous to $\cZ^{{\rm II}}_{(g,g)}$ in the $\bz_2$ case.

\subsubsection{Closure of OPEs}\label{sub_closure}

As we have seen, the proposal for the action of $g$ on fixed points in the twisted Hilbert spaces for class II models reproduces
the $\cZ^{{\rm II}}_{(g^r,g^p)}$ constructed independently by requiring modular invariance.
We now want to confirm that the structure of these Hilbert spaces is consistent with the operator product expansion.

In practice we will check that the product of two vertex operators associated to physical states yields a third vertex operator of another 
physical state present in the spectrum. We refer to \cite{Dixon:1986qv} for the construction of vertex operators in orbifold twisted sectors.
One important fact to recall is that vertex operators of NS and R states appear with canonical ghost charge, $(-1)$ and
$(-\frac12)$, respectively.  Thus, picture changing (p.c.) might be necessary to identify the state corresponding to the product.
For example, for the product of NS states in the vector class,  
${\rm V}_{-1} \otimes {\rm V}_{-1} = {\rm{Sc}}_{-2} \xrightarrow{{\rm{p.c.}}} {\rm V}_{-1}$.
In the following we will drop the subscript for the ghost charge and simply write the result after picture changing (if needed), e.g.
${\rm V} \otimes {\rm V}=  {\rm V}$,  ${\rm V} \otimes   {\rm{Sc}} = {\rm{Sc}}$, and so on.

In general we consider products $A \otimes B = C$, where $A$ and $B$ represent states in twisted sectors which are
attached to definite fixed points and lattices, as in \eqref{statesz2}. To determine the resulting $C$ we need to take into account
the orbifold selection rule for the equivalent 3-point amplitude. For instance, for the $\bz_2$ orbifold, if $A$ and $B$ are in the
$g$-twisted sector at fixed points $f_A$ and $f_B$, then $C$ is in the untwisted sector with a particular winding $v_C$. 
Labelling untwisted states as $U$ and states in the $g^r$-twisted sector as $T^r$,  the 3-point amplitude in question is denoted 
$\langle T_A T_B U_C \rangle$. 
The selection rule says that the product of space group elements for $A$, $B$ and $C$ must include the identity $(1,0)$.{\footnote{We follow the notation of \cite{Dixon:1986qv} and denote the elements of the space group as $(g,v)$, where $g$ and $v$ correspond to rotations and translations in the torus lattice, respectively --- see \cite[section 3]{Dixon:1986qv}.} 
More precisely, in the example at hand
\be\label{selectionrule}
(g, (1-g)(f_A + v_A)) (g, (1-g)(f_B + v_B)) (1, v_C) = (1,0) \ \, \Rightarrow \ \,
v_C = (1-g)(f_B-f_A + v) \, , 
\ee
where $v_A$, $v_B$ and $v_C$ are vectors in the $\bt^2$ lattice and $v=v_B-v_A$. Notice that for $\bz_2$, $(1-g)v=2v$.

Let us now investigate the class II $\bz_2$ model with spin structures $(s_1,s_2)=(1,0)$,
as taken in \eqref{statesz2}. Consider the OPE depicted in Figure~\ref{fig_ope}.a, equivalent to $\langle T_A T_B U_C \rangle$.
In this example, $A$ is a state of type (V,V) in the $g$-twisted sector at fixed point $f_1$, which has grading $d_1=1$ so that the lattice is
$\Gamma_{(1,1)}^{(e)}$. The state $B$ is also chosen to be (V,V) but the lattice is $\Gamma_{(1,1)}^{(o)}$ because 
it sits instead at fixed point $f_2$ with grading $d_2=-1$. According to the space group selection rule, the product yields a state in the 
untwisted sector, with $\bt^2$ winding $v_C=(2m + 1)e_1 + 2\ell e_2$, $m,\ell \in \bz$.
The winding $w_3$ in $C$ is odd because it comes from the sum of $w_3$ even in $A$ and odd in $B$.
Therefore, the full (3,3) lattice of the untwisted state $C$ is even. This is consistent with the fact that untwisted
(V,V) states must appear with $\Gamma_{(3,3)}^{(e)}$, cf. \eqref{G1p}. The OPE in Figure~\ref{fig_ope}.b, and many others involving fermions as well, can be worked out in a similar way.

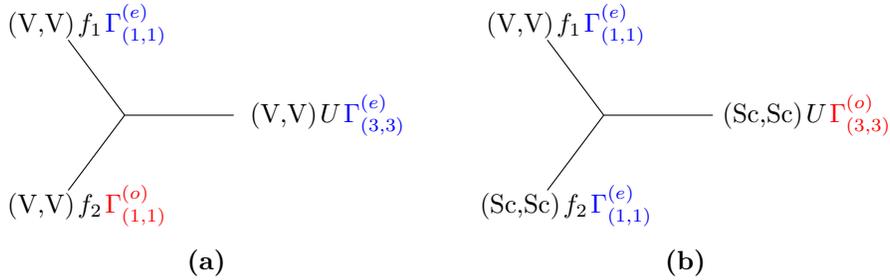
\begin{figure}[htb]
\centering
\subfloat[{}]{
\begin{tikzpicture}
[scale=1,
	back/.style={thin},
	edge/.style={color=blue!95!black, thick},
         point/.style={inner sep=.75pt,circle,draw=red,fill=red,thick,anchor=base}]
\draw[back](0.0,0)--(1.45,0)
node [shift={(1.25,0)}] {\mfn{\text{(V,V)}\op U\op \blue{\Gamma_{(3,3)}^{(e)}} }};
\draw[back](-0.75,1)--(0.0,0)
node [shift={(-0.5,1.2)}] {\mfn{\text{(V,V)}\op  f_1 \op \blue{\Gamma_{(1,1)}^{(e)} }}};
\draw[back](-0.75,-1)--(0,0)
node [shift={(-0.5,-1.2)}] {\mfn{\text{(V,V)}\op f_2 \op \red{\Gamma_{(1,1)}^{(o)}} }};
\end{tikzpicture}
}
\hspace{3mm}
\subfloat[{}]{
\begin{tikzpicture}
[scale=1,
	back/.style={thin},
	edge/.style={color=blue!95!black, thick},
         point/.style={inner sep=.75pt,circle,draw=red,fill=red,thick,anchor=base}]
\draw[back](0.0,0)--(1.45,0)
node [shift={(1.25,0)}] {\mfn{\text{(Sc,Sc)}\op U \op \red{\Gamma_{(3,3)}^{(o)}} }};
\draw[back](-0.75,1)--(0.0,0)
node [shift={(-0.5,1.2)}] {\mfn{\text{(V,V)}\op f_1 \op \blue{\Gamma_{(1,1)}^{(e)}} }};
\draw[back](-0.75,-1)--(0,0)
node [shift={(-0.5,-1.2)}] {\mfn{\text{(Sc,Sc)}\op f_2 \op \blue{\Gamma_{(1,1)}^{(e)}} }};
\end{tikzpicture}
}
\caption{3-point vertices in class II ${\mathbb Z}_2$ orbifold.}
\label{fig_ope}
\end{figure}

In the $\bz_4$ orbifold the space group selection rule allows 3-point couplings involving twisted sectors which are of the form 
$\langle T_A T^3_B U_C \rangle$, $\langle T^2_A T^2_B U_C \rangle$ and $\langle T_A T_B T^2_C \rangle$.
We will only discuss the last one, since the first two containing untwisted states are analogous to the previous $\bz_2$ example. 
For $A$ and $B$ in the $g$-sector,
the fixed points $f_A$ and $f_B$ can be either $f_1$ or $f_4$, and the space group element is as in eq.~\eqref{selectionrule}. 
For $C$ in the $g^2$-sector,  the fixed point could be any of $f_1$, $f_2$, $f_3$ or $f_4$, and the space group element is of the form
$(g^2, (1-g^2)(g^k f_C + v_C))$, with $k=0,1$ when $f_C$ is $f_2$ or $f_3$. The distinction is necessary because 
$f_2 \leftrightarrow f_3$ under $g$. Imposing that the product of space group elements includes $(1,0)$ leads to
\be\label{srule2}
(1-g^2)(g^k f_C - f_B) = (1-g)(f_A-f_B + v) + 2u \, ,
\ee
where $v=v_A-v_B$ and $u=v_B-v_C$ are vectors in the $\bt^2$ lattice.  It is easy to prove that when $f_A=f_B$,
$f_C$ is necessarily $f_1$ or $f_4$, but when $f_A\not=f_B$, in which case we can take $f_A=f_1$ and $f_B=f_4$,
then $f_C$ must be equal to $f_2$ or $f_3$.
We can now select particular bosonic or fermionic states in the $A$ and $B$ channels, with lattices assigned according to the
gradings of the fixed points, and then determine the state in $C$ from $A \otimes B$. Some examples are
\be
\begin{tabular}{ccc}
\msm{T_A} \ &  \msm{T_B}  & \msm{T^2_C} \\
\msm{{\rm (V,V)}\, f_4\, \Gamma_{(1,1)}^{(o)}} &  \msm{{\rm (Sc,Sc)}\, f_4\, \Gamma_{(1,1)}^{(e)}} &  
\msm{{\rm (Sc,Sc)}\, f_1, f_4\,  \Gamma_{(1,1)}^{(o)}}, \\
\msm{{\rm (V,V)}\, f_1\,  \Gamma_{(1,1)}^{(e)}} &  \msm{{\rm (V,Sp)}\, f_1\, \hat\Gamma_{(1,1)}^{(e)}} &  
\msm{{\rm (V,Sp)}\, f_1, f_4\,  \hat\Gamma_{(1,1)}^{(e)}}, \\
\msm{{\rm (Sc,Sc)}\,  f_1\, \Gamma_{(1,1)}^{(o)}} &  \msm{({\rm Sc,Sp}')\, f_4\, \hat\Gamma_{(1,1)}^{(e)}} &  
\msm{{\rm (V,Sp)}\, f_2, f_3\, \hat\Gamma_{(1,1)}^{(o)}}.
\end{tabular}
\ee
The results for $C$ are consistent with the content of the Hilbert space for the $g^2$-sector, which can be read from
\eqref{statesz2} exchanging $f_4$ and $f_3$.

\subsection{Spectrum}
\label{sub_T3_spec}

We have so far computed the partition function for cyclic orbifold models with non-standard spin structures. 
In each of these theories we now want to determine the spectrum, which is encoded in the full partition function \eqref{T3ZNclass12} 
for class I and class II models. 
In particular, we are interested in the spectrum of tachyonic states.
We will see that the existence of tachyons depends on the $S^1$ radius, namely $R_3:=R$.

The masses can be read off from the exponents of $q$ and  $\bar q$ upon expanding 
$\cZ^{{\rm I}}_{(g^r, g^p)}$ and $\cZ^{{\rm II}}_{(g^r, g^p)}$. The contribution of worldsheet fermions to the partition functions \eqref{ZrpI} and \eqref{ZrpII} is of the form, see eq. \eqref{sumform}:
\be\label{wsh_fmn}
\Big(\sum_{p_{L_i}}q^{\frac12(p_{L_1}^{2}+p_{L_2}^{2}+p_{L_3}^{2}+(p_{L_4}+\frac rN)^{2})}
e^{2\pi i(p_{L_4}+\frac rN)\frac pN}\Big)
\Big(\sum_{p_{R_i}}\bar q^{\frac12(p_{R_1}^{2}+p_{R_2}^{2}+p_{R_3}^{2}+(p_{R_4}+\frac rN)^{2})}
e^{-2\pi i(p_{R_4}+\frac rN)\frac pN}\Big)\ ,
\ee
where $p_{L}=(p_{L_1},p_{L_2},p_{L_3},p_{L_4})$ and $p_{R}=(p_{R_1},p_{R_2},p_{R_3},p_{R_4})$ are $SO(8)$ weights associated
to the left and right moving sectors, respectively. $p_{L_a},p_{R_a}\in\bz$ or $\bz+\tf12$ correspond to the NS or Ramond sectors, 
respectively. In the untwisted sector the modified GSO projection \eqref{gso} imposes the constraint that $\sum_a p_{L_a}$ 
and $\sum_a p_{R_a}$ are odd or even depending on the grading of the winding numbers $\sum_i s_i w_i$ along the cycles of the torus, 
see eq.~\eqref{G1p}.
In the twisted sectors the GSO projection further depends on the spin structure $s_g$, see eq.~\eqref{ABdef}. 
There are also contributions to the $q$ and $\bar q$ expansions from the bosonic part of the partition functions \eqref{ZrpI} 
and \eqref{ZrpII}, namely the $\vartheta_{(\frac{1}{2},\frac{1}{2})}$ and $\eta$ in the denominator, as well as the lattice sums.

All in all, the masses of the NS sector states in the left moving $g^r$ twisted sector are given by \cite{Font:2005td}:
\be
\alpha'm_L^2(g^r)=\tf{\alpha'}4\left(\tf{k_3}{R}+(w_3+\tf{r}{N})\frac{R}{\alpha'}\right)^2+
\frac12\left(p_{L_1}^2+p_{L_2}^2+p_{L_3}^2+(p_{L_4}+\tf{r}N)^2\right)+N_L+E_0-\frac12\ ,
\label{leftmass_NS}
\ee
where $N_L=\tf nN$, $n\in\bz_{\ge0}$ is the oscillator number, the normal ordering factor $(-\tf12)$ arises from the $q$ expansion of the 
$\eta$ and $\vartheta$ functions in the denominator of $\cZ_{(g^r, g^p)}$, and
\be\label{E0}
E_0=\tf12\tf{r}{N}(1-\tf{r}{N})
\ee
is the twisted sector vacuum energy. In the untwisted sector $E_0=0$, $N_L\in\bz_{\ge0}$, and an additional term $\frac12 P_L^2$ for the momenta of the $\bt^2$ lattice has to be added to the mass formula. A similar expression holds for the right moving NS sector:
\be
\alpha'm_R^2(g^r)=\tf{\alpha'}4\left(\tf{k_3}{R}-(w_3+\tf{r}{N})\frac{R}{\alpha'}\right)^2+
\frac12\left(p_{R_1}^2+p_{R_2}^2+p_{R_3}^2+(p_{R_4}+\tf{r}N)^2\right)+N_R+E_0-\frac12\ .
\label{rightmass_NS}
\ee
The Ramond sector masses are given by shifting the momenta $k_3\to k_3+\tf{s_3}2$ in the above expressions. 

Allowed states must satisfy the level matching condition $m_L^2=m_R^2$ and 
have to be invariant under the $\bz_N$ action. The orbifold projection in the 
sector twisted by $g^r$ follows from the sum in $p$ in \eqref{T3ZNclass12}.
Taking into account the $p$-dependent phases accompanying powers of $q$ and 
$\bar q$, and summing over $p$, we find that
\be\label{invcond}
\frac1{N}\left(p_{L_4} - p_{R_4} + k_3 + \Delta^{{\rm I 
(II)}} + m_{\rm{osc}}\right) \in 
\bz \, .
\ee
Here $m_{\rm{osc}}/N$, $m_{\rm{osc}} \in \bz$, arises when oscillators from 
$\bt^2$ bosons are present. The term $\Delta^{{\rm I (II)}}$ is non-vanishing only in fermionic (NS,R) and (R,NS) sectors,
and its form is different for class I and class II models. For class I models we 
have  $\Delta^{{\rm I}} =\frac12(s_3 + N s_g)$. Interestingly enough, this implies that for fermions, since $p_{L_4} \in \bz$ 
and  $p_{R_4} \in \bz + \frac12$, or viceversa, the invariance requirement can be satisfied only if $(s_3 + N s_g) =1\!\mod\! 2$. 
Therefore, the condition \eqref{condition} is necessary to have fermions in the spectrum. Since class II models exist only for $\bz_2$ and 
$\bz_4$, $\Delta^{{\rm II}}$ can be simply read directly from \eqref{r1II} and \eqref{r2II}.  

We have in total 24 orbifold models with various choices of the orbifold group 
and spin structures, see Table \ref{table3bobby}. 
The spectrum follows from the above mass formulae after imposing level matching and 
the orbifold projection.  
Below we shall first discuss one specific example in some detail in subsection 
\ref{sub_T3_spec_Z3} and then summarise the main results for the spectrum of 
other theories in subsection \ref{sub_T3_spec_ZN}.

\subsubsection{$(\bt^2\times S^1)/\bz_3$}
\label{sub_T3_spec_Z3}

The particular model we examine in some detail here is type IIB theory on the 
$\bz_3$ orbifold background. The reason for this choice is that 
the $\bz_3$ orbifold is the only model that allows both periodic and anti-periodic spin structures along the $S^1$ 
(of course, apart from the purely toroidal models). 
All other orbifolds have anti-periodic spin structure along the circle because $N$ is even, see 
Table \ref{table3bobby}. Additionally, $s_1=s_2=0$ for $\bz_3$.

\vskip15pt
\noindent{\bf{Periodic spin structure, $s_i=0$}}

\vskip 10pt
\noindent In this case we have $s_3=0$ and $s_g=1$. We compute the mass spectrum 
using eqs. (\ref{leftmass_NS}) and (\ref{rightmass_NS}) with $N=3$, and 
$r=\{0,1,2\}$. Since the $s_i$ are all zero, the odd lattice sum 
$Z^{(o)}_{\Gamma}$ drops out and
$Z^{(e)}_{\Gamma}=Z_{\Gamma}$. 

\medskip
{\bf{Untwisted sector.}} From \eqref{G1p}, as well as from \eqref{Zr0explicit}, 
we see that in the untwisted 
sector ($r=0$), the terms that appear in (NS,NS) are only (V,V) and in (R,R) 
only (Sp,Sp). Hence,
there are no tachyons. There are massless bosons in the (NS,NS) sector which  
include the metric, antisymmetric tensor, and dilaton in 7 dimensions, as well 
as vectors and scalars. The massless states in the (R,R) sector comprise 
antisymmetric tensors, vectors, and scalars in 7 dimensions.

An important feature of the $\bt^3/\bz_3$ orbifold with this spin structure 
is that it admits massless spacetime fermions in the (NS,R) and (R,NS) untwisted 
 sectors. Even though the Dirac operator has no zero modes, the
existence of massless fermions in this case can be expected on the following grounds:
the gravitino fields in ten dimensions are sections of the spin bundle of the 3-manifold 
with values in the tangent bundle. Because the manifold is flat, the structure group of the spin bundle
reduces to $\bz_3$. For this particular choice of spin structure both vectors and spinors transform in the same real 
reducible representation of $\bz_3$ and, hence their tensor product contains two $\bz_3$ singlets. Hence,
there are two modes of each gravitino field which are effectively scalars on the 3-manifold.

\medskip
{\bf{Twisted sector.}} It is enough to consider the $g^r$-twisted sector with 
$r=1$, since the sector with $r=2$ is analogous.
Eq. \eqref{ABdef} shows that in there are tachyons in the (NS,NS) sector, from the 
$|\cA|^2$ term, which appear below radius 
$\frac R{\sqrt{\alpha'}}<2\sqrt3$. In the (R,R) sector there are no massless nor 
tachyonic states. Finally, in the (NS,R) and (R,NS) sectors massless fermions 
could exist but that requires tuning $R$. For instance, in the (R,NS) sector for 
$p_L^2=1$, $p_{L_4}=-\tf12$, $p_R^2=0$,  $N_L=N_R=0$, $k_3=-1$, and $w_3=0$ we 
find
\be
m_L^2 = m_R^2=\frac{\alpha'}4\Big(\frac{R}{3\alpha'} - \frac{1}{R}\Big)^2 \, .
\ee
However, to have $m_L=m_R=0$, requires $R$ below the tachyon bound $\frac 
R{\sqrt{\alpha'}}<2\sqrt3$. 

The conclusion is that tachyons appear in the twisted sectors below a particular 
radius. For arbitrary large $R$ there are neither tachyons nor massless states 
in the twisted sectors.

\vskip15pt
\noindent{\bf{Anti-periodic spin structure}}

\vskip 10pt
\noindent We now have $s_3=1$ and $s_g=0$. The GSO projection depends on 
whether the winding number $w_3$ is even or odd, see section \ref{sub_T3}. 
Moreover, modular invariance requires that the 
quantised momentum $k_3$ is shifted by $\tf12$ in the Ramond sector.

\medskip
{\bf{Untwisted sector.}} For \underline{even windings} there are no tachyons. 
The (NS,NS) and (R,R) sectors have massless spacetime bosons for $N_L=N_R=0$, 
$k_3=w_3=0$, and $P_L=P_R=0$. There can be no massless spacetime fermions 
because in (NS,R) and (R,NS) sectors the quantised momentum $k_3$ is shifted by 
$\tf12$: fine-tuning $R$ cannot give $m_L^2=m_R^2=0$. However, the orbifold invariance and 
level-matching does allow towers of massive spacetime fermions in (NS,R) and (R,NS) 
sectors: they correspond to massive Rarita-Schwinger fields and fermions in 7 
dimensions.

For \underline{odd windings} there can be tachyons in the (NS,NS) sector 
appearing below the bound $\frac R{\sqrt{\alpha'}}=\sqrt2$. In (R,R) sector 
there are neither tachyons nor massless bosons. To have massless fermions in the 
(NS,R) and (R,NS) sectors would again require special values of $R$.

To recap, in the untwisted sector there are massless bosons including the 
graviton, antisymmetric tensors, and scalars. Massless fermions could exist for 
particular values of $R$ for which, however, there would be tachyons.

\medskip
{\bf{Twisted sectors.}} The $g$ and $g^2$ sectors are analogous. For 
\underline{even windings}, the (NS,NS) sector has tachyons below the bound 
$\frac R{\sqrt{\alpha'}}<\sqrt6$. The (R,R) sector has no tachyons nor massless 
states. In (NS,R) and (R,NS) sectors the level-matching condition prevents 
tachyons. There could be massless states for special values of $R$ which are 
below the tachyon bound.

For \underline{odd windings}, there are tachyons in (NS,NS) sector. In the (R,R) 
sector all states are massive. In the (NS,R) and (R,NS) sectors there could 
again be massless fermions for special values of $R$  which are, however, below 
the tachyon bound.

\medskip
We observe that the periodic and anti-periodic spin structures models mainly 
differ in the untwisted sector. 
In the former, tachyons are absent and there are massless fermions. 
In the latter, there are tachyons, towers of massive fermions and massless 
fermions only for a particular value of the circle radius. 
The matter content of both models in the twisted sectors is similar: tachyons 
occur and massive fermions can become massless at special 
radii which, however, lie at a value for which tachyons are present. 
Tachyons are absent at a
sufficiently large radius.

\subsubsection{$(\bt^2\times S^1)/\bz_N$}\label{sub_T3_spec_ZN}
We now discuss the spectrum of $\bz_N$ orbifold backgrounds.

\medskip
\noindent
{\bf{Class I.}} The analysis for the remaining class I models is completely 
analogous to the $\bz_3$ models discussed in the previous subsection. In all 
cases tachyons appear below a bound on the size of the radius of the circle. Let 
us first consider the untwisted sector.
Among the (NS,NS) states in (Sc,Sc), which couple to the odd winding lattice 
$Z^{(o)}_{\Gamma}$, tachyons will always be present for
\begin{equation}
\frac{R}{\sqrt{\alpha'}} < \sqrt2\ .
\label{uttachyons}
\end{equation}
In the $g^r$ twisted sectors, the possible tachyons can be determined from the 
explicit form of the partition function
in \eqref{Zr0explicit} and \eqref{ABdef}. Now (Sc, Sc) terms couple as 
$|\cA'|^2$ to $Z^{(o)}_{\Gamma}$ if $r s_g={\rm even}$, but 
as $|\cA|^2$ to $Z^{(e)}_{\Gamma}$ if $r s_g={\rm odd}$. Tachyons will emerge at 
radii
\begin{eqnarray}
\frac{R}{\sqrt{\alpha'}} & < & \sqrt{\frac{2N}{N-r}}\ ,\qquad\qquad\qquad rs_g= 
{\rm even}\ ,\,w={\rm odd}\ , \\
\frac{R}{\sqrt{\alpha'}} & <& \frac{\sqrt{2N(N-r)}}{r}\ ,\qquad\qquad rs_g={\rm 
odd}\ , \,w={\rm even}\ .\nonumber
\end{eqnarray}
Contrary to the untwisted sector, tachyons can also arise from (V,V) terms. 
Specifically, tachyons are present at
\begin{eqnarray}
s_g={\rm odd}\ , \,w={\rm odd}\ ,\\
\frac{R}{\sqrt{\alpha'}} & < &\frac{\sqrt{2 N r}}{N-r}\ ,\qquad\qquad r s_g={\rm 
odd}\ , \,w={\rm odd}\ ,\\
\frac{R}{\sqrt{\alpha'}} & < &\sqrt{\frac{2N}{r}}\ , \qquad\qquad\;  r s_g= {\rm 
even}\ ,\,w={\rm even}\ .\nonumber
\end{eqnarray}
The results for the tachyon bounds in class I models are summarised in Table 
\ref{tachyon_all_ZN}. Since the partition function in 
the $g^r$ and $g^{N-r}$ twisted sectors are equivalent (see the discussion at 
the end of subsection \ref{T3_ZN_1gr}), we only include  
$r\le\lfloor{\tf{N}{2}}\rfloor$. 

It is straightforward to check that in the (R,R) sector there are no tachyons; in the 
untwisted sector there are massless as well as massive bosonic 
states and in the twisted sectors only massive bosons.
In the (NS,R)/(R,NS) sectors there are untwisted and twisted massive fermions which could 
become massless   
at special radii lying below the tachyon bound.

\begin{table}[ht]
\renewcommand{\baselinestretch}{1.45}
\footnotesize{
\setlength\tabcolsep{3.75pt}
\begin{tabular}{c|cc|cc||cc|cc||ccc|ccc||cccc|cccc||}
\cline{2-23}
{}& \multicolumn{4}{|c||}{$\bz_2$}   &  \multicolumn{4}{|c||}{$\bz_3$} &  
\multicolumn{6}{|c||}{$\bz_4$} & \multicolumn{8}{|c||}{$\bz_6$}\\
\hline
\multicolumn{1}{|c|}{$s_g$} & 
\multicolumn{2}{|c|}{0} &   \multicolumn{2}{|c||}{1} &  
\multicolumn{2}{|c|}{0} &   \multicolumn{2}{|c||}{1}  &   
\multicolumn{3}{|c|}{0} &   \multicolumn{3}{|c||}{1}  &   
\multicolumn{4}{|c|}{0} &   \multicolumn{4}{|c||}{1}     
\\
\hline
\multicolumn{1}{|c|}{$r$} & 
\multicolumn{1}{|c|}{0} &   \multicolumn{1}{|c|}{1}  &   \multicolumn{1}{|c|}{0} 
 &   \multicolumn{1}{|c||}{1} & 
\multicolumn{1}{|c|}{0} &   \multicolumn{1}{|c|}{1}  &   \multicolumn{1}{|c|}{0} 
 &   \multicolumn{1}{|c||}{1} &  
\multicolumn{1}{|c|}{0} &   \multicolumn{1}{|c|}{1}  &   \multicolumn{1}{|c|}{2} 
 &  
\multicolumn{1}{|c|}{0}  &   \multicolumn{1}{|c|}{1}  &   
\multicolumn{1}{|c||}{2} & 
\multicolumn{1}{|c|}{0} &   \multicolumn{1}{|c|}{1}  &   \multicolumn{1}{|c|}{2} 
 &  \multicolumn{1}{|c|}{3}  &
\multicolumn{1}{|c|}{0}  &   \multicolumn{1}{|c|}{1}  &   
\multicolumn{1}{|c|}{2} & \multicolumn{1}{|c||}{3}  
\\  
 \hline
\multicolumn{1}{|c|}{$\tf R{\sqrt{\alpha'}}$} & 
\multicolumn{1}{|c|}{$\sqrt2$} & \multicolumn{1}{|c|}{2} & 
\multicolumn{1}{|c|}{$\sqrt2$} &  \multicolumn{1}{|c||}{2} & 
\multicolumn{1}{|c|}{$\sqrt2$} & \multicolumn{1}{|c|}{$\sqrt6$} & 
\multicolumn{1}{|c|}{$-$} & \multicolumn{1}{|c||}{$2\sqrt3$} &
\multicolumn{1}{|c|}{$\sqrt2$} &   \multicolumn{1}{|c|}{2$\sqrt2$}  &   
\multicolumn{1}{|c|}{2}  &   
\multicolumn{1}{|c|}{$\sqrt2$}  &   \multicolumn{1}{|c|}{2$\sqrt6$}  &   
\multicolumn{1}{|c||}{2} &
\multicolumn{1}{|c|}{$\sqrt2$} &   \multicolumn{1}{|c|}{2$\sqrt3$}  &   
\multicolumn{1}{|c|}{$\sqrt6$}  &  
\multicolumn{1}{|c|}{2}  &
\multicolumn{1}{|c|}{$\sqrt2$}  &   \multicolumn{1}{|c|}{2$\sqrt{15}$}  &   
\multicolumn{1}{|c|}{$\sqrt6$} & 
\multicolumn{1}{|c||}{2}  
 \\
\hline
\end{tabular}
}
\caption{Tachyon bounds on the radius of $S^1$ for $\bz_N$ class I orbifold 
models. In $\bz_2$, $\bz_4$ and $\bz_6$, $s_3=1$, while
 $s_3=(1 + s_g)\!\!\!\mod 2$ in $\bz_3$.}
\label{tachyon_all_ZN}
\end{table}

\medskip
\noindent
{\bf{Class II.}} These models arise only for  $\bz_2$ and $\bz_4$
. Recall that the possible spin structures are $s_3=1$ and 
$(s_1,s_2)=(1,0), (0,1),( 1,1)$ for $\bz_2$ and $(s_1,s_2)=( 1,1)$ for $\bz_4$.
Let us begin by discussing the spectrum of the untwisted sector where there are contributions 
from Kaluza-Klein momenta $k_i$ and winding modes $w_i$ from all  $\bt^2$ and $S^1$ directions ($i=1,2,3$).
Even though the expressions for the masses are the same  for both class I and class II models, the constraint on winding modes  
$s_1w_ 1+s_2w _2+ w_3={\rm odd} (\rm even)$ depending on the sector considered (see \eqref{G1p}), leads to different spectra. 
The lowest masses values arise from the ${\rm(Sc,Sc)}\, Z^{(o)}_{\Gamma_{(3,3)}}$
sector  with $p_{L_a}=p_{R_a}=0$ ($a=1,\ldots, 4$) and $N_L=N_R=0$. For  these states the level matching condition reads 
$k_1w_ 1+k_2w _2+ k_3w_3=0$, while the odd sum selects $s_1w_ 1+s_2w _2+ w_3={\rm odd}$. 
The  solution  $k_i=0$ leads to tachyonic states for radii values satisfying
\be
w_1^2 R_1^2+w_2^2 R_2^2+w_3^2 R^2 < 2\ap \, ,
\ee
where $R:= R_3$ as before. In particular, notice that if say, $s_1=1$, we can choose $w_2=w_3=0$ and tachyons will arise for 
$R_1<\sqrt{2\ap}$, independently of  the $R_2$ and $R$ values. This is different from class I models where there are 
no tachyons for $R\ge \sqrt{2\ap}$, independently of $R_1$ and $R_2$. 

Another interesting result regarding the untwisted sector is that no massless fermions are allowed, in contrast with class I models. 
For instance, in $({\rm Sc,Sp}')\, \hat Z^{(o)}_{\Gamma_{(3,3)}}$, since  $s_i=1$ for $i=1\, {\rm and/or}\  2$ and $s_3=1$, 
the Kaluza-Klein momenta will be shifted as $k_3+\frac12$, and $k_i+\frac{s_i}2$. It is easy to check that these shifts, and the condition 
$s_1w_ 1+s_2w _2+ w_3={\rm odd}$, force the masses to be positive.

For the analysis of the spectrum in twisted sectors it proves helpful to recall \eqref{average01}. In particular, for
the $\bz_2$ orbifold we observed that the partition function in the $g$-twisted sector coincides in class II and class I models,
implying that in class II tachyons are avoided for $\frac{R}{\sqrt{\ap}} > 2$, as in class I.
For the $\bz_4$ class II, from \eqref{average01} we conclude that the partition function in $g^r$-twisted sectors,
$r=1,3$, is the average of class I models with $s_g=0$ and $s_g=1$. As a consequence, 
in these sectors, tachyons will appear at radius  $\frac{R}{\sqrt\ap}<2\sqrt6$ (see Table \ref{tachyon_all_ZN}). 
In the $g^2$-twisted sector, cf. \eqref{r1II} and \eqref{r2II}, we find that tachyons 
appear in the (NS,NS) sector for $ \frac{R}{\sqrt\ap}<2$, whereas only massive bosons arise in the (R,R) sector. 
We also find that in this sector massless fermions can arise at the specific value of the radius 
$\frac{R}{\sqrt\ap}=1$ in  both ${\rm(V,Sp)}+{\rm(Sp,V)}$ and ${\rm(Sc,Sp')}+{\rm(Sp',Sc)}$.


\section{1-loop Vacuum Energy}\label{section_lambda}

As we have seen, the $(\bt^2 \times S^1)/\bz_N$ orbifolds are non-supersymmetric
and have tachyons unless the circle radius is above a certain bound. Thus,
the 1-loop vacuum energy, essentially a potential for the moduli and denoted $\Lambda$, is expected to be non-zero
and finite in the regime where tachyons are avoided.
In this section we will compute $\Lambda$, which up to normalization is 
given by
\be\label{lambda}
\Lambda=-\cal Z\ .
\ee
Here $\cZ$ is the full partition function in \eqref{T3ZNclass12}, namely the sum 
over contributions from all 
$(g^r, g^p)$ sectors.  When $(r,p)\not=(0,0)$, the summands in class I and class II models are 
$\cZ^{{\rm I}}_{(g^r,g^p)}$, cf. \eqref{ZrpI}, and $\cZ^{{\rm II}}_{(g^r,g^p)}$,  
cf. \eqref{ZrpII}. Besides, both $\cZ^{{\rm I}}_{(1,1)}$ and $\cZ^{{\rm II}}_{(1,1)}$ are read from
$\cZ_{\bt^3}$ in \eqref{ZTd}. We will explicitly consider class I models with $s_1=s_2=0$ and drop 
the labels in the following. Some features of $\Lambda$ in class II can be extracted taking into account relations 
such as \eqref{average01}. 
 
To begin, the integral over the fundamental domain will be approximated analytically
in the limit where the size of the circle is much larger than the size of the torus. 
The outcome will be an estimate for the leading behaviour of $\Lambda$ as a function of 
the circle radius $R$. We will then report on the results of a numerical estimate of $\Lambda$. In particular, the numerical findings will show the divergences at the values of $R$ where tachyons 
emerge and confirm the large $R$ behaviour found analytically.

\subsection{Analytical results}
\label{lambda_analyt}

We want to estimate the leading behaviour of $\Lambda$ in the limit where the radius of the circle $R$ is much larger than the 
torus radii, $R_i$, $i=1,2$, which in turn are kept to be of the order of the string length $\sqrt{\alpha'}$.
The assumption on the $R_i$ ensures that neither the Kaluza-Klein nor the winding modes on $\bt^2$ approach the continuum limit. On the other hand, in the limit of large $R$ the winding modes on $S^1$ can be neglected.
The reason is that their contribution is of the form $e^{-2\pi\tau_2w_3^2R^2}$ and is exponentially suppressed in the large 
$R$ limit, since in the fundamental domain $\tau_2\ge \tf{\sqrt3}2$. 
Now, in the $g^r$-twisted sectors the winding modes are shifted by $w_3\to w_3+\tf rN$, $r\ne0$.
Therefore, the twisted sector states in the limit are very massive and drop out at leading order. 
By the same argument untwisted states with non-zero winding can be neglected.
The upshot is that the leading approximation to $\Lambda$ comes purely from states with zero winding in the untwisted sector, namely
\be\label{lambda_ii}
\Lambda\big|_{{\rm large}\,R}=-\frac1N\sum_{p=0}^{N-1}\cZ_{(1,g^p)}\big|_{w_i=0} \ .
\ee
These are essentially the contributions which arise in the field theory limit from Type II supergravity.
Furthermore, notice that, if present, the odd lattice sums, $Z_{\Gamma_o}$ and $\hat Z_{\Gamma_o}$, will vanish 
since they run over windings with $\sum_is_iw_i= \text{odd}$.

Let us first consider the contribution from $\cZ_{(1,1)}$, which is given by the purely toroidal partition function  in eq.~\eqref{ZTd}. 
After restricting to zero winding, we perform a Poisson resummation and evaluate the integrals. 
The $\tau_1$ integral imposes the level matching condition $m_L^2=m_R^2$. 
To do the integral over $\tau_2$ we split the integration over two regions: one above and one below $R^a$, with $1 < a < 2$.
The region above gives the leading contribution in the large $R$ limit. Moreover,  the terms in the momentum sum dominate over 
terms involving positive integer powers of $e^{-2\pi \tau_2}$. In this way we find that
for the periodic spin structure the integral over the fundamental domain is approximated by
\be\label{Z11s0int}
\cZ_{(1,1)}^{s_3=0}=\frac{2\sqrt2\op R\op\Gamma(4)}{(2\pi R^2)^4}\,(N_b-N_f)\zeta(8)\ ,
\ee
where $N_b$ and $N_f$ are the number of massless bosonic and fermionic modes, respectively, 
$\zeta(a)$ is the Riemann zeta function, and $\Gamma(4)= 3!$.
For the anti-periodic spin structure we instead obtain
\be\label{Z11s1int}
\cZ_{(1,1)}^{s_3=1}=\frac{2\sqrt2\op R\op \Gamma(4)}{2^8\op (2\pi R^2)^4}\,\Big[(N_b-N_f)\zeta(8)+(N_b+N_f)\zeta(8,\tf12)\Big]\ ,
\ee
where $\zeta(a,b)$ is the Hurwitz zeta function. In the compactification of type IIB theories that we have been studying, 
$N_b=N_f=128$. Thus $\cZ_{(1,1)}^{s_3=0}=0$, as expected since for the periodic spin structure the toroidal theory is supersymmetric.
On the other hand,
\be\label{Z11s1int2}
\cZ_{(1,1)}^{s_3=1}=\frac{2\sqrt2\op R\op\Gamma(4)}{(2\pi R^2)^4}\,\zeta(8,{\tf12}) \, .
\ee
This non-zero result reflects supersymmetry breaking for the anti-periodic spin structures, as expected. 

The leading order of $\cZ_{(1,g^p)}$, cf.  \eqref{Z0r}, can be obtained using the same procedure to approximate the
$\tau_2$ integral. The general expression for class I $\bz_N$ orbifolds turns out to be
\begin{align}\label{Z1ps0int}
&\cZ_{(1,g^p)}=\frac{128 \sqrt2 R \,\Gamma(4)}{(2\pi R^2)^4}\,\bigg\{\!
\Big(\! \sin^8(\tf{\pi p}{2N})+\cos^8(\tf{\pi p}{2N})\! \Big)\Big(\! \zeta(8,\tf pN)+\zeta(8,1\!-\!\tf pN) \!\Big)  \\
& + \tfrac{(-1)^{ps_g}}{2^8}
\Big(\! \sin^8(\tf{\pi p}{2N})-\cos^8(\tf{\pi p}{2N})\!\Big)\! \bigg[\zeta(8,\tf p{2N})+\zeta(8,1\!-\!\tf p{2N})+ 
(-1)^{s_3}\Big(\!\zeta(8,\tf12\!+\!\tf p{2N})+\zeta(8,\tf12\!-\!\tf p{2N})\!\Big)\!\bigg]\!\bigg\}. \notag
\end{align}
Note that $\cZ_{(1,g^p)}$ is invariant under $p\to N-p$, since 
the spin structures $s_3$ and $s_g$ are related through the condition \eqref{condition}, namely
$(-1)^{N s_g + s_3}=-1$. Observe also that the first term 
is due to spacetime bosons, while the second with the prefactor $(-1)^{p s_g}$, is due to spacetime fermions.
It is easy to check that in the limit $p \to 0$ we recover the previous results for $\cZ_{(1,1)}$. 
The dependence on $R$ is also consistent with dimensional analysis.

We now specialize to the $\bz_3$ model, which is the only theory admitting both periodic and anti-periodic spin 
structures along $S^1$, i.e. $s_3=0$ or $s_3=1$. Our motivation is  
to examine whether there is a different behaviour of $\Lambda$ depending on $s_3$. 
All in all, for the $\bz_3$ orbifold we then find the large $R$ behaviour of $\Lambda$ to be
\be\label{Z1pZ3s0int}
\Lambda_{\bz_3}^{s_3=0}=
- \frac{2\sqrt2 \op  R\, \Gamma(4)}{3 (2\pi R^2)^4}  \op 81 \left[ \zeta(8,\ts{\frac13}) + \zeta(8,\ts{\frac23}) \right] .
\ee
for the periodic spin structure and
\be\label{Z1pZ3s1int}
\Lambda_{\bz_3}^{s_3=1}=- \frac{2\op \sqrt2 \op R\, \Gamma(4)}{3\op (2\pi R^2)^4} 
\left\{\zeta(8,\ts{\frac12}) + \frac{1317}{32} \left[ \zeta(8,\ts{\frac13}) + \zeta(8,\ts{\frac23})\right] 
- \frac{5}{32} \left[ \zeta(8,\ts{\frac16}) + \zeta(8,\ts{\frac56})\right] \right\}.
\ee
for the anti-periodic spin structure. 
The analytical results in the large $R$ limit show that the 1-loop cosmological constant is negative for both models, 
tends to zero as $R^{-7}$, and
\be\label{lambdaratio}
\frac{\Lambda_{\bz_3}^{s_3=0}}{\Lambda_{\bz_3}^{s_3=1}}\approx59\, .
\ee
As such, there is no qualitative difference between the two models with periodic and anti-periodic spin structures on the $S^1$.

For other class I $\bz_N$ orbifolds the leading order of $\Lambda$ for large $R$ can be easily obtained from the previous results,
setting $s_3=1$. In all cases $\Lambda=-c/R^7$, where the constant $c$ is a function of $N$ and $s_g$. 
The negative sign is a consequence of having more massless bosons than fermions in the spectrum. In fact, only the $\bz_3$
with $s_3=0$ has massless fermions for large $R$.
For $\bz_2$ there is no dependence at all on $s_g$, as expected from the fact that the $\bz_2$ orbifolds with
$s_g=0$ and $s_g=1$ are equivalent. 

For class II models we can make the same assumptions to estimate $\Lambda$, i.e. winding states are neglected and only the
untwisted sector is included. The contributions from $\cZ^{{\rm II}}_{(1, g^p)}$ can be read from \eqref{Z1ps0int}, discarding
the term with prefactor $(-1)^{p s_g}$ because fermions are absent. 
In $\cZ^{{\rm II}}_{(1, 1)}$ we can use techniques similar to those in class I to approximate the
$\tau_2$ integral in the large torus volume limit. However, to probe the effect of anti-periodic spin structures $(s_1,s_2)$
we need to keep the ratios $R_1/R$ and $R_2/R$ finite; this leads to the appearance of Epstein zeta 
functions depending on these ratios, as opposed to Hurwitz zeta functions. We did not pursue this approach to greater extent since 
we are mostly interested in qualitative features that can be determined by numerical integration as discussed below.

\subsection{Numerical results}\label{lambda_num}

The integration over the 1-loop fundamental domain can be implemented numerically, as has been done for
other non-supersymmetric toroidal string compactifications in order to obtain the dependence of $\Lambda$ on
the moduli \cite{Sakai:1985cs,Ginsparg:1986wr,Itoyama:1986ei,Blum:1997gw}. In our case the relevant
modulus is the circle radius $R$. The radii of the $\bt^2$ are kept fixed at values of the order of the string length.

\begin{figure}[h!]
\captionsetup[subfigure]{labelformat=empty, labelsep=none}
\centering
\subfloat[\mbox{\hspace*{-1cm} $s_g=0$}] {} \hspace{7cm}  \subfloat[\mbox{$s_g=1$}] {} \\[-3mm]
\subfloat[]
{\includegraphics[scale=.45]{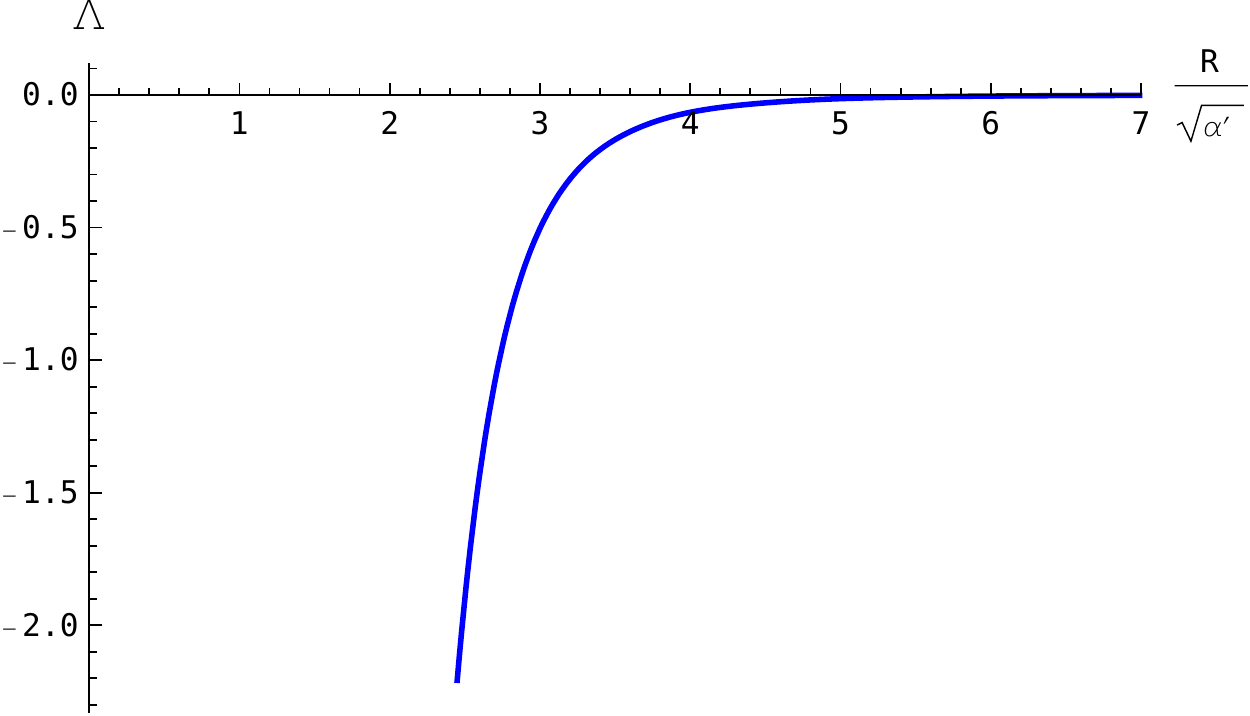}} \hspace{2cm}
\subfloat[]
{\includegraphics[scale=.45]{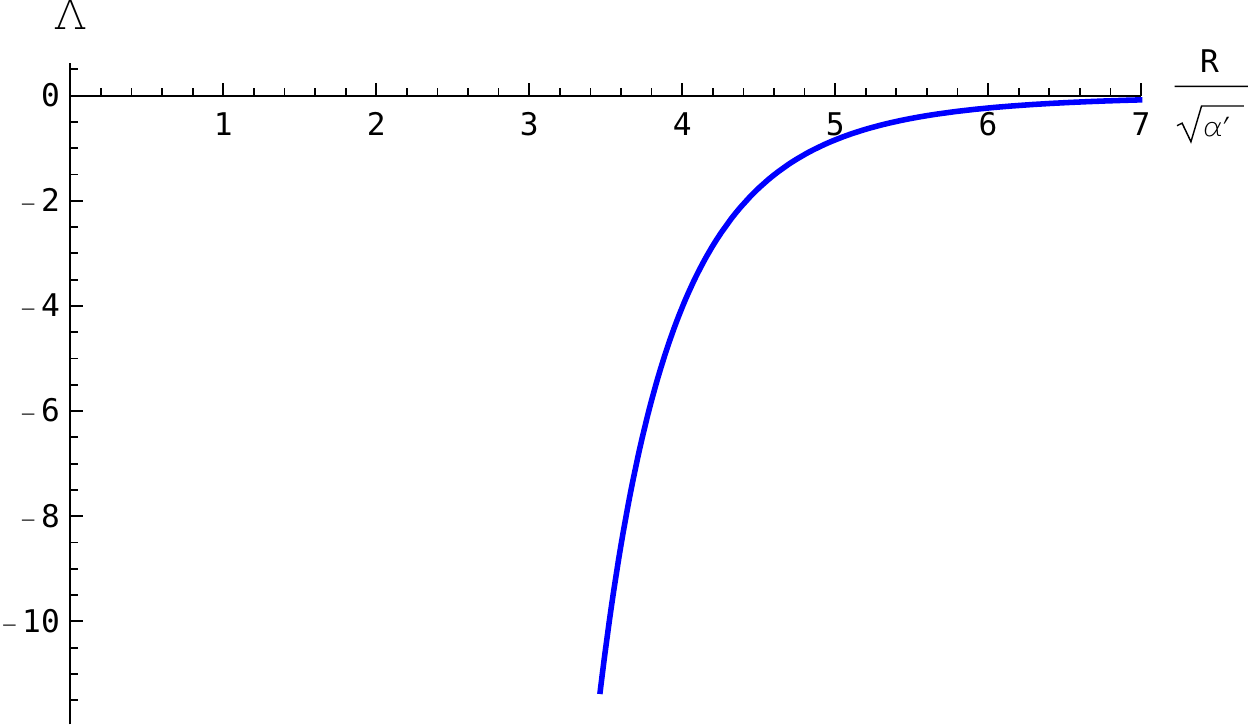}}
\\[-10mm]
\subfloat[\hspace{-5mm} \mbox{(a) $\bz_3$}]{} 
\\[-3mm]
\subfloat[]
{\includegraphics[scale=.45]{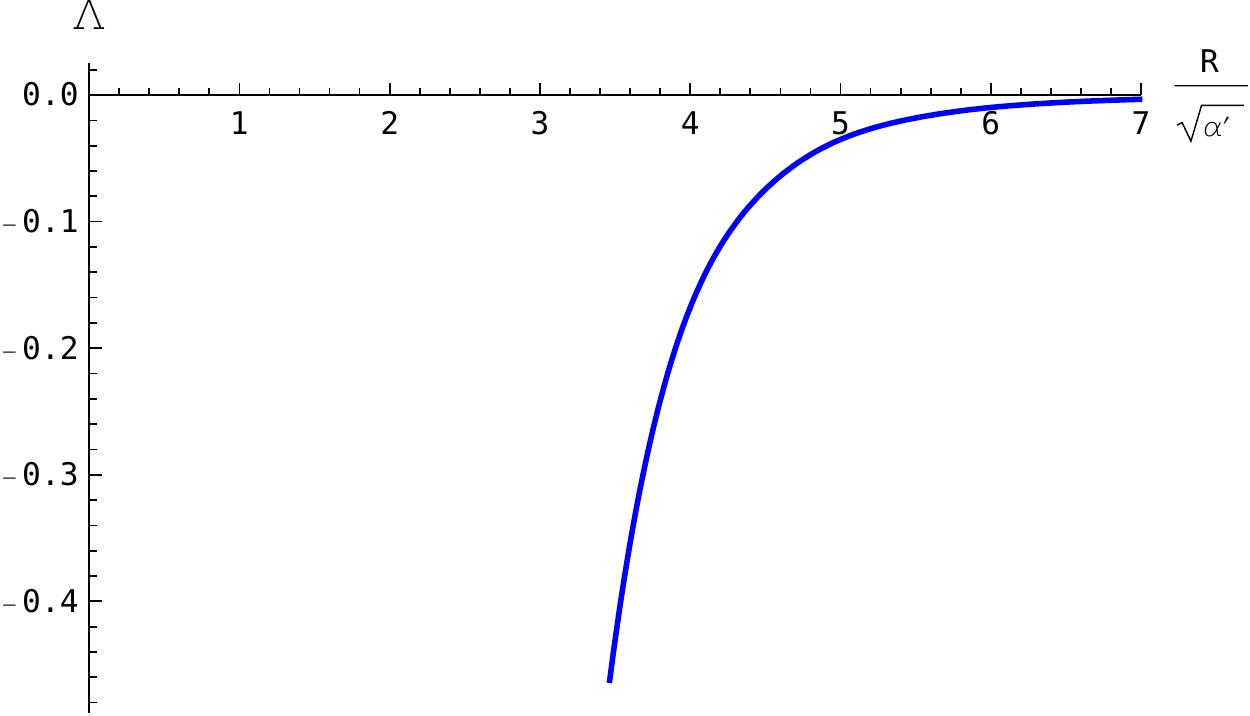}}\hspace{2cm}
\subfloat[]
{\includegraphics[scale=.45]{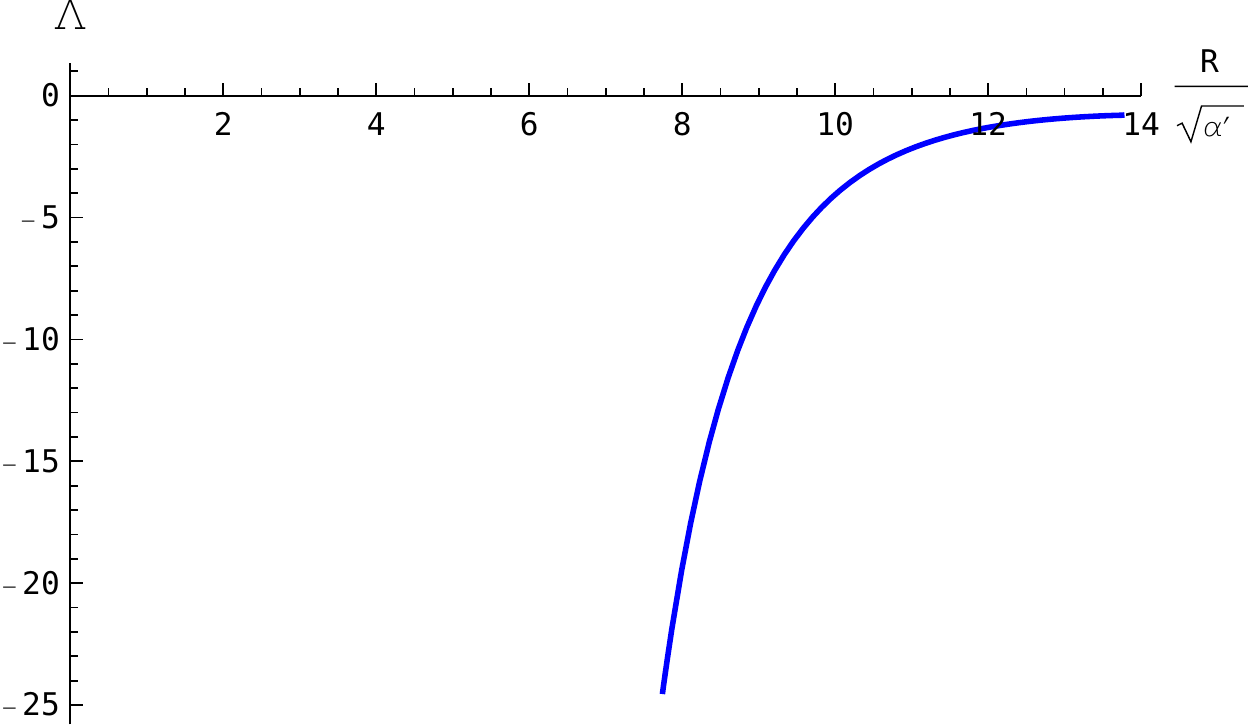}}
\\[-10mm]
\subfloat[\hspace{-5mm} \mbox{(b) $\bz_6$}]{} 
\caption{1-loop $\Lambda$ for $N=3,6$, class I $(\bt^2\times S^1)/\bz_N$ orbifolds with spin structures $s_g=0$ and $s_g=1$.
The vertical and horizontal axis correspond respectively to $\Lambda$ and the circle radius in units of $\sqrt{\alpha'}$. 
}
\label{fig_Z3_Z6}
\end{figure}

Figure \ref{fig_Z3_Z6} depicts the results, depending on whether the spin
structure $s_g$ is 0 or 1, for the models that only admit class I with $(s_1,s_2)=(0,0)$, 
namely the $\bz_3$ and $\bz_6$ orbifolds.  
The results for $\bz_2$ and $\bz_4$ orbifolds, that admit both class I and class II with $(s_1,s_2)\not=(0,0)$,
are shown in Figure \ref{fig_Z2_Z4}. For the $\bz_2$ orbifold there is only one curve for $s_g=0$ because in class I the model
with $s_g=1$ is equivalent, and in class II, as remarked below eq.~\eqref{average01}, 
$\cZ^{{\rm II}}_{(1, g)}=\cZ^{{\rm I}, s_g=0}_{(1, g)}$.
The purely toroidal contributions are different in class I and class II,
but the disparity is numerically insignificant. For the $\bz_4$ orbifold there are curves for class II and each of the two class I cases.

\begin{figure}[h!]
\centering
\subfloat[$\bz_2$]{\includegraphics[scale=.5]{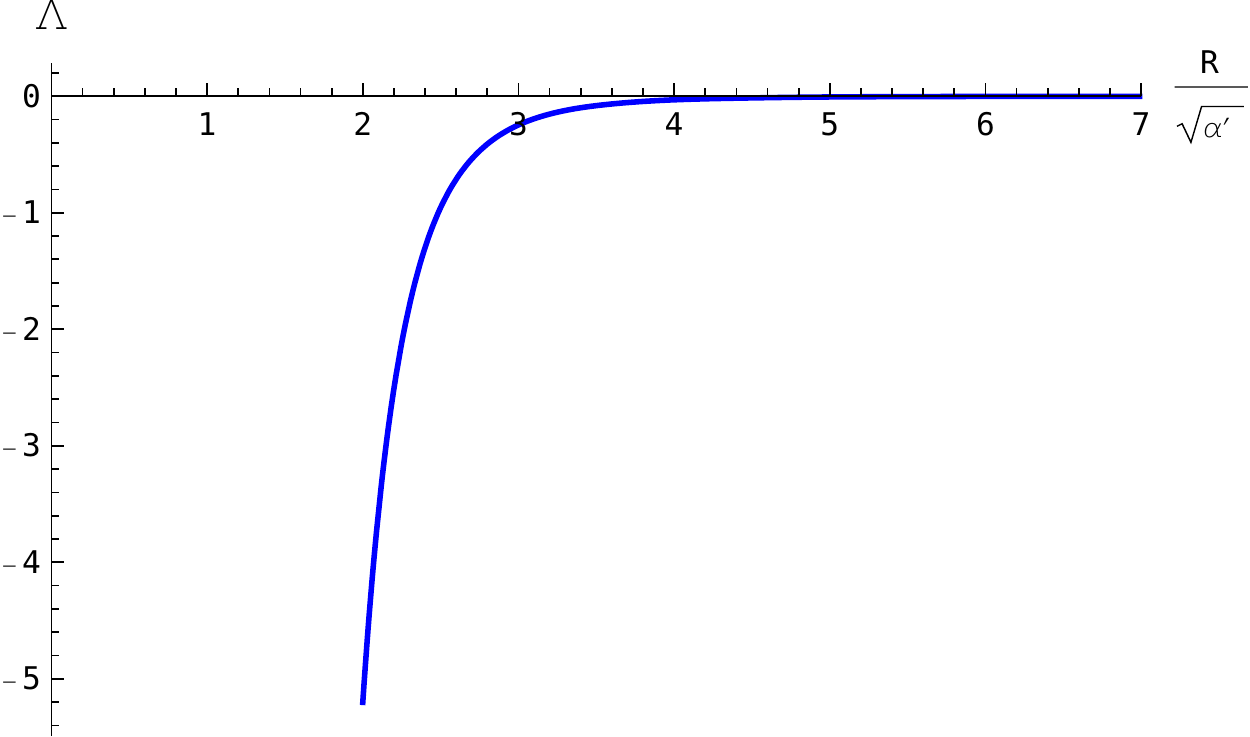}}\hspace{2cm}
\subfloat[$\bz_4$]{\includegraphics[scale=.5]{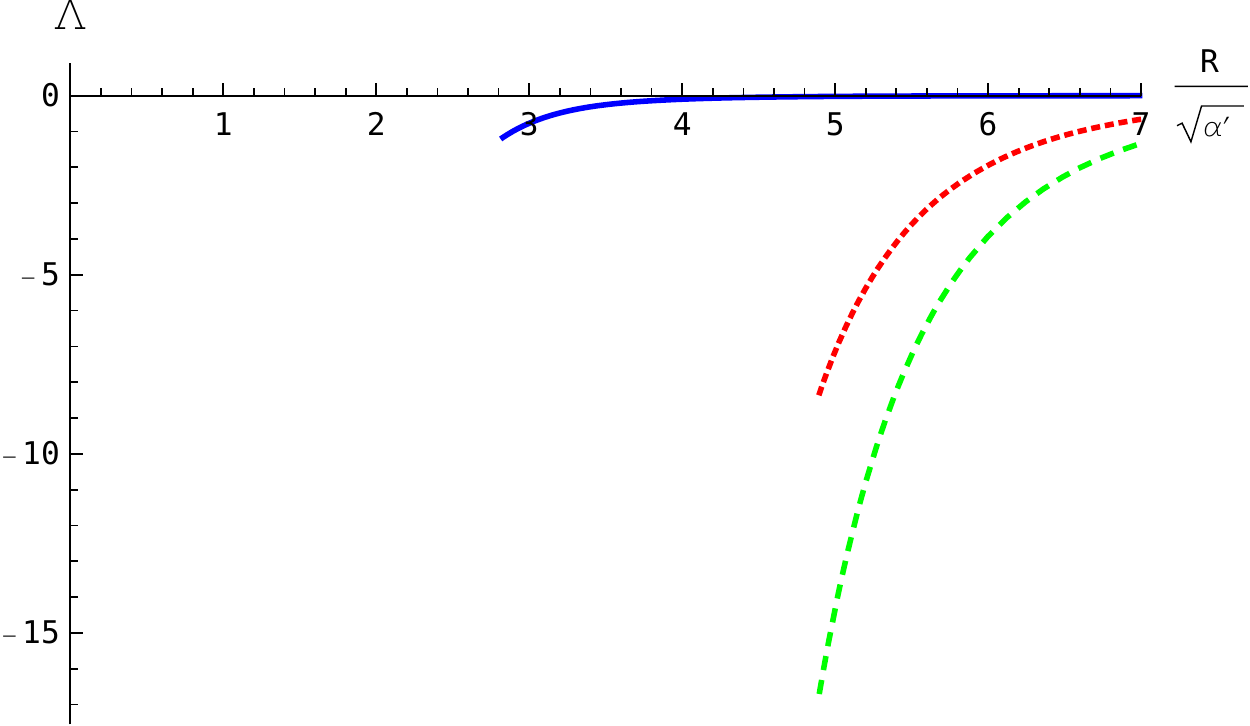}}
\caption{1-loop $\Lambda$ for $N=2,4$, $(\bt^2\times S^1)/\bz_N$ orbifolds. For $\bz_2$ the curve represents $\Lambda$
in the class I model with $s_g=0$. For $\bz_4$, solid blue, the large dashed green and short dashed red curves correspond respectively to 
class I $s_g=0$, class I $s_g=1$ and class II  with $(s_1,s_2)=(1,1)$.
}
\label{fig_Z2_Z4}
\end{figure}

The numerical results are found by integrating at fixed $R$ and then interpolating to obtain a smooth curve.
They confirm the large $R$ behaviour found in the analytical approach.
In all orbifolds $\Lambda$ tends to zero from below as $R^{-7}$ at large $R$. The negative sign is expected because in the 
$(\bt^2\times S^1)/\bz_N$ orbifolds there are more massless bosons than fermions. 
In Figure \ref{fig_Z2_Z4}, $\Lambda$ for the class II $\bz_4$ models can be seen as the average of the class I results
for $s_g=0$ and $s_g=1$, in agreement with exact properties of $\cZ^{{\rm II}}_{(1, g^p)}$ discussed
previously  

The plots clearly show that $\Lambda$ diverges at small $R$ when tachyons emerge in the spectrum. The precise
values of $R$ match the results in Table \ref{tachyon_all_ZN}, determined from the analysis of the spectrum.
To compare contributions from untwisted and twisted sectors, we plot $\Lambda$ for each sector separately for the $\bz_6$ model 
(other orbifolds show a similar behaviour). 
This is presented in Figure \ref{fig_lambda_Z6_sg01} for $s_g=0$ and $s_g=1$.
Since the twisted sectors $g^r$ and $g^{N-r}$ have the same partition functions we only included $r=1,2,3$ as well as the untwisted $r=0$.
In each sector the position of the vertical asymptotes agrees with the bound on $R$ given in Table \ref{tachyon_all_ZN}.
We also observe that for large $R$ the untwisted sector dominates, as we argued in the analytic approach. The twisted sectors are 
suppressed because they always host winding modes that become infinitely massive as $R \to \infty$.

\begin{figure}[h!]
\captionsetup[subfigure]{labelformat=empty, labelsep=none}
\centering
\subfloat[\mbox{\hspace*{-1cm} $s_g=0$}] {} \hspace{7cm}  \subfloat[\mbox{$s_g=1$}] {} \\[-3mm]
\subfloat[]
{\includegraphics[scale=.45]{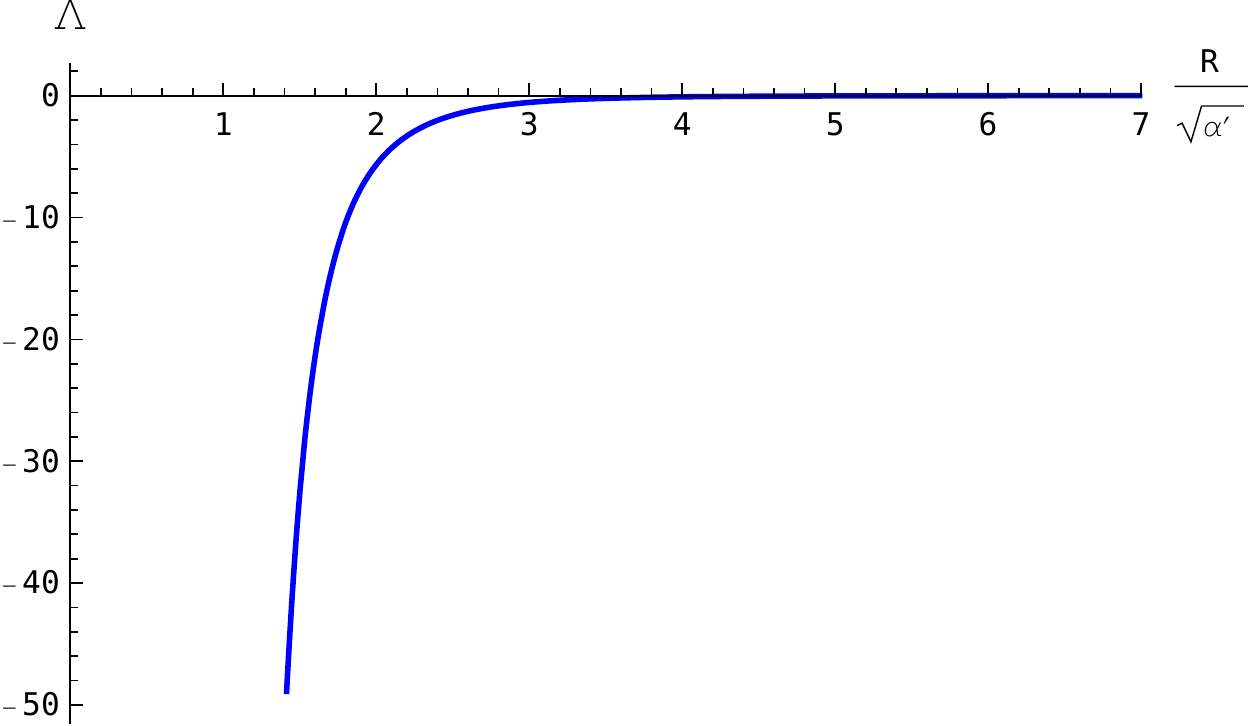}\label{Z6Sg0unt}}\hspace{2cm}
\subfloat[]
{\includegraphics[scale=.45]{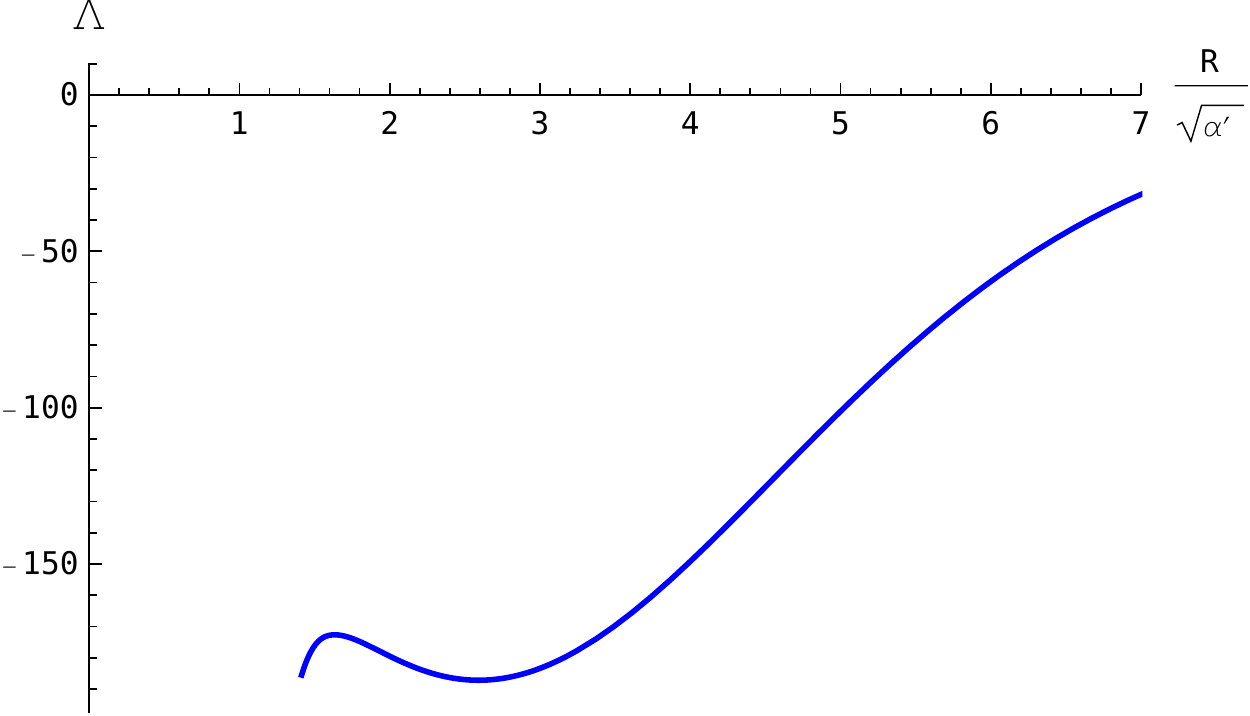}\label{Z6Sg1unt}}
\\[-10mm]
\subfloat[\mbox{\hspace*{-10mm} (a) $r=0$}]{} 
\\[-3mm]
\subfloat[]
{\includegraphics[scale=.45]{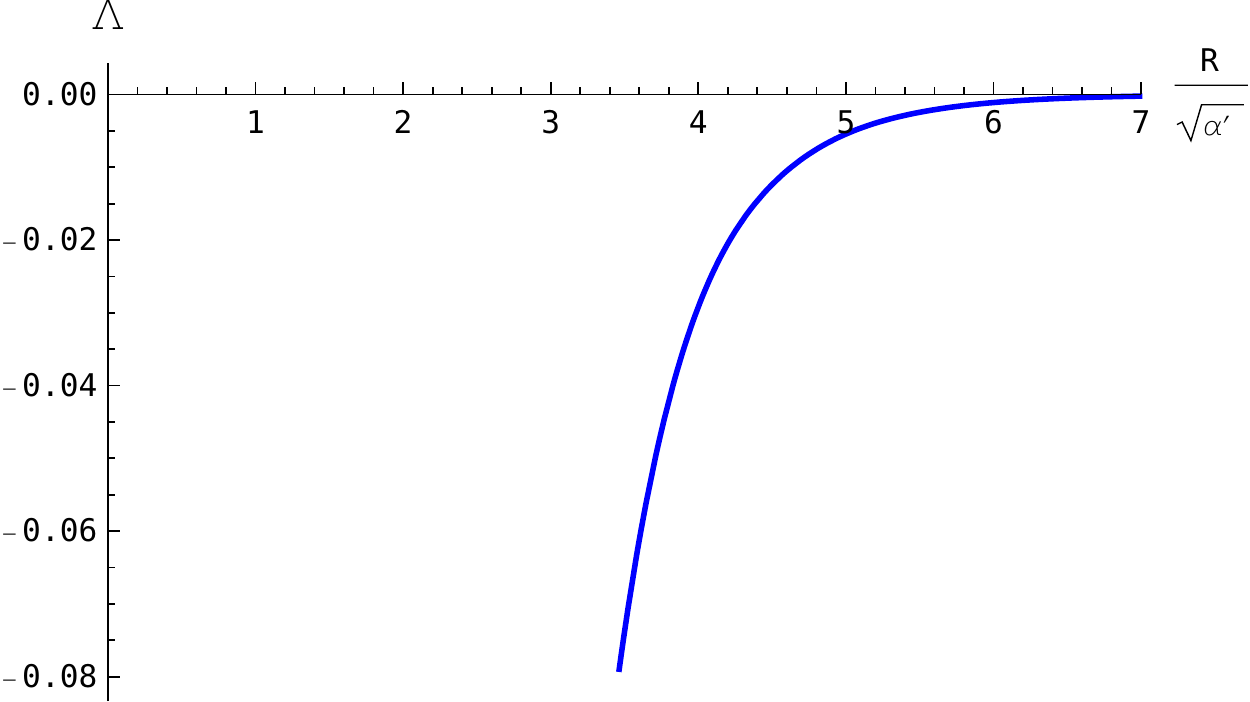}\label{Z6Sg0g}}\hspace{2cm}
\subfloat[]
{\includegraphics[scale=.45]{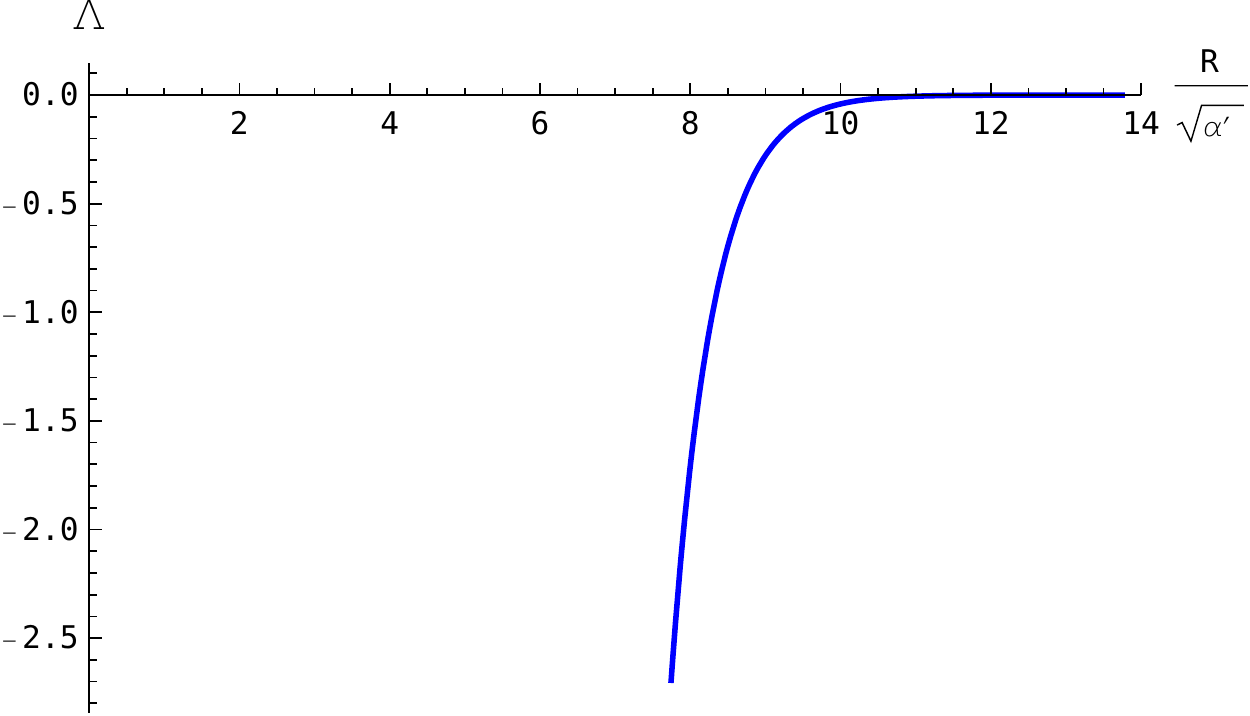}\label{Z6Sg1g}}
\\[-10mm]
\subfloat[\mbox{\hspace*{-10mm} (b) $r=1$}]{} 
\\[-3mm]
\subfloat[]
{\includegraphics[scale=.45]{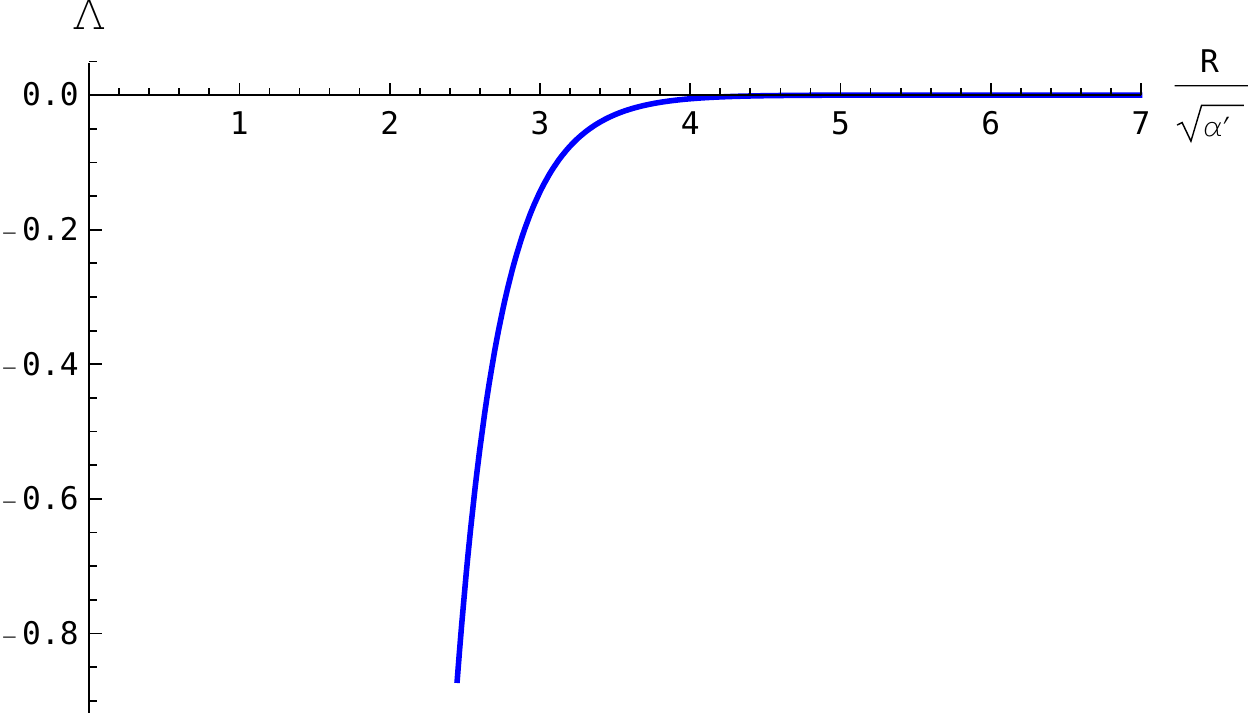}\label{Z6Sg0g2}}\hspace{2cm}
\subfloat[]
{\includegraphics[scale=.45]{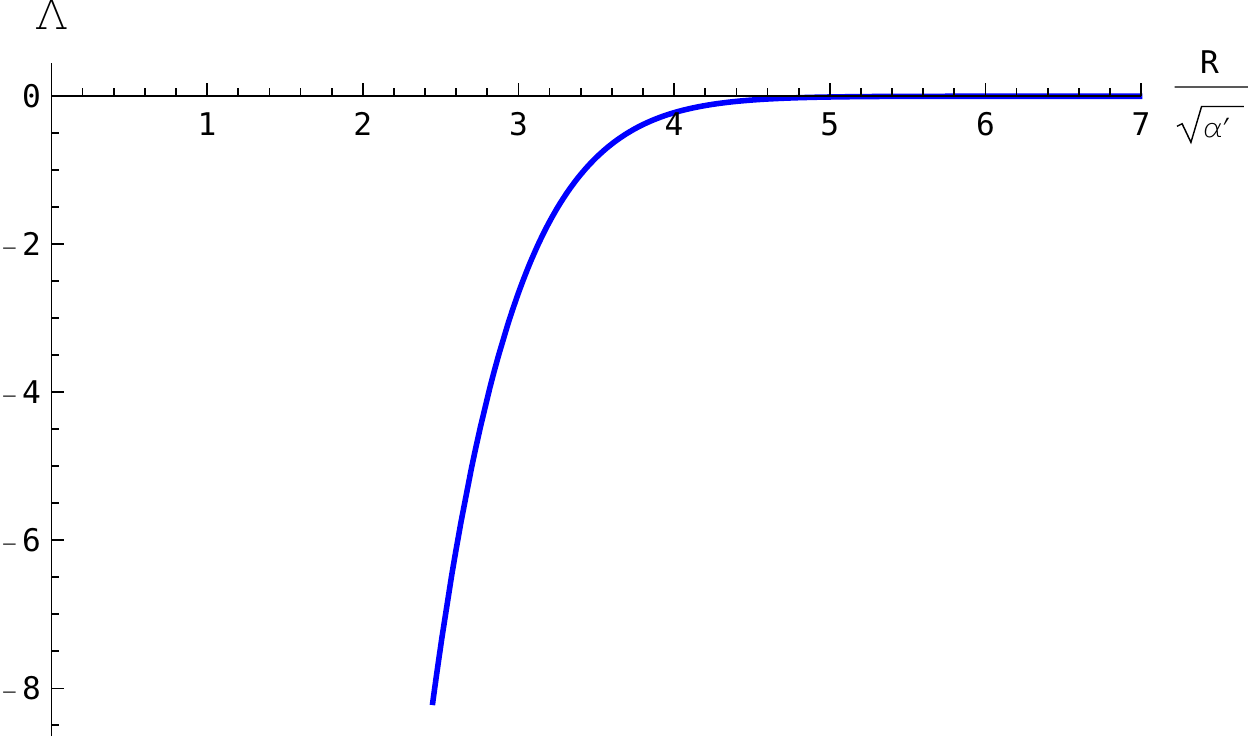}\label{Z6Sg1g2}}
\\[-10mm]
\subfloat[\mbox{\hspace*{-10mm} (c) $r=2$}]{} 
\\[-3mm]
\subfloat[]
{\includegraphics[scale=.45]{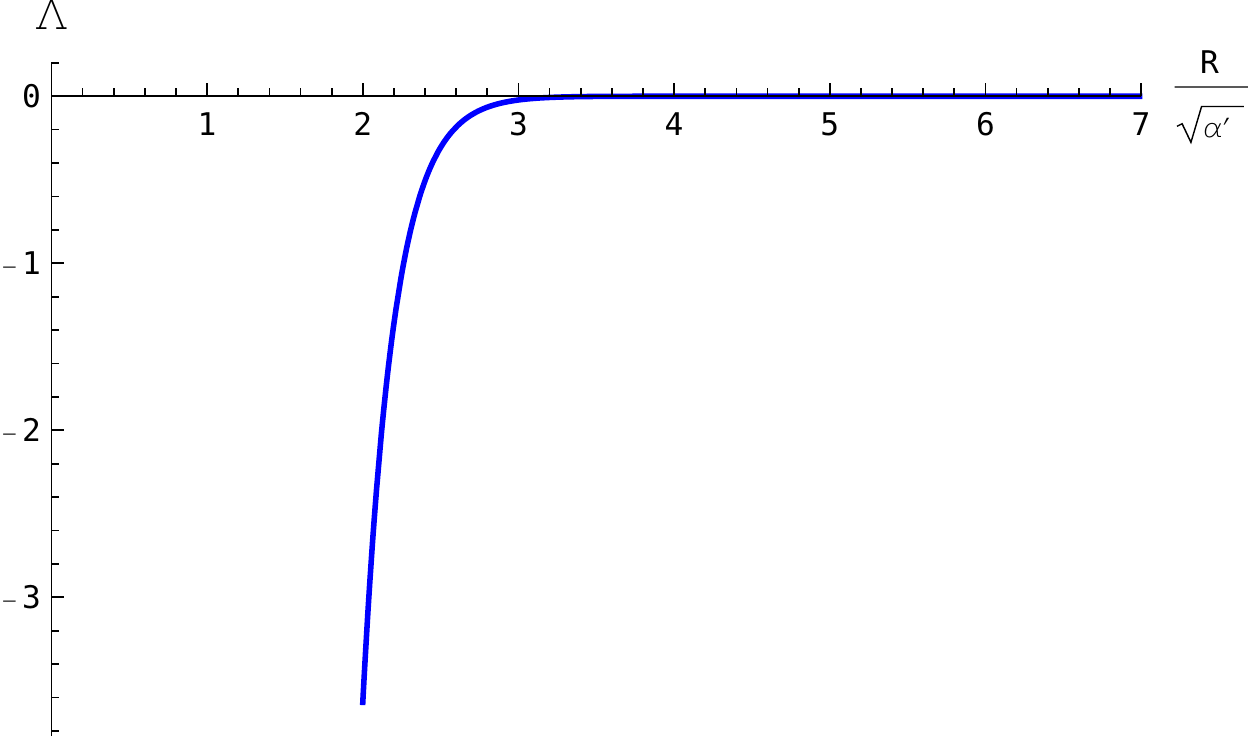}\label{Z6Sg0g3}}\hspace{2cm}
\subfloat[]
{\includegraphics[scale=.45]{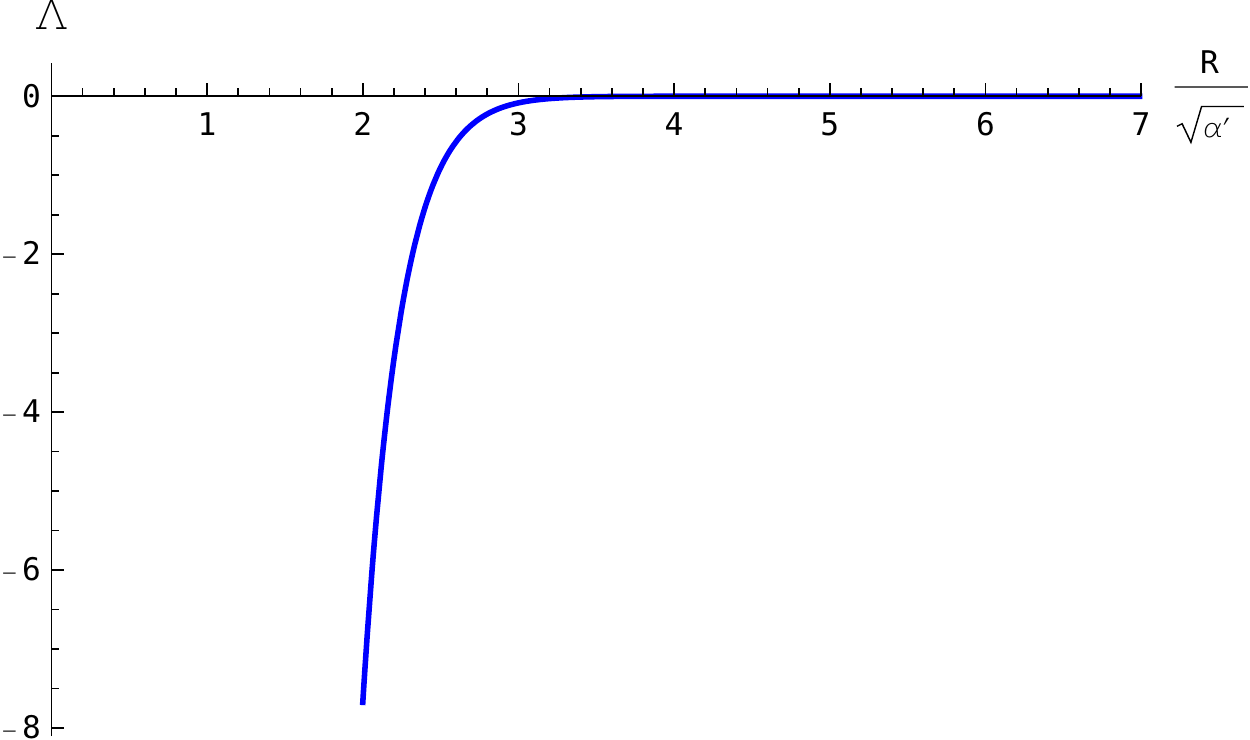}\label{Z6Sg1g3}}
\\[-10mm]
\subfloat[\mbox{\hspace*{-10mm} (d) $r=3$}]{} 
\caption{Contribution from $g^r$-twisted sectors to 1-loop $\Lambda$ in class I $\bz_6$ orbifold with $s_g=0,1$.}
\label{fig_lambda_Z6_sg01}
\end{figure}

\section{\texorpdfstring{$(\bt^d\times S^1)/\bz_N$}{(TdxS1)/ZN} backgrounds}
\label{section_Td}

In this section we briefly comment on the generalisation of these models to higher dimensions. A detailed study of these theories will be presented in \cite{upcoming}. We consider $\bt^d$, $d=4,6$, with one extra circle.. 
Let $g = ({\mathcal R}_g, \mathcal{T}_g)$ be the generator of $\bz_N$ which acts as rotations (${\mathcal R}_g$) 
on various planes of $\bt^4$ or $\bt^6$ and a translation{\footnote{If we set the translation equal to zero, then we get an orbifold with fixed points and in that case $S^1$ is a spectator and can be decompactified (after choosing the periodic spin structure on  $S^1$),
without affecting the results.}}
($\mathcal{T}_g$) on $S^1$. 

We consider $\bt^6$ and diagonalise ${\mathcal R}_g$ by using appropriate complex coordinates:
\begin{equation}
{\mathcal R}_g :(z_1,z_2,z_3) \rightarrow (e^{2\pi i\frac{n_1}N}z_1 ,e^{2\pi i\frac{n_2}N}z_2, e^{2\pi i\frac{n_3}N}z_3)
\label{ztrans}
\end{equation}
where $n_i$ are integers with $0 \leq n_i <N$,  and satisfying  $p (\tf{n_1}N, \tf{n_2}N, \tf{n_3}N) \notin \bz^3$, for $p \in \bz$ and 
$0 < p <N$. This last condition just says that the orbifold group is really $\bz_N$, and not a subgroup thereof, because there is no $p$
such that ${\mathcal R}_g^p=1$.
We also allow one or two of the $n_i$ to be possibly zero, so that the general formula we will get below for the partition functions 
can also describe $\bt^4/\bz_N$ and $\bt^2/\bz_N$, respectively. If we set  $n_2=n_3=0$, then we get the $\bt^2$ orbifolds of section \ref{section_T3}. If we set, say,  $n_3=0$ then we get $\bt^4$ orbifolds. In this case we can decompactify the last $\bt^2$ defined by $z_3$ (after setting the standard spin structure on  this $\bt^2$) to get ($\bt^4 \times S^1)/\bz_N$ model. 

The order $N$ and the integers $n_i$ are constrained by the requirement that there must exist a torus lattice where 
the rotation ${\mathcal R}_g$ acts crystallographically. It is well known that the order must be $N=2,3,4,....$ through $10,12,14,15,18,20,24,30$
\cite{Dixon:1986jc}, see also \cite{Erler:1992ki} where the possible $n_i$ are tabulated. Imposing that supersymmetry is unbroken
would give additional constraints but here we keep the discussion general. Given an allowed ${\mathcal R}_g$ one still has to
specify the lattice where it acts as an automorphism. In the supersymmetric case such lattices have been classified \cite{Fischer:2012qj}.
To simplify our analysis we will assume that the lattice has some properties to be stated shortly.

Finally, we take the translation part $\mathcal{T}_g$ to be the shift by $\frac{t}{N}R$  on $S^1$ where $R$ is the radius of $S^1$ and $t$ is an integer with $0\leq t < N$. If we want a truly freely acting orbifold then $t$ should be non-zero and, moreover, coprime with respect to $N$. The reason for the latter condition is, that if $(t,N)$ are not coprime then there exist integers $p$ and $q$ with $p < N$ such that $p t= q N$. This would then mean that $g^p$ sector will have fixed point singularities. Note that we could have also included translations along  the 
planes rotated by $\mathcal{R}_g$, but they are irrelevant, since in all the twisted sectors $(g^r,g^p)$ where $(r,p) \neq (0,0)$, the momenta along these directions are zero.

The action of $g$ in momentum space is given by:
\begin{equation}
g\,|k;W_1,W_2,W_3,W_4\rangle = e^{\frac{2\pi i}{N}t k}
e^{\frac{2\pi i}{N}(n_1 W_1+ n_2 W_2+n_3 W_3)} |k;W_1,W_2,W_3,W_4\rangle\ ,\quad W_i \in\bz
\label{gB}
\end{equation}
for $SO(8)$ bosons and
\be
\begin{split}
&\hspace*{-3mm} g\,|k+\tf{s_7}{2};W_1,W_2,W_3,W_4\rangle = \\
&(-1)^{s_g}~e^{\frac{2\pi i}{N}t( k+\frac{s_7}{2})} 
e^{\frac{2\pi i}{N}(n_1 W_1+ n_2 W_2+n_3 W_3)}|k+\frac{s_7}{2};W_1,W_2,W_3,W_4\rangle\ ,\quad W_i \in\bz +\tf{1}{2} 
\end{split}
\label{gF}
\ee
for $SO(8)$ fermions. Here $W_i$ are $SO(8)$ weights, $s_i$, $i=1,\cdots,7$, correspond to spin structures along 
$\bt^6 \times S^1$, and $s_g$ is the spin structure associated with the $g$-action on fermions.
The $g^N=1$ condition acting on bosons is clearly satisfied, as all the weights $W_i$ are integers. 
On fermions, since weights are shifted by half, imposing $g^N=1$  gives the constraint
\begin{equation}
(-1)^{N s_g + t s_7+n_1+n_2+n_3}=1\ .
\label{ZNcondition}
\end{equation}
This is exactly the same constraint as \eqref{condition} that we obtained for $\bt^3$ orbifolds for $n_1=t=1$ and $n_2=n_3=0$.

We next consider the partition function in the $(g^r,g^p)$ sectors for class I models, i.e. $s_i=0$ for all $i$ corresponding to planes that are rotated by $g$. To simplify the analysis we will assume that the $\bt^6$ lattice is such that for all sectors the invariant sublattices of $\Gamma$
(including winding and momenta) that appear are even and self-dual. There could still be shifts and phases encoded in the arguments
$(x,y;u,v)$ of the lattice sum $Z_\Gamma$, as in eq.~\eqref{ZGamma}. The steps to construct the partition function are then the same
as in the $(\bt^2 \times S^1)/\bz_N$ orbifolds. 
If ${\rm gcd}(r,p) \tf{n_i}{N}\notin\bz$, $i=1,2,3$, we find the partition function:
\be
\begin{split}
&\cZ_{(g^r,g^p)}=\frac{1}{4}\int_{\cal F}  \frac{d^2 \tau}{\tau_2^{\frac52} }
\bigg|\frac1{\eta^3(\tau)}\prod_{ i=1}^3\frac{2\sin({\rm gcd}(r,p) \pi \frac{n_i}{N})}
{\vartheta_{(\frac{1}{2},\frac{1}{2})}(\frac{r n_i}{N}\tau+\frac{p n_i}{N},\tau)}\bigg|^2
\sum_{\alpha_L, \beta_L,\alpha_R ,\beta_R=\{0,\frac12\}}  \hspace*{-10mm} 
{\mathcal C}(\alpha_L, \beta_L,\alpha_R ,\beta_R) \times
\\
&\times \; 
(-1)^{2s_g\big(p(\alpha_L+\alpha_R)+r(\beta_L+\beta_R)\big)}(-1)^{-2\pi i(\beta_L+\beta_R)\frac rNs_7}\times \\
&\times\;\vartheta_{(\alpha_L,\beta_L)}(\tau)\Big(\prod_{i=1}^3\vartheta_{(\alpha_L,\beta_L)}(\tf{(r\tau+p)n_i}{N},\tau)\Big)\;\;
\bar\vartheta_{(\alpha_R,\beta_R)}(\bar\tau)\;\Big(\prod_{i=1}^3\bar\vartheta_{(\alpha_R,\beta_R)}(\tf{(r\bar{\tau}+p)n_i}{N},\bar{\tau})\Big)\times \\
&\hspace*{5cm}\times\;Z_{\Gamma_{(r,p)}}\big(\tau,(\alpha_L+\alpha_R)s_7,\tf{rt}{N};\tf{pt}{N},(\beta_L+\beta_R)s_7\big)\, , 
\end{split}
\label{ZrpclassI} 
\ee
where $\Gamma_{(r,p)}$ is the sublattice of $\Gamma$ that is invariant under $g^{r}$ and $g^p$ simultaneously. If, as it is assumed above, 
${\rm gcd}(r,p)\tf{n_i}{N} \notin\bz$ for all $i=1,2,3$, then $\Gamma_{(r,p)}$ is just the lattice of momenta and windings along $S^1$. If on the other hand, ${\rm gcd}(r,p) \frac{n_i}{N}\in\bz$ for some $i$, then the $i^{\rm th}$ plane will not be rotated by both $g^r$ and $g^p$, and the lattice $\Gamma_{(r,p)}$ will include momenta and windings along those $i^{\rm th}$ planes as well. From our assumption on the 
$\bt^6$ lattice, $\Gamma_{(r,p)}$  will still be even and self-dual. 
Actually, the above formula holds even if the spin structures along those $i^{\rm th}$ planes are non-zero, provided these
$s_i$ are taken into account in the lattice sum.

The partition functions for class II models are given by eq. \eqref{ZrpclassI} upon inserting the projector 
${\mathcal P}(\alpha_L,\alpha_R,\beta_L,\beta_R; r,p)$ defined in \eqref{projdef}, see eq. (\ref{ZrpII}). 
As a result, in class II models, $s_g$ drops out completely just as we observed before for $\bt^2\times S^1$ orbifolds.

\section{Conclusions}
\label{conc}

In this paper we studied compactification of type II string theory on 3-dimensional non-simply connected compact Ricci flat spin manifolds which are all smooth quotients of the form $(\bt^2\times S^1)/\bz_N$. These manifolds admit multiple spin structures which can be both periodic and ant-periodic along the 1-cycles of $\bt^2$ and $S^1$. Except from the original toroidal compactification on $\bt^3$ (with totally periodic spin structure), all the remaining models provide instances of non-supersymmetric compactifications of string theory.

One goal was to develop the worldsheet conformal field theory of string/M-theory compactification on manifolds with spin structures in various dimensions, in order to understand the nature of instabilities at the quantum level and its connection to geometrical properties of the manifold. In this paper, we focused on the lowest dimension where such manifolds exist, namely dimension three \cite{Pfaeffle,Acharya:2019mcu}. (In dimensions one and two the only orientable compact manifolds are $S^1$ and $\bt^2$, respectively). We constructed models which explicitly realise the spacetime spin structures in the worldsheet conformal field theory. We computed the 1-loop partition function and spectrum of these theories and found that, except for the supersymmetric $\bt^3$, all of them contain tachyons in their spectrum, and that the tachyons emerge only for sufficiently small circle radii.

We classified these models into two groups: those which have anti-periodic spin structure only along the $S^1$, which we dubbed
class I models, and those which have anti-periodic spin structures along the 1-cycles of the $\bt^2$, which are termed class II models --- see Table \ref{table3bobby}. Class I models are the higher dimensional generalisation of circle compactifications constructed in earlier works
\cite{Rohm:1983aq, Atick:1988si, Kounnas:1989dk}. On the other hand, class II models are not studied in detail in the literature 
and we analysed them carefully.  In particular, we studied the structure of their twisted Hilbert spaces and examined the closure of their 
operator algebras.

In order to probe these non-supersymmetric orbifold theories to greater extent, we computed the 1-loop vacuum energy density, $\Lambda$,
both analytically and numerically. We find that all orbifold backgrounds $(\bt^2\times S^1)/\bz_N$ have the same qualitative behaviour in developing instabilities for type II compactifications. In particular, there is no qualitative difference between the two families of 
$\bz_3$ orbifolds with periodic and anti-periodic spin structure along the circle, which is the only model with this property in the 
constructions of \cite{Pfaeffle,GarciaEtxebarria:2020xsr}.
The numerical calculations confirmed the analytical estimates at large radius
and showed that all class I orbifold models have the same instability behaviour at small radius.

One next natural step is to study higher dimensional generalisations of these models. This is briefly discussed in section \ref{section_Td} and 
will be extended in \cite{upcoming}. Higher dimensional orbifolds offer many families of non-supersymmetric compactifications of string 
and M-theory and understanding their instability behaviour is of great interest in particular in the context of compactifications into 
four spacetime dimensions.

Another interesting direction to explore is the dynamics of tachyon condensation in these theories. The worldsheet theory has enhanced superconformal symmetry ($\cN=2$ or higher) for particular values of $R$. In this case one can study tachyon condensation analytically 
through analysing the RG flow of the worldsheet theory from UV to IR. Switching on tachyon condensation deforms the worldsheet theory 
with relevant operators and as such, breaks the conformal symmetry. However, the $\cN=2$ (or higher) worldsheet supersymmetry is preserved and 
provides a concrete way of analysing the RG flow by studying the chiral ring of the theory. In particular cases, the tachyon might become massless under some conditions, e.g. at specific radii, and the perturbation yields an exactly marginal deformation of the original CFT. 
References \cite{Vafa:2001ra} and \cite{Harvey:2001wm} developed this technique and studied deformations of 
non-compact orbifolds, e.g. $\mathbb C^d/\bz_n$, through tachyon condensation of closed strings.{\footnote{Other approaches to tachyon condensation of non-compact orbifolds are studied in, e.g. \cite{Sen:1998tt,Adams:2001sv,David:2001vm}.}} Moreover, \cite{Vafa:2001ra} studied compact orbifolds and realised the RG flow in the mirror description of the theory. It would be interesting to study the mirror 
description of these theories and to understand tachyon condensation at least at special radii where the worldsheet supersymmetry is accidentally enhanced.

\section*{Acknowledgments}

We are grateful to L.~Ib\'a\~nez, M.~Montero, F.~Quevedo, S.~Theisen, E.~Witten
for interesting comments and valuable insights. 
A.~Font and G. Aldazabal  acknowledge hospitality and support from ICTP, and 
IFT UAM-CSIC via the Centro de Excelencia Severo Ochoa Program under Grant SEV-2016-0597.
This work is partially supported  by CONICET grant PIP-11220110100005  and   
PICT-2016-1358. The work of BSA is supported by a grant from the Simons Foundation (\#488569, Bobby Acharya).

\appendix

\section{Theta functions and lattice sums}\label{app_thetas_lattices}

The basic ingredients in the partition function of strings on toroidal orbifolds are
Jacobi Theta functions and lattice sums. In this appendix we offer
a handy summary of definitions and fundamental properties.

\subsection{Theta functions}\label{app_thetas}
The Jacobi Theta functions $ \vartheta_{(\alpha,\beta)}(z,\tau)$ are defined as
\be
 \vartheta_{(\alpha,\beta)}(z,\tau)= \sum_{n=-\infty}^{\infty} e^{i\pi\tau(n+\alpha)^2 + 2 i \pi (n+\alpha)(z+\beta)}\, .
\label{sumform}
\ee 
There is also a product form 
 \be
 \vartheta_{(\alpha,\beta)}(z,\tau) =e^{2\pi i \alpha(z+\beta)} q^{\frac{\alpha^2}2}
 \prod_{n=1}^\infty(1-q^n)(1+e^{2\pi i (z+\beta)} q^{n+\alpha-\frac{1}{2}})
 (1+e^{-2\pi i (z+\beta)} q^{n-\alpha-\frac{1}{2}}) \, ,
 \label{prodform}
 \ee
where $q=e^{2i\pi\tau}$. In particular, 
\be
\vartheta_{(\frac12,\beta)}(z,\tau) = 2 \cos \pi (z+\beta)\, q^{\frac18} \prod_{n=1}^\infty(1-q^n)
(1+e^{2\pi i (z+\beta)} q^{n})(1+e^{-2\pi i (z+\beta)} q^{n}) \, .
\label{prodform2}
\ee
It is conventional to introduce the special cases
\be\label{theta1234}
\vartheta_{(0,0)}(0,\tau):=\vartheta_3(\tau)\ ,\quad\vartheta_{(0,\frac{1}{2})}(0,\tau):=\vartheta_4(\tau)\ ,\quad
\vartheta_{(\frac{1}{2},0)}(0,\tau):=\vartheta_2(\tau)\ ,\quad\vartheta_{(\frac{1}{2},\frac{1}{2})}(0,\tau):=\vartheta_1(\tau)\ .
\ee
They satisfy the `abtruse' identity
\begin{equation}
\vartheta_3^4(\tau) - \vartheta_4^4(\tau) - \vartheta_2^4(\tau) \mp \vartheta_1^4(\tau) =0 \ .
\label{abtruse}
\end{equation} 
Although $\vartheta_1(\tau)$ vanishes identically, it is convenient to keep it throughout.

We will need several properties under transformations of the characteristics $(\alpha, \beta)$ and the arguments $(z,\tau)$.
From the definition we can easily derive
\begin{align}
 \vartheta_{(-\alpha,-\beta)}(z,\tau)&= \vartheta_{(\alpha,\beta)}(-z,\tau)\, , 
 \label{abchsgn} \\[2mm]
 \vartheta_{(\alpha+m,\beta+m')}(z,\tau)&= e^{2 \pi i m' \alpha} \vartheta_{(\alpha,\beta)}(z,\tau) \, .
\label{abintshift}
\end{align}
Under $z \rightarrow z+ m_1 \tau+ m_2$ for $m_1,m_2 \in {\bz}$, the $\vartheta$ functions transform as
  \begin{equation}
    \vartheta_{(\alpha,\beta)}(z+m_1 \tau+m_2,\tau)  = e^{-\pi i m_1^2 \tau -2 \pi i m_1 z+2 \pi i(m_2 \alpha-m_1 \beta)}
    \vartheta_{(\alpha,\beta)}(z,\tau) \, .
    \label{monodromy}
  \end{equation}
Under a modular transformation by an $SL(2,\bz)$ element $h=\begin{pmatrix}
 a& b\\
  c&d
 \end{pmatrix}$, and with  $\tau'= h(\tau)= \frac{a \tau+b}{c \tau+d}$ and $z'=h(z)= \frac{z}{c\tau+d}$, we have
\begin{equation}
  \vartheta_{(\alpha,\beta)}(z,\tau)=\epsilon((\alpha,\beta),h)\frac{1}{\sqrt{c\tau+d}} 
  e^{-\pi i \frac{c z^2}{c\tau+d}} \vartheta_{(\alpha',\beta')}(z',\tau')
  \label{thetamod}
\end{equation}
where $\alpha'=h(\alpha)= d \alpha-c \beta + \frac{1}{2} (c-d+1)$ and $\beta'=h(\beta)= -b \alpha+a \beta+ \frac{1}{2}(b-a+1)$,
and $\epsilon((\alpha,\beta),h)$ is a phase to be discussed shortly.
Note that for $\alpha, \beta \in \{0,\frac12\}$, $\alpha'$ and $\beta'$ 
can also be brought to take values $0$ or $\frac{1}{2}$  by using the relation \eqref{abintshift}

An important property of equation (\ref{thetamod}) is that it satisfies
the $SL(2,\bz)$ group composition rule, namely if we express $\vartheta_{(\alpha',\beta')}(z',\tau')$ on the RHS in terms of another 
transformation by $h'$, using the same relation with $h$ replaced by $h'$,
then the result is the same as starting from $\vartheta_{(\alpha,\beta)}(z,\tau)$ and using the transformation $h'h$, 
since the phase $\epsilon$ satisfies
\begin{equation}
  \epsilon((\alpha,\beta),h) \epsilon(h(\alpha,\beta),h')= \epsilon((\alpha,\beta),h'h) \, .
  \label{phaserel}
\end{equation}
For our purposes we do not need the explicit form of the phase $\epsilon((\alpha,\beta),h)$ but rather
the results for standard $T$ and $S$ transformations given by \\
\noindent 1) for $T:\, \tau' = \tau+1$, which corresponds to $a=b=d=1$ and $c=0$ in \eqref{thetamod}:
\begin{equation}
 \vartheta_{(\alpha,\beta)}(z,\tau)= e^{\pi i \alpha(\alpha-1)}\vartheta_{(\alpha,\beta-\alpha+\frac{1}{2})}(z,\tau')
 \label{thetaT}
 \end{equation}  \\
 \noindent 2) for $S:\, \tau'=-1/\tau$, which corresponds to $a=d=0$ and $b=-c=-1$  in \eqref{thetamod}:
 \begin{equation}
 \vartheta_{(\alpha,\beta)}(z,\tau)=\sqrt{-i\tau'} e^{\pi i  z^2 \tau'}e^{2\pi i \alpha \beta} \vartheta_{(\beta,-\alpha)}(z \tau',\tau') 
 \label{thetaS}
\end{equation}\\
Note that in these two transformations the phase is determined explicitly. One can in principle obtain $\epsilon((\alpha,\beta),h) $ for an 
arbitrary $SL(2,\bz)$ transformation by repeatedly applying $S$ and $T$ transformations.
However, we really need only the $(\alpha, \beta)$ dependence of the fourth power of this phase  
since every term  in the partition function appears with 4 powers of $\vartheta$ functions, namely of the form
$\prod_{i=1}^4 \vartheta_{(\alpha, \beta)}(z_i, \tau)$. 
The correct expression for the  fourth power of this phase turns out to be \cite{AlvarezGaume:1986es, Brezhnev:2013}
\begin{equation}
  \epsilon((\alpha,\beta),h)^4 = e^{4\pi i ( b d \alpha^2-2 c b \alpha \beta + a c \beta^2 -ab(d\alpha - c\beta))}\tilde\epsilon(h)
  \label{epsilon}
  \end{equation}
where $\tilde\epsilon(h)$ is $\pm 1$ and depends only on the $SL(2,\bz)$ element, which will again not be needed, since the same 
$\pm 1$ appears both in the left and right moving sectors and therefore will cancel out.
One can check that \eqref{epsilon} agrees with the result for $S$ and $T$ transformations given above. 
Furthermore, it furnishes a representation of the $SL(2,\bz)$ group in the  sense of (\ref{phaserel}).

To conclude we recall the definition of the Dedekind $\eta$-function
\be
\eta(\tau) = q^{1/24} \prod_{n=1}^\infty(1- q^n) \, .
\label{etadef}
\ee
The $T$ and $S$ transformation rules are
\be
\eta(\tau) = e^{-i\pi/12} \, \eta(\tau'), \ \tau'=\tau+1, \quad
\eta(\tau) = \sqrt{-i \tau'}\, \eta(\tau'), \ \tau'=-\tf1\tau\, .
\label{etaTS}
\ee

\subsection{Lattice sums}\label{app_lattice}

In section \ref{sub_T3} we introduced the lattice sum $ Z_{\Gamma}(\tau,x,y;u,v)$ defined as
\be\label{ZGamma2}
 Z_{\Gamma}(\tau,x,y;u,v):=\sum_{k,w\in\bz^d} q^{\frac{1}{2}\sum\limits_{i=1}^d\big(\frac{k_i+x_i}{2 R_i}+(w_i+y_i)R_i\big)^2}
 \bar{q}^{\frac{1}{2}\sum\limits_{i=1}^d\big(\frac{k_i+x_i}{2 R_i}-(w_i+y_i)R_i\big)^2}e^{2\pi i \sum\limits_{i=1}^d\big(u_i(k_i+x_i)+v_i(w_i+y_i)\big)}\ ,
\ee
where $x,y,u,v$ are $d$-dimensional real vectors with components $x_i$, $y_i$, $u_i$ and $v_i$. 
Shifting $x$ and $y$ by integers can be undone by redefining $k$ and $w$, so $x,y \in\R^d/\bz^d$. Shifting $u$ and $v$ by integers, however, multiplies the lattice partition function by overall phases. One can show easily that for $m_i \in\bz^d$, $i=\{1,\ldots,4\}$:
\begin{equation}
  Z_{\Gamma}(\tau,x+m_1,y+m_2;u+m_3,v+m_4)=  e^{2 \pi i(m_3.x+m_4.y)}Z_{\Gamma}(\tau,x,y;u,v)\ .
  \label{period}
\end{equation}
Furthermore, under $SL(2,\bz)$ modular transformations $\tau'=\frac{a \tau+ b}{c \tau +d}$,
\begin{equation}
  (\tau_2)^{\frac{d}{2}}  Z_{\Gamma}(\tau,x,y;u,v) =e^{2 \pi i(a c u v - b c u x-b c v y +b d x y)}(\tau'_2)^{\frac{d}{2}} Z_{\Gamma}(\tau',x',y';u',v')\ ,
  \label{Zmod}
\end{equation}
where $\tau=\tau_1+i\tau_2$ and
\begin{equation}
\begin{pmatrix}x'\\v'\end{pmatrix}=
\begin{pmatrix}d&-c\\-b&a\end{pmatrix}
\begin{pmatrix}x\\v\end{pmatrix}\ ,\qquad
\begin{pmatrix}y'\\u'\end{pmatrix}=
\begin{pmatrix}d&-c\\-b&a\end{pmatrix}
\begin{pmatrix}y\\u\end{pmatrix}\ .
\end{equation}
This  equation can be derived by first doing a Poisson resummation over $k$ to go to the Lagrangian description where the modular 
transformation is manifest. Afterwards one again Poisson resums the  windings along the new $t$ cycle to get (\ref{Zmod}). 
Observe that under modular transformations $(x,v)$ mix only among themselves, and likewise $(y,u)$ mix only among themselves.
Note also that $SL(2,\bz)$ descends to $PSL(2,\bz)$ since the element with $a=d=-1$, $b=c=0$ leaves 
$Z_{\Gamma}(\tau,x,y;u,v)$ invariant due to the identity
\begin{equation}
Z_{\Gamma}(\tau,x,y;u,v) =  Z_{\Gamma}(\tau,-x,-y;-u,-v)\ .
\end{equation}
Two special modular transformations of interest are 
\begin{alignat}{3}
 T&: \tau'= \tau+1, \quad
&&Z_{\Gamma}(\tau,x,y;u,v)= e^{2 \pi i x y} Z_{\Gamma}(\tau+1,x,y;u-y,v-x)\ ,
\label{TonZ} \\
 S&: \tau'=-\frac1{\tau}, \quad
&&Z_{\Gamma}(\tau,x,y;u,v)=e^{2 \pi i ( u x+v y)}  \frac{1}{(\tau  \bar{\tau})^{\frac{d}{2}}}   Z_{\Gamma}(-\tf{1}{\tau},v,u;-y,-x)\ .
\label{SonZ}
\end{alignat}

\small\baselineskip=.87\baselineskip
\let\bbb\bibitem\def\bibitem{\itemsep1.5pt\bbb}

\bibliographystyle{utphys}
\bibliography{Z_N_orbifolds}

\providecommand{\href}[2]{#2}\begingroup\raggedright\begin{thebibliography}{10}

\bibitem{Acharya:2019mcu}
B.~S. Acharya, ``{Supersymmetry, Ricci Flat Manifolds and the String
  Landscape},'' \href{http://arxiv.org/abs/1906.06886}{{\tt arXiv:1906.06886
  [hep-th]}}.

\bibitem{Pfaeffle}
F.~Pf{\"a}ffle, ``{The Dirac spectrum of Bieberbach manifolds},'' {\em J. Geom.
  Phys.} {\bf 35} (2000)  367.

\bibitem{McInnes:1998wq}
B.~McInnes, ``{Existence of parallel spinors on nonsimply connected Riemannian
  manifolds},'' \href{http://dx.doi.org/10.1063/1.532293}{{\em J. Math. Phys.}
  {\bf 39} (1998)  2362--2366}.

\bibitem{Witten:1981gj}
E.~Witten, ``{Instability of the Kaluza-Klein Vacuum},''
  \href{http://dx.doi.org/10.1016/0550-3213(82)90007-4}{{\em Nucl. Phys. B}
  {\bf 195} (1982)  481--492}.

\bibitem{GarciaEtxebarria:2020xsr}
I.~Garc{\'{\i}}a~Etxebarria, M.~Montero, K.~Sousa, and I.~Valenzuela,
  ``{Nothing is certain in string compactifications},''
  \href{http://arxiv.org/abs/2005.06494}{{\tt arXiv:2005.06494 [hep-th]}}.

\bibitem{upcoming}
B.~S. Acharya, G.~Aldazabal, E.~Andr\'es, A.~Font, K.~S. Narain, and I.~G.
  Zadeh. {Work in progress}.

\bibitem{Rohm:1983aq}
R.~Rohm, ``{Spontaneous Supersymmetry Breaking in Supersymmetric String
  Theories},'' \href{http://dx.doi.org/10.1016/0550-3213(84)90007-5}{{\em Nucl.
  Phys. B} {\bf 237} (1984)  553--572}.

\bibitem{Kounnas:1989dk}
C.~Kounnas and B.~Rostand, ``{Coordinate Dependent Compactifications and
  Discrete Symmetries},''
  \href{http://dx.doi.org/10.1016/0550-3213(90)90543-M}{{\em Nucl. Phys. B}
  {\bf 341} (1990)  641--665}.

\bibitem{Atick:1988si}
J.~J. Atick and E.~Witten, ``{The Hagedorn Transition and the Number of Degrees
  of Freedom of String Theory},''
  \href{http://dx.doi.org/10.1016/0550-3213(88)90151-4}{{\em Nucl. Phys. B}
  {\bf 310} (1988)  291--334}.

\bibitem{Seiberg:1986by}
N.~Seiberg and E.~Witten, ``{Spin Structures in String Theory},''
  \href{http://dx.doi.org/10.1016/0550-3213(86)90297-X}{{\em Nucl. Phys. B}
  {\bf 276} (1986)  272}.

\bibitem{Dixon:1986iz}
L.~J. Dixon and J.~A. Harvey, ``{String Theories in Ten-Dimensions Without
  Space-Time Supersymmetry},''
  \href{http://dx.doi.org/10.1016/0550-3213(86)90619-X}{{\em Nucl. Phys. B}
  {\bf 274} (1986)  93--105}.

\bibitem{AlvarezGaume:1986jb}
L.~Alvarez-Gaum\'e, P.~H. Ginsparg, G.~W. Moore, and C.~Vafa, ``{An O(16) x
  O(16) Heterotic String},''
  \href{http://dx.doi.org/10.1016/0370-2693(86)91524-8}{{\em Phys. Lett. B}
  {\bf 171} (1986)  155--162}.

\bibitem{Nair:1986zn}
V.~Nair, A.~D. Shapere, A.~Strominger, and F.~Wilczek, ``{Compactification of
  the Twisted Heterotic String},''
  \href{http://dx.doi.org/10.1016/0550-3213(87)90112-X}{{\em Nucl. Phys. B}
  {\bf 287} (1987)  402--418}.

\bibitem{Ginsparg:1986wr}
P.~H. Ginsparg and C.~Vafa, ``{Toroidal Compactification of Nonsupersymmetric
  Heterotic Strings},''
  \href{http://dx.doi.org/10.1016/0550-3213(87)90387-7}{{\em Nucl. Phys. B}
  {\bf 289} (1987)  414}.

\bibitem{Itoyama:1986ei}
H.~Itoyama and T.~Taylor, ``{Supersymmetry Restoration in the Compactified
  ${\rm O(16)}\times {\rm O(16)}'$ Heterotic String Theory},''
  \href{http://dx.doi.org/10.1016/0370-2693(87)90267-X}{{\em Phys. Lett. B}
  {\bf 186} (1987)  129--133}.

\bibitem{Blum:1997gw}
J.~D. Blum and K.~R. Dienes, ``{Strong / weak coupling duality relations for
  nonsupersymmetric string theories},''
  \href{http://dx.doi.org/10.1016/S0550-3213(97)00803-1}{{\em Nucl. Phys. B}
  {\bf 516} (1998)  83--159}, \href{http://arxiv.org/abs/hep-th/9707160}{{\tt
  arXiv:hep-th/9707160}}.

\bibitem{Kachru:1998hd}
S.~Kachru, J.~Kumar, and E.~Silverstein, ``{Vacuum energy cancellation in a
  nonsupersymmetric string},''
  \href{http://dx.doi.org/10.1103/PhysRevD.59.106004}{{\em Phys. Rev. D} {\bf
  59} (1999)  106004}, \href{http://arxiv.org/abs/hep-th/9807076}{{\tt
  arXiv:hep-th/9807076}}.

\bibitem{Harvey:1998rc}
J.~A. Harvey, ``{String duality and nonsupersymmetric strings},''
  \href{http://dx.doi.org/10.1103/PhysRevD.59.026002}{{\em Phys. Rev. D} {\bf
  59} (1999)  026002}, \href{http://arxiv.org/abs/hep-th/9807213}{{\tt
  arXiv:hep-th/9807213}}.

\bibitem{Kachru:1998yy}
S.~Kachru and E.~Silverstein, ``{Selfdual nonsupersymmetric type II string
  compactifications},''
  \href{http://dx.doi.org/10.1088/1126-6708/1998/11/001}{{\em JHEP} {\bf 11}
  (1998)  001}, \href{http://arxiv.org/abs/hep-th/9808056}{{\tt
  arXiv:hep-th/9808056}}.

\bibitem{Kachru:1998pg}
S.~Kachru and E.~Silverstein, ``{On vanishing two loop cosmological constants
  in nonsupersymmetric strings},''
  \href{http://dx.doi.org/10.1088/1126-6708/1999/01/004}{{\em JHEP} {\bf 01}
  (1999)  004}, \href{http://arxiv.org/abs/hep-th/9810129}{{\tt
  arXiv:hep-th/9810129}}.

\bibitem{Font:2002pq}
A.~Font and A.~Hern\'andez, ``{Nonsupersymmetric orbifolds},''
  \href{http://dx.doi.org/10.1016/S0550-3213(02)00336-X}{{\em Nucl. Phys. B}
  {\bf 634} (2002)  51--70}, \href{http://arxiv.org/abs/hep-th/0202057}{{\tt
  arXiv:hep-th/0202057}}.

\bibitem{Blaszczyk:2014qoa}
M.~Blaszczyk, S.~Groot~Nibbelink, O.~Loukas, and S.~Ramos-S\'anchez,
  ``{Non-supersymmetric heterotic model building},''
  \href{http://dx.doi.org/10.1007/JHEP10(2014)119}{{\em JHEP} {\bf 10} (2014)
  119}, \href{http://arxiv.org/abs/1407.6362}{{\tt arXiv:1407.6362 [hep-th]}}.

\bibitem{Abel:2015oxa}
S.~Abel, K.~R. Dienes, and E.~Mavroudi, ``{Towards a nonsupersymmetric string
  phenomenology},'' \href{http://dx.doi.org/10.1103/PhysRevD.91.126014}{{\em
  Phys. Rev. D} {\bf 91} (2015) no.~12, 126014},
  \href{http://arxiv.org/abs/1502.03087}{{\tt arXiv:1502.03087 [hep-th]}}.

\bibitem{Aaronson:2016kjm}
B.~Aaronson, S.~Abel, and E.~Mavroudi, ``{Interpolations from supersymmetric to
  nonsupersymmetric strings and their properties},''
  \href{http://dx.doi.org/10.1103/PhysRevD.95.106001}{{\em Phys. Rev. D} {\bf
  95} (2017) no.~10, 106001}, \href{http://arxiv.org/abs/1612.05742}{{\tt
  arXiv:1612.05742 [hep-th]}}.

\bibitem{Kaidi:2019tyf}
J.~Kaidi, J.~Parra-Martinez, Y.~Tachikawa, and w.~a. m. a. b.~A. Debray,
  ``{Topological Superconductors on Superstring Worldsheets},''
  \href{http://dx.doi.org/10.21468/SciPostPhys.9.1.010}{{\em SciPost Phys.}
  {\bf 9} (2020)  10}, \href{http://arxiv.org/abs/1911.11780}{{\tt
  arXiv:1911.11780 [hep-th]}}.

\bibitem{Itoyama:2020ifw}
H.~Itoyama and S.~Nakajima, ``{Stability, enhanced gauge symmetry and
  suppressed cosmological constant in 9D heterotic interpolating models},''
  \href{http://dx.doi.org/10.1016/j.nuclphysb.2020.115111}{{\em Nucl. Phys. B}
  {\bf 958} (2020)  115111}, \href{http://arxiv.org/abs/2003.11217}{{\tt
  arXiv:2003.11217 [hep-th]}}.

\bibitem{Dixon:1985jw}
L.~J. Dixon, J.~A. Harvey, C.~Vafa, and E.~Witten, ``{Strings on Orbifolds},''
  \href{http://dx.doi.org/10.1016/0550-3213(85)90593-0}{{\em Nucl. Phys. B}
  {\bf 261} (1985)  678--686}.

\bibitem{Dixon:1986jc}
L.~J. Dixon, J.~A. Harvey, C.~Vafa, and E.~Witten, ``{Strings on Orbifolds.
  2.},'' \href{http://dx.doi.org/10.1016/0550-3213(86)90287-7}{{\em Nucl. Phys.
  B} {\bf 274} (1986)  285--314}.

\bibitem{Vafa:1986wx}
C.~Vafa, ``{Modular Invariance and Discrete Torsion on Orbifolds},''
  \href{http://dx.doi.org/10.1016/0550-3213(86)90379-2}{{\em Nucl. Phys. B}
  {\bf 273} (1986)  592--606}.

\bibitem{AlvarezGaume:1986es}
L.~Alvarez-Gaum\'e, G.~W. Moore, and C.~Vafa, ``{Theta Functions, Modular
  Invariance and Strings},'' \href{http://dx.doi.org/10.1007/BF01210925}{{\em
  Commun. Math. Phys.} {\bf 106} (1986)  1--40}.

\bibitem{Dixon:1986qv}
L.~J. Dixon, D.~Friedan, E.~J. Martinec, and S.~H. Shenker, ``{The Conformal
  Field Theory of Orbifolds},''
  \href{http://dx.doi.org/10.1016/0550-3213(87)90676-6}{{\em Nucl. Phys. B}
  {\bf 282} (1987)  13--73}.

\bibitem{Font:2005td}
A.~Font and S.~Theisen, ``{Introduction to string compactification},''
\href{http://dx.doi.org/10.1007/11374060_3}{{\em Lect. Notes Phys.} {\bf 668}
  (2005)  101--181}.

\bibitem{Sakai:1985cs}
N.~Sakai and I.~Senda, ``{Vacuum Energies of String Compactified on Torus},''
  \href{http://dx.doi.org/10.1143/PTP.75.692}{{\em Prog. Theor. Phys.} {\bf 75}
  (1986)  692}. [Erratum: Prog.Theor.Phys. 77, 773 (1987)].

\bibitem{Erler:1992ki}
J.~Erler and A.~Klemm, ``{Comment on the generation number in orbifold
  compactifications},'' \href{http://dx.doi.org/10.1007/BF02096954}{{\em
  Commun. Math. Phys.} {\bf 153} (1993)  579--604},
  \href{http://arxiv.org/abs/hep-th/9207111}{{\tt arXiv:hep-th/9207111}}.

\bibitem{Fischer:2012qj}
M.~Fischer, M.~Ratz, J.~Torrado, and P.~K. Vaudrevange, ``{Classification of
  symmetric toroidal orbifolds},''
  \href{http://dx.doi.org/10.1007/JHEP01(2013)084}{{\em JHEP} {\bf 01} (2013)
  084}, \href{http://arxiv.org/abs/1209.3906}{{\tt arXiv:1209.3906 [hep-th]}}.

\bibitem{Vafa:2001ra}
C.~Vafa, ``{Mirror symmetry and closed string tachyon condensation},''
  \href{http://arxiv.org/abs/hep-th/0111051}{{\tt arXiv:hep-th/0111051}}.

\bibitem{Harvey:2001wm}
J.~A. Harvey, D.~Kutasov, E.~J. Martinec, and G.~W. Moore, ``{Localized
  tachyons and RG flows},'' \href{http://arxiv.org/abs/hep-th/0111154}{{\tt
  arXiv:hep-th/0111154}}.

\bibitem{Sen:1998tt}
A.~Sen, ``{SO(32) spinors of type I and other solitons on brane - anti-brane
  pair},'' \href{http://dx.doi.org/10.1088/1126-6708/1998/09/023}{{\em JHEP}
  {\bf 09} (1998)  023}, \href{http://arxiv.org/abs/hep-th/9808141}{{\tt
  arXiv:hep-th/9808141}}.

\bibitem{Adams:2001sv}
A.~Adams, J.~Polchinski, and E.~Silverstein, ``{Don't panic! Closed string
  tachyons in ALE space-times},''
  \href{http://dx.doi.org/10.1088/1126-6708/2001/10/029}{{\em JHEP} {\bf 10}
  (2001)  029}, \href{http://arxiv.org/abs/hep-th/0108075}{{\tt
  arXiv:hep-th/0108075}}.

\bibitem{David:2001vm}
J.~R. David, M.~Gutperle, M.~Headrick, and S.~Minwalla, ``{Closed string
  tachyon condensation on twisted circles},''
  \href{http://dx.doi.org/10.1088/1126-6708/2002/02/041}{{\em JHEP} {\bf 02}
  (2002)  041}, \href{http://arxiv.org/abs/hep-th/0111212}{{\tt
  arXiv:hep-th/0111212}}.

\bibitem{Brezhnev:2013}
Y.~V. Brezhnev, ``{Non-canonical extension of $\vartheta$-functions and modular
  integrability of $\vartheta$-constants},''
  \href{http://dx.doi.org/10.1017/S0308210512001023}{{\em {Proceedings of the
  Royal Society of Edinburgh A}} {\bf 143} (2013) no.~4, 689--738},
  \href{http://arxiv.org/abs/1011.1643}{{\tt arXiv:1011.1643 [math.CA]}}.

\end{thebibliography}\endgroup

\end{document}